\documentclass[aps,prd,onecolumn,showpacs,10pt,superscriptaddress,preprintnumbers,nofootinbib]{revtex4-2}
\usepackage{amsmath,amssymb,bm}\allowdisplaybreaks[1]
\usepackage{graphicx}
\usepackage{physics}
\usepackage{xcolor}
\usepackage[colorlinks, linkcolor=blue!80!black, citecolor=green!50!black, urlcolor=magenta]{hyperref}
\usepackage[normalem]{ulem}
\usepackage{adjustbox}
\usepackage{orcidlink}
\usepackage{tikz}
\usetikzlibrary{arrows.meta}
\usepackage{enumitem}

\renewcommand{\sec}[1]{Sec.~\ref{#1}}
\newcommand{\fig}[1]{Fig.~\ref{#1}}

\newcommand{\eq}[1]{Eq.~\eqref{#1}}
\newcommand{\eqs}[2]{Eqs.~\eqref{#1} and~\eqref{#2}}
\newcommand{\eqss}[3]{Eqs.~\eqref{#1}~\eqref{#2} and~\eqref{#3}}
\newcommand{\refcite}[1]{Ref.~\cite{#1}}
\newcommand{\refs}[1]{Refs.~\cite{#1}}
\newcommand*{\Scale}[2][4]{\scalebox{#1}{\ensuremath{#2}}}

\newcommand{\pp}[1]{\left(#1\right)}
\newcommand{\bb}[1]{\left[#1\right]}
\newcommand{\cc}[1]{\left\{#1\right\}}
\newcommand{\vv}[1]{\left\langle #1 \right\rangle}
\newcommand{\bigpp}[1]{\big(#1\big)}
\newcommand{\bigbb}[1]{\big[#1\big]}
\newcommand{\bigcc}[1]{\big\{#1\big\}}
\newcommand{\bigvv}[1]{\big\langle #1 \big\rangle}
\newcommand{\biggpp}[1]{\bigg(#1\bigg)}

\newcommand{\beq}[1][]{\begin{equation}\label{#1}}
\newcommand{\eeq}{\end{equation}}
\newcommand{\bse}[1][]{\begin{subequations}\label{#1}}
\newcommand{\ese}{\end{subequations}}
\newcommand{\nn}{\nonumber}

\renewcommand{\slash}[1]{#1 \hspace{-0.45em} / }
\newcommand{\Slash}[1]{#1 \hspace{-0.66em} / }

\renewcommand{\P}{\mathcal{P}}
\renewcommand{\O}{\mathcal{O}}
\newcommand{\wt}[1]{\widetilde{#1}}
\newcommand{\sgn}[1]{{\rm sgn}\left(#1\right)}

\newcommand{\M}{\mathcal{M}}

\newcommand{\F}{\mathcal{F}}
\renewcommand{\H}{\mathcal{H}}
\newcommand{\Ft}{\widetilde{\mathcal{F}}}

\def\A{\mathcal{A}}

\def\M{\mathcal{M}}

\def\P{\mathcal{P}}
\def\F{\mathcal{F}}
\def\Ft{\widetilde{\mathcal{F}}}
\newcommand{\G}{\mathcal{G}}

\def\H{\mathcal{H}}
\def\Ht{\widetilde{\mathcal{H}}}
\def\E{\mathcal{E}}
\def\Et{\widetilde{\mathcal{E}}}
\def\l{\ell}

\begin{document}

\preprint{
	{\vbox {			
		\hbox{\bf JLAB-THY-25-4591}
}}}
\vspace*{0.2cm}

\title{New framework for extracting GPDs from exclusive photon electroproduction}

\author{Jian-Wei Qiu\,\orcidlink{0000-0002-7306-3307}}
\email{jqiu@jlab.org}
\affiliation{Theory Center, Jefferson Lab,
Newport News, Virginia 23606, USA}
\affiliation{Department of Physics, William \& Mary,
Williamsburg, Virginia 23187, USA}

\author{Nobuo Sato\,\orcidlink{0000-0002-1535-6208}}
\email{nsato@jlab.org}
\affiliation{Theory Center, Jefferson Lab,
Newport News, Virginia 23606, USA}

\author{Zhite Yu\,\orcidlink{0000-0003-1503-5364}}
\email{yuzhite@jlab.org (corresponding author)}
\affiliation{Theory Center, Jefferson Lab, 
Newport News, Virginia 23606, USA}

\date{\today}

\begin{abstract}
Recently, a new framework for studying generic $2 \to 3$ hard exclusive reactions, referred to as single-diffractive hard exclusive processes, has been introduced to provide a cleaner separation of the underlying physical mechanisms. In this work, we expand this formalism to the case of exclusive real-photon electroproduction off a nucleon, $e(\ell) + N(p) \to e(\ell') + N(p') + \gamma(q')$, which represents the classical channel for accessing generalized parton distributions (GPDs) in nucleons and nuclei. This extension enables a more systematic and physically transparent formulation of the reaction dynamics, paving the way for improved extractions of GPDs from experimental data as compared to existing approaches. 
\end{abstract}

\maketitle
\tableofcontents

\section{Introduction}
\label{sec:intro}

Understanding how quarks and gluons are confined inside hadronic systems remains one of the most fascinating and challenging open problems in modern nuclear physics. 
Although a complete theoretical explanation of confinement is still out of reach, major experimental advances over the past several decades have made it possible to probe the internal quark and gluon structure of hadrons through hard exclusive reactions. 
Such information is formulated in terms of generalized parton distributions (GPDs), first introduced in \refs{Muller:1994ses, Ji:1996ek, Radyushkin:1997ki}, which describe light-cone parton correlations in off-forward hadron kinematics (for reviews, see \refs{Goeke:2001tz, Diehl:2003ny, Belitsky:2005qn, Boffi:2007yc}). 
GPDs provide a combined description of transverse spatial and longitudinal momentum distributions of partons, thereby enabling a three-dimensional tomography of the nucleon and, more generally, of confined hadronic systems. 
They also open a unique window into emergent hadron properties, offering direct connections to the partonic interpretation of the proton’s mass~\cite{Ji:1994av, Ji:1995sv, Lorce:2017xzd, Metz:2020vxd}, spin~\cite{Ji:1996ek}, and internal pressure and shear force distributions~\cite{Polyakov:2002yz, Polyakov:2018zvc, Burkert:2018bqq}.

Decades of theoretical and experimental efforts have established the accessibility of GPDs across a broad class of exclusive reactions~\cite{Ji:1996nm, Radyushkin:1997ki, Brodsky:1994kf, Frankfurt:1995jw, Berger:2001xd, Guidal:2002kt, Belitsky:2002tf, Belitsky:2003fj, Kumano:2009he, Kumano:2009ky, ElBeiyad:2010pji, Pedrak:2017cpp, Pedrak:2020mfm, Siddikov:2022bku, Siddikov:2023qbd, Boussarie:2016qop, Duplancic:2018bum, Duplancic:2022ffo, Duplancic:2023kwe, Qiu:2022bpq, Qiu:2023mrm, Qiu:2024mny}. 
This progress has been made possible by the development of QCD factorization theorems~\cite{Collins:1996fb, Collins:1998be, Ji:1998xh, Qiu:2022pla}, which provide a systematic framework to express the scattering amplitudes in terms of universal GPDs convoluted with perturbatively calculable hard coefficients.
Historically, the deeply virtual Compton scattering (DVCS)~\cite{Ji:1996nm, Radyushkin:1997ki} has been regarded as the ``golden channel'' for accessing GPDs 
among all the exclusive processes. 
It has been extensively measured at HERA~\cite{H1:2001nez, H1:2005gdw, H1:2007vrx, H1:2009wnw, ZEUS:2003pwh, ZEUS:2008hcd, HERMES:2001bob, HERMES:2006pre, HERMES:2008abz, HERMES:2009cqe, HERMES:2010dsx, HERMES:2011bou, HERMES:2012gbh}, 
COMPASS~\cite{COMPASS:2018pup}, 
and Jefferson Lab~\cite{CLAS:2001wjj, CLAS:2006krx, JeffersonlabHallA:2006prd, JeffersonlabHallA:2007jdm, CLAS:2007clm, CLAS:2008ahu, CLAS:2014qtk, CLAS:2015bqi, JeffersonlabHallA:2015dwe, CLAS:2015uuo, Defurne:2017paw, CLAS:2017udk, Benali:2020vma, CLAS:2021ovm, JeffersonlabHallA:2022pnx, CLAS:2022syx, CLAS:2024qhy}, 
and will be a central measurement program at the future Electron-Ion Collider (EIC)~\cite{Accardi:2012qut} 
and potentially also the Electron-ion collider in China (EicC)~\cite{Anderle:2021wcy}.

A key challenge in reconstructing GPDs from processes such as DVCS is the presence of background contributions, most notably from the Bethe-Heitler (BH) subprocess. 
In experimental analyses, these backgrounds are typically subtracted from the measured cross section~\cite{Belitsky:2001ns, Shiells:2021xqo, Kriesten:2019jep, Kriesten:2020apm} in order to isolate the DVCS signal, which is then expressed in terms of azimuthal modulations that can be related to GPDs. 
However, because the BH contribution itself induces nontrivial azimuthal structure in general, the extracted harmonic coefficients can inherit distortions that are not purely associated with the DVCS amplitude. 
This mixing of kinematic and dynamical effects ultimately limits the precision and model independence of GPD extractions.

Recently, in \refcite{Qiu:2024reu} we introduced a general framework for analyzing all $2 \to 3$ single-diffractive hard exclusive processes (SDHEPs), based on an optimized choice of reference frame that permits the measurement of azimuthal modulations directly sensitive to GPDs. 
This formulation incorporates a systematic power expansion in the two hierarchical scales intrinsic to SDHEPs and enables a clean separation of the contributions arising from different underlying dynamics, including those directly related to GPDs and those that are not. 
In this work, we specialize the SDHEP formalism to the case of exclusive photon electroproduction, 
which is a physically measured process containing the BH and DVCS as two subprocesses. 
We also contrast this framework with the traditional approach formulated in the Breit frame, highlighting how the SDHEP framework provides a cleaner and more systematic pathway for GPD extraction and phenomenological analysis.

The rest of this paper is organized as follows.
Before delving into details, we first give a general consideration of the kinematics and frame choice in \sec{sec:frame} of the exclusive photon electroproduction process.
After reviewing in \sec{sec:review} the conventional approach formulated in the Breit frame, 
we will move on to reformulate it in the SDHEP framework in \sec{sec:dvcs-sdhep}
and present the leading-order (LO) amplitude calculation in \sec{sec:calc-sdhep} for both the BH and the DVCS subprocesses. 
Combining and squaring these two amplitudes in obtaining the cross section gives eight polarization asymmetries associated with a set of well defined azimuthal modulations,
which will be the main practical result of this paper.
In \sec{sec:unique}, we will make a further comparison between the Breit frame and SDHEP frame,
which will provide more insights of the azimuthal structure and demonstrate the uniqueness of the both frames. 
In \sec{sec:conclusion}, we summarize the main results of this paper and give our conclusion.

\begin{figure}[tbp]
\centering
	\begin{tabular}{cc}
	\includegraphics[scale=0.55]{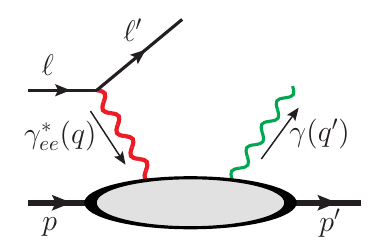} &
		\includegraphics[scale=0.55]{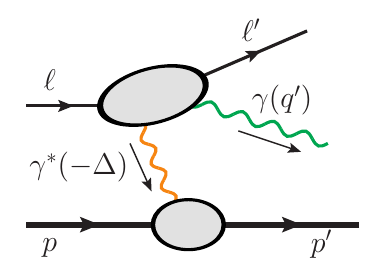} \\
	(a) & (b)
	\end{tabular}
	\caption{(a) DVCS and (b) BH subprocesses that make up the photon electroproduction at the LO of QED.}
\label{fig:breit}
\end{figure}

\section{General overview of the kinematic frame}
\label{sec:frame}

We first give a general discussion of the kinematics of the exclusive photon electroproduction process,
\beq[eq:dvcs-sdhep]
	N(p) + e(\ell) \to N(p') + e(\ell') + \gamma(q')\,.
\eeq
Here, $p$ and $p'$ denote the four-momenta of the incoming and outgoing (diffracted) nucleon, respectively, and $\ell$ and $\ell'$ are the four-momenta of the incoming and scattered lepton. The vector $q'$ represents the momentum of the radiated final-state photon. 
Without loss of generality, the process can be decomposed into two contributing subprocesses: 
the DVCS and the BH mechanisms, as illustrated in Fig.~\ref{fig:breit}.

Focusing on the DVCS subprocess shown in Fig.~\ref{fig:breit}(a), the reaction can be viewed as occurring in two stages,
\bse\label{eq:one-photon}
\begin{align}
	&e(\ell) \to \gamma_{ee}^*(q = \ell - \ell') + e(\ell'),  \\
    &\hspace{10ex} 
    \begin{tikzpicture}
        \node[inner sep=0pt] (arrow) at (0, 0) {
            \tikz{\draw[->, >={Stealth}, double, double distance=1pt, line width=1pt] (0, 0.28) to [out=-90, in=180] (0.8, 0);}
        };
    \end{tikzpicture}
    \hspace{1ex}
    \gamma_{ee}^*(q) + N(p) \to \gamma(q') + N(p'),  \label{eq:compton}
\end{align}
\ese
where the virtual photon $\gamma_{ee}^*(q)$ is emitted by the scattered electron and carries a large spacelike virtuality, 
$Q^2 = -q^2 \gg \Lambda_{\rm QCD}^2$. 
Following its kinematic similarity to the inclusive lepton-hadron deep inelastic scattering (DIS) or semi-inclusive DIS (SIDIS),
the DVCS has been conventionally analyzed in the Breit frame,
where the $\gamma_{ee}^*$ moves along the $-z$ direction to hit the nucleon $N(p)$, which is at rest or moves along the $+z$ direction.
This frame is convenient for proofs of factorization~\cite{Collins:1996fb, Collins:1998be}, with a natural hard scale provided by the virtuality $Q$.
It also allows for the Compton form factor (CFF) decomposition~\cite{Belitsky:2001ns, Belitsky:2005qn}
and helicity amplitude description~\cite{Kroll:1995pv, Diehl:1997bu, Diehl:2005pc, Belitsky:2008bz, Belitsky:2012ch, Braun:2012hq, Goldstein:2010gu, Kriesten:2019jep}, 
which offer a powerful analysis method to employ the azimuthal modulations associated with the nucleon and virtual photon polarizations~\cite{Diehl:1997bu},
thus allowing to disentangle different types of GPDs.
There has been intensive study on extracting the CFFs from DVCS in this frame~\cite{Belitsky:2001ns, Guidal:2008ie, Guidal:2009aa, Guidal:2010ig, Guidal:2010de, Guidal:2013rya, Boer:2014kya, Kumericki:2013br, Kumericki:2007sa, Kumericki:2009uq, Kumericki:2011rz, Goldstein:2010gu, Kumericki:2016ehc, Kriesten:2019jep, Kriesten:2020wcx, Kriesten:2020apm, Almaeen:2022imx, Almaeen:2024guo, Guo:2021gru, Shiells:2021xqo, Grigsby:2020auv, Moutarde:2019tqa, Cuic:2020iwt}.

We argue, however, that the use of the Breit frame is not optimal for studying exclusive photon electroproduction processes, owing to the difficulty in describing the interference between the DVCS and BH amplitudes. In the Breit frame, the coordinate system is defined around the virtual photon momentum of the DVCS subprocess, which does not align event by event with the virtual photon exchanged in the BH subprocess. In the latter, the electron $e(\ell)$ emits the real photon $\gamma(q')$ \emph{and} a low-virtuality photon $\gamma^*(p' - p)$ that interacts with the nucleon. This kinematic misalignment between the two subprocesses complicates the data analysis and interpretation of azimuthal modulations in terms of GPDs.

The core of the difficulty arises when evaluating the BH contribution in the Breit frame, which induces intricate azimuthal dependence in both the numerators and denominators of the BH amplitude~\cite{Belitsky:2001ns}. Because the BH process typically dominates over the DVCS contribution, experimental access to GPDs relies on isolating their interference. Following the conventional procedure~\cite{Belitsky:2001ns, Shiells:2021xqo, Kriesten:2019jep, Kriesten:2020apm}, one subtracts the theoretically computed BH cross section from the measured data, fits the remaining signal to a harmonic expansion in the azimuthal angle $\varphi$, and extracts the GPD information from the corresponding Fourier coefficients. The complex $\varphi$ dependence of the BH term therefore poses major challenges for experimental analyses and can introduce model-dependent assumptions at each stage of the extraction.

We can provide an intuitive understanding for the underlying cause of the problem.
In DVCS, the virtual photon $\gamma^*_{ee}(q)$ connects the leptonic and hadronic planes, with the final-state real photon $\gamma(q')$ lying in the hadronic plane as part of the Compton scattering amplitude. It is thus natural to define the azimuthal angle $\varphi$ between these two planes. In contrast, the BH process involves no such virtual photon $\gamma^*_{ee}$ to establish a fixed geometric relation between the leptonic and hadronic planes, which leads to  a nontrivial $\varphi$ dependence that emerges from the BH subprocess.

\begin{figure}[htbp]
	\def\sc{1.5}
	\centering
	\begin{align*}
		\adjincludegraphics[valign=c, scale=0.55]{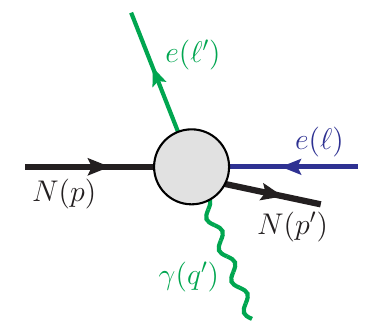}	
			\quad\Scale[\sc]{=}\quad
		\adjincludegraphics[valign=c, scale=0.55]{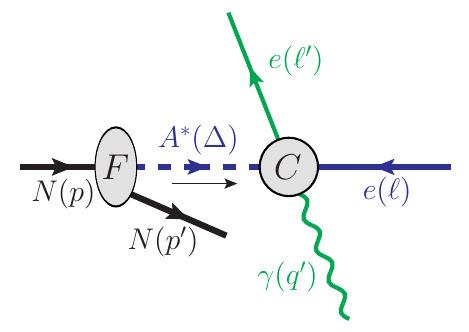} 
	\end{align*}
	\caption{Two-stage picture of the photon electroproduction process.
	}
	\label{fig:dvcs-sdhep}
\end{figure}

On the other hand, as introduced in \refcite{Qiu:2024reu}, the SDHEP frame provides a better choice of frame that allows to bypass the issues mentioned by considering kinematic regions where two distinct momentum scales $|(\ell-\ell')^2| \gg |(p-p')^2|$ are present (\refs{Qiu:2022pla, Qiu:2023mrm, Qiu:2024mny}).
Specifically, when the final-state electron and photon are produced from a hard scattering at the scale $|(\ell-\ell')^2|$,  the nucleon only undergoes a diffraction at a soft scale $|(p-p')^2|$.
This sets a clear separation of the timescales in the full process, motivating us to picture it as happening in two stages,
\footnote{Note that our convention for $\Delta$ differs from the normal one (e.g., in \refcite{Diehl:2003ny}) by a minus sign.
This change is to better coincide with the two-stage picture.}
\bse\label{eq:two-stage}\begin{align}
	&N(p) \to A^*(\Delta = p - p') + N(p'), 	\label{eq:diffractive}\\
    &\hspace{10ex} \begin{tikzpicture}
        \node[inner sep=0pt] (arrow) at (0, 0) {
            \tikz{\draw[->, >={Stealth}, double, double distance=1pt, line width=1pt] (0, 0.28) to [out=-90, in=180] (0.8, 0);}
        };
    \end{tikzpicture} \hspace{1ex}
    A^*(\Delta) + e(\l) \to e(\l') + \gamma(q'),  \label{eq:hard 2to2}
\end{align}\ese
as sketched in \fig{fig:dvcs-sdhep}.
First, the nucleon diffraction emits a virtual state $A^*$ with momentum $\Delta = p - p'$, which
has a high energy but low virtuality $t = \Delta^2$ in the center-of-mass (c.m.) frame of the collision.  
Then we can treat $A^*$ as a quasireal particle 
that propagates for a {\it long} lifetime before participating in the 
{\it short} hard collision with the electron beam to produce a real photon.
This picture directly encodes the factorization of the two subprocesses in \eq{eq:two-stage},
because quantum interference from soft gluon interactions between these two subprocesses 
is suppressed by powers of $|(p-p')^2|/|(\ell-\ell')^2|$,
as fully justified in \refs{Ji:1998xh, Collins:1998be, Collins:1996fb, Qiu:2022pla}. 
The resulting theoretical treatment is expected to be a good approximation when the transverse components of $\l'$ and $q'$ 
with respect to the $N$-$e$ collision axis
are much greater than the invariant mass of $A^*$,
i.e., $\l'_{T} \sim q'_{T} \gg \sqrt{-t}.$ 

With this two-stage picture in mind, the full amplitude of the electroproduction process in Eq.~\eqref{eq:dvcs-sdhep} can be expressed as a sum over all possible intermediate channels in $A^*$.
The leading channel corresponds to $A^* = \gamma^*$, which is simply the BH subprocess,
while in the subleading channels, $A^*$ starts with a quark-antiquark pair $[q\bar{q}]$ or a gluon-gluon pair $[gg]$,
which corresponds to the ``leading-twist'' contribution of the DVCS. 
Since the virtual photon $\gamma^*_{ee}$ in the DVCS has a high virtuality, we include it in the hard part.
In this way, instead of focusing on the high-virtuality $\gamma^*_{ee}$ of the DVCS, 
the new two-stage picture centers around the low-virtuality exchange state $A^*$ and 
sets the BH and DVCS on the same coherent framework. 
The distribution of the azimuthal angle $\phi$ between the nucleon diffraction plane [of \eq{eq:diffractive}] and the hard scattering plane [of \eq{eq:hard 2to2}]
will then be determined by the interference of amplitudes corresponding to different spin states of $A^*$.  
Since the quantum numbers of the $A^*$ effectively define the GPDs, 
the observed $\phi$ distributions in this frame are directly related to contributions from various GPDs, 
and the azimuthal modulations entail clear physical interpretations.
In particular, we will see that the interference of BH and DVCS at leading power (LP) contributes eight polarization asymmetries 
with $\cos\phi$ or $\sin\phi$ modulations, which exactly correspond to the eight real degrees of freedom of the GPD integrals.
Therefore, the new framework retains the original Breit frame advantage of 
using azimuthal modulations to disentangle different GPDs.
Furthermore, as we will see, such a coherent description of the photon electroproduction
is a unique feature of the SDHEP frame~\cite{Qiu:2024reu}.

Having discussed the frame choices for studying DVCS and BH, in the next Section we will review the conventional theory treatment to connect GPDs and data using the Breit-frame approach. We will transition into the SDHEP frame in \sec{sec:dvcs-sdhep}.

\section{Review of the conventional approach}
\label{sec:review}

\begin{figure}[htbp]
\centering
	\includegraphics[scale=0.55]{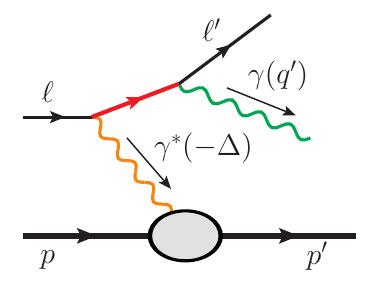}
	\includegraphics[scale=0.55]{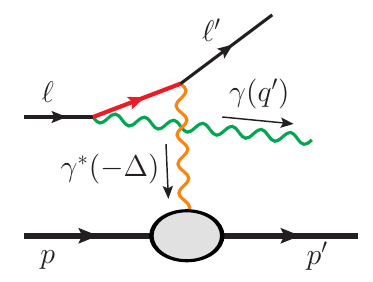}
	\caption{Diagrams of the BH subprocess at the LO of QED.}
\label{fig:breit-bh}
\end{figure}

Conventional study of the real photon electroproduction process in \eq{eq:dvcs-sdhep}
has been maximizing the analogy to DIS or SIDIS
and classifies the whole process into the DVCS and BH subprocesses,
as shown in \fig{fig:breit}.
The amplitudes of both subprocesses can be written as a leptonic tensor contracted with a hadronic one.
Let us write the full amplitude of \eq{eq:dvcs-sdhep} as
\beq[eq:dvcs-bh-amp]
	\M = \M^{\rm BH} + \M^{\rm DVCS}.
\eeq
The BH part probes the elastic scattering of the nucleon,
\beq[eq:BH-amplitude]
	\M^{\rm BH}
	= \frac{i e^3}{t} \, L^{\mu\nu} \, \epsilon^*_{\nu}(q') 
		F_{\mu}(p, p'),
\eeq
where the leptonic tensor is given in \fig{fig:breit-bh},
\begin{align}
	L^{\mu\nu} = \bar{u}(\l') 
		\bb{ \gamma^{\mu} \frac{\gamma\cdot (\l - q')}{(\l - q')^2} \gamma^{\nu} 
			+ \gamma^{\nu} \frac{\gamma\cdot (\l + \Delta)}{(\l + \Delta)^2} \gamma^{\mu} 
		} u(\l).
\label{eq:BH-lepton-tensor}
\end{align}
The hadronic vector $F_{\mu}$ is given by the matrix element of the 
electromagnetic (EM) current $J_{\mu} = \sum_{q = u, d, \ldots} e_q \bar{\psi}_q \gamma_{\mu} \psi_q$,
which is normalized by $(e_u, e_d) = (2/3, -1/3)$,
and is parametrized by the EM form factors $F_1(t)$ and $F_2(t)$,
\begin{align}
	F_{\mu}(p, p') = \langle N(p', \lambda_N') | J_{\mu}(0) | N(p, \lambda_N) \rangle	
		= \bar{u}(p', \lambda_N') \bb{ F_1(t) \, \gamma_{\mu} - F_2(t) \, \frac{i\sigma_{\mu\nu} \Delta^{\nu}}{2m} } u(p, \lambda_N),
\label{eq:EM-form-factor}
\end{align}
with $m$ being the nucleon mass and $\lambda_N$ and $\lambda_N'$ labeling the spin states.
The DVCS part probes the virtual Compton scattering $N(p) + \gamma_{ee}^*(q) \to N(p') + \gamma(q')$,
\beq[eq:DVCS-amp]
	\M^{\rm DVCS}
	= \frac{e^3}{Q^2} \, \bar{u}(\l') \gamma_{\mu} u(\l) \, \epsilon^*_{\nu}(q') \, i T^{\mu\nu}(p, p', q, q'),
\eeq
where we have included a $-1$ factor for the electron charge and $T^{\mu\nu}$ is the virtual Compton tensor, 
\beq[eq:compton-amp]
	i T^{\mu\nu}(p, p', q, q')
	= \int d^4 z \, e^{i (q + q') \cdot z / 2} \vv{ N(p', \lambda_N') | \mathcal{T} J^{\nu}(z / 2) J^{\mu}(-z / 2) | N(p, \lambda_N) }.
\eeq
In the above and in what follows, we define
momenta of the two virtual photons, $\gamma^*$ and $\gamma^*_{ee}$,
\beq[eq:breit-Q-t]
	\Delta = p - p', \quad
	q = \l - \l', \quad
	t = \Delta^2, \quad
	Q^2 = - q^2.
\eeq

Both the EM form factors and virtual Compton tensor are nonperturbative objects.
In the kinematic limit of our interest, $Q^2 \gg |t|$,
the EM form factors probe the global properties of the nucleon.
It is the virtual Compton tensor that encodes internal partonic structures of the nucleon.
Extracting this is therefore the central goal in experimental studies of the photon electroproduction process.

\subsection{Compton form factors in the DVCS}
\label{ssec:cffs}

Aside from being exclusive, the field operator in \eq{eq:compton-amp} for DVCS is the same as that of DIS or SIDIS.
So similarly to \refs{Bacchetta:2006tn, Bacchetta:2008xw}, 
we can decompose the virtual Compton tensor $T^{\mu\nu}$
into independent gauge-invariant structure functions, or CFFs.
This can be achieved by Lorentz and parity symmetries and gauge invariance, 
\beq[eq:T-conserve]
	q_{\mu} T^{\mu\nu} = T^{\mu\nu} q'_{\nu} = 0,
\eeq 
resulting in 18 CFFs, $(\H_i, \E_i)$ for $i = 1, \ldots, 5$, and $(\wt{\H}_i, \wt{\E}_i)$ for $i = 1, \ldots, 4$,
\begin{align}
	T^{\mu\nu}
	& = \frac{1}{2 P \cdot R} \bar{u}(p', \lambda_N') 
		\bb{ \sum_{i = 1}^5 
				\pp{ \H_i \, \gamma\cdot R - \E_i \frac{i \sigma^{R \Delta}}{2m} }
				\tau_i^{\mu\nu}
			+ \sum_{i = 1}^4
				\pp{ \wt{\H}_i \, \gamma\cdot R \gamma_5 - \wt{\E}_i \, \gamma_5 \frac{R \cdot \Delta}{2m} }
				\wt{\tau}_i^{\mu\nu}  
		} u(p, \lambda_N).
\label{eq:CFF-decomp}
\end{align}
These CFFs are dimensionless scalars that can only depend on three independent momentum variables. 
By choosing them as
\beq[eq:kin-P-R]
	P = (p + p') / 2, \quad
	\Delta = p - p' = q' - q, \quad
	R = (q + q') / 2,
\eeq
with the corresponding scalar products,
\begin{align}
	P^2 = m^2 - t / 4, \quad
	\Delta^2 = t, \quad
	R^2 = - (2 Q^2 + t) / 4, \quad
	\Delta \cdot R = (q' - q) \cdot (q' + q) / 2 = Q^2 / 2, \quad
	\mbox{and }
	P \cdot R,
\label{eq:cff-args}
\end{align}
we can parametrize the CFFs as functions of $t / m^2$, $t / Q^2$, and
\beq
    \eta = \Delta \cdot R / (2 P \cdot R) = Q^2 / (4 P \cdot R),
\eeq
so that
\beq[eq:cffs-exprs]
	\bigcc{ \H_i, \E_i, \wt{\H}_i, \wt{\E}_i } = \bigcc{ \H_i, \E_i, \wt{\H}_i, \wt{\E}_i }(\eta, t / m^2, t / Q^2).
\eeq

The detailed derivation of \eq{eq:CFF-decomp} is given in Appendix~\ref{app:cffs}, generalizing the method in \refs{Diehl:2001pm, Meissner:2009ww}.
The independent tensors $\tau_i^{\mu\nu}$ and $\wt{\tau}_i^{\mu\nu}$ are given in \eqs{eq:p-even-tau}{eq:p-odd-tau}, respectively,
using the ``transverse basis'' constructed on top of $q$ and $q'$ in \eq{eq:trans-q-q2} (see also \refs{Braun:2012bg, Braun:2012hq, Braun:2014sta}).
Each of them satisfies the current conservation in \eq{eq:T-conserve} individually.
Only 12 of these 18 structures, with $i = 1, 2, 3$, contribute to physical observables, 
because $\tau_4$, $\tau_5$, and $\wt{\tau}_4$ are proportional to $q^{\prime \nu}$,
which vanishes when contracting with the physical polarization vector $\epsilon^*_{\nu}(q')$.

The general decomposition of $T^{\mu\nu}$ was first done in \refcite{Tarrach:1975tu}. 
\refs{Belitsky:2001ns, Belitsky:2005qn} also gave different decompositions inspired by perturbative calculations.
Our strategy is more similar to \refcite{Tarrach:1975tu} but the result 
differs by being explicitly gauge invariant and having a direct correspondence to the GPD decompositions in \eq{eq:GPD-def-q} below.
Following \eq{eq:CFF-decomp}, the definitions of CFFs are purely nonperturbative,
without mixing with the argument of twist and perturbation expansions such as in \refs{Belitsky:2001ns, Guo:2021gru, Kriesten:2019jep}.
This result does not require introducing the auxiliary vectors $n$ and $\bar{n}$, and is valid to all powers of $t / Q^2$ and all orders in $\alpha_s$.

\subsection{Factorization of CFFs and perturbative calculation}
\label{ssec:cffs-calc}

\begin{figure}[htbp]
	\def\sc{1.5}
	\centering
	\begin{align*}
		\adjincludegraphics[valign=c, scale=0.55]{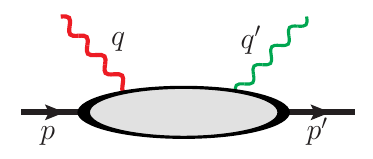}	
		&\Scale[\sc]{=}
			\adjincludegraphics[valign=c, scale=0.55]{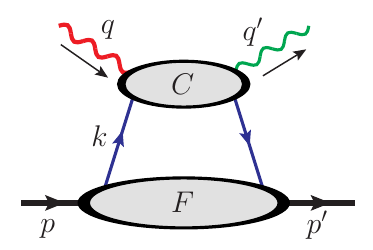} 
			\Scale[\sc]{+}
			\adjincludegraphics[valign=c, scale=0.55]{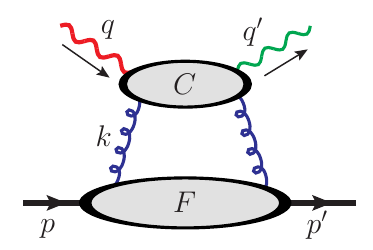} 
		\\
		& \Scale[\sc]{+}
			\adjincludegraphics[valign=c, scale=0.55]{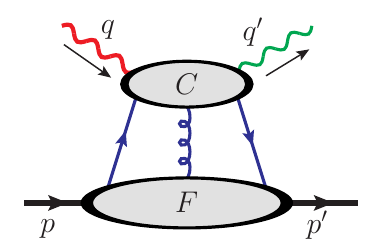}
			\Scale[\sc]{+}
			\adjincludegraphics[valign=c, scale=0.55]{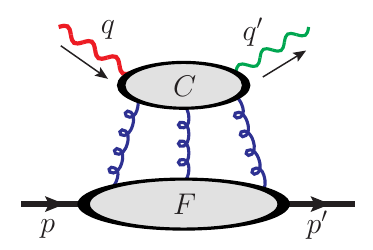}
			\Scale[\sc]{+ \cdots}
	\end{align*}
	\caption{Region decomposition of the virtual Compton amplitude.
	Partons connecting the collinear part $F$ to the hard part $C$ are collinear to the nucleons. 
	The ``$\cdots$'' stands for regions with more than three partons connecting $C$ and $F$. 
    We have implicitly chosen a physical gauge such that all the gluons displayed are transversely polarized. In a covariant gauge, each diagram can contain an arbitrary number of longitudinally polarized gluons.
	}
	\label{fig:breit-dvcs-twist}
\end{figure}

Analogous to the structure functions in DIS, the nonperturbative CFFs can 
be extracted experimentally without reliance on a specific theoretical treatment of the hadronic part.
On the other hand, in the deeply virtual kinematic region, $Q^2 \gg |t|$, 
leading contributions to the CFFs correspond to the dynamics with the partons interacting with the photons.
With the large momentum transfer $Q$, we can apply the short-distance operator product expansion to 
the two-current operator that defines the DVCS in \eq{eq:compton-amp}, 
which decomposes the virtual Compton amplitude to 
a hard part $C$ associated with the photons and a collinear part $F$ associated with the nucleons,
joined by a set of parton lines collinear to the nucleons,
as shown in \fig{fig:breit-dvcs-twist}. 
These two parts can be factorized to yield a set of universal parton correlation functions,
which encode information of partonic structures.

The factorization is most conveniently done 
in the ``photon frame'' where both $q$ and $q'$ move along the $-\hat{z}$ direction,
with the partons exchanged between $C$ and $F$ carrying a total momentum $\Delta$ boosted along the $\hat{z}$ direction.
The two transverse directions $X^{\mu} = (0, \hat{x})$ and $Y^{\mu} = (0, \hat{y})$ are given by 
$X^{\mu} \propto P_{\perp}^{\mu}$ and $Y^{\mu} \propto \epsilon^{\mu P q q'}$,
with $P_{\perp}$ defined in \eq{eq:trans-q-q2} that satisfies $P_{\perp} \cdot q = P_{\perp} \cdot q' = 0$.

The photon frame coincides with the $q$-$q'$ transverse basis used in the tensors 
$\tau_i^{\mu\nu}$ and $\wt{\tau}_i^{\mu\nu}$ [\eqs{eq:p-even-tau}{eq:p-odd-tau}]
and grants the latter with a physical understanding in terms of the helicity structures of the two photons.
Using the explicit forms of the polarization vectors along the $-z$ direction,
\footnote{Throughout most of this paper, we use lightfront coordinates for Lorentz vectors,
$V^{\mu} = (V^+, V^-, \bm{V}_T)$,
with $V^\pm = (V^0 \pm V^z) / \sqrt{2}$ and $\bm{V}_T = (V^x, V^y)$.
With two light-cone unit vectors, $\bar{n}^\mu = (1,0,\bm{0}_T)$ and $n^\mu = (0,1,\bm{0}_T)$, 
we have $V^+=V\cdot n$ and $V^- = V\cdot \bar{n}$.
When potential ambiguity arises, we explicitly denote Cartesian coordinates with a subscript ``c'',
writing $V^{\mu} = (V^0, V^x, V^y, V^z)_{\rm c}$.}
\beq
	\epsilon^{\mu}_{\pm}(q) = \epsilon^{\mu}_{\pm}(q') = \frac{1}{\sqrt{2}}(0, \mp 1, i, 0)_{\rm c}, \quad
	\epsilon^{\mu}_0(q) = \frac{1}{Q} \pp{ q^{\mu} - \frac{q^2}{q \cdot q'} q^{\prime \mu} },
\eeq
we can express the nonvanishing six tensors $\tau_i^{\mu\nu}$ and $\wt{\tau}_i^{\mu\nu}$ ($i = 1, 2, 3$) as
\begin{align}
	&\tau_1^{\mu\nu} 
		= X^{\mu} X^{\nu} + Y^{\mu} Y^{\nu} 
		= \epsilon^{\mu *}_{+}(q) \, \epsilon^{\nu}_{+}(q') + \epsilon^{\mu *}_{-}(q) \, \epsilon^{\nu}_{-}(q'), 
	&&\wt{\tau}_1^{\mu\nu} 
		= -i \pp{ X^{\mu} Y^{\nu} - Y^{\mu} X^{\nu} }
		= \epsilon^{\mu *}_{+}(q) \, \epsilon^{\nu}_{+}(q') - \epsilon^{\mu *}_{-}(q) \, \epsilon^{\nu}_{-}(q'), \nn\\
	&\tau_2^{\mu\nu} 
		= - X^{\mu} X^{\nu} + Y^{\mu} Y^{\nu} 
		= \epsilon^{\mu *}_{+}(q) \, \epsilon^{\nu}_{-}(q') + \epsilon^{\mu *}_{-}(q) \, \epsilon^{\nu}_{+}(q'), 
	&&\wt{\tau}_2^{\mu\nu} 
		= -i \pp{ X^{\mu} Y^{\nu} + Y^{\mu} X^{\nu} }
		= \epsilon^{\mu *}_{+}(q) \, \epsilon^{\nu}_{-}(q') - \epsilon^{\mu *}_{-}(q) \, \epsilon^{\nu}_{+}(q'), \nn\\
	&\tau_3^{\mu\nu} \propto \epsilon_0^{\mu}(q) \, X^{\nu}
		= \epsilon^{\mu}_{0}(q) \, \bigbb{ \epsilon^{\nu}_{-}(q') - \epsilon^{\nu}_{+}(q') } / \sqrt{2}, 
	&&\wt{\tau}_3^{\mu\nu}  \propto i \epsilon_0^{\mu}(q) \, Y^{\nu}
		= \epsilon^{\mu}_{0}(q) \, \bigbb{ \epsilon^{\nu}_{-}(q') + \epsilon^{\nu}_{+}(q') } / \sqrt{2}.
\label{eq:cff-helicity-structure}
\end{align}
In the photon frame, the collinear set of partons and the two photons all lie along the $z$ axis, 
so helicities should conserve in the hard collision $C$.
The total helicity of the partons is characterized by different GPDs that they get factorized into.
Apparently from \eq{eq:cff-helicity-structure}, 
the $\tau_1^{\mu\nu}$ and $\wt{\tau}_1^{\mu\nu}$ do not flip the photon helicities, 
so the associated CFFs $(\H_1, \E_1, \Ht_1, \Et_1)$ should receive contributions at the twist-2 level 
only from the unpolarized GPDs $F^{q, g}$ and polarized GPDs $\wt{F}^{q, g}$.
The $\tau_2^{\mu\nu}$ and $\wt{\tau}_2^{\mu\nu}$ flip the photon helicities by two units,
so at twist-2 level only the gluon transversity GPD $F^g_T$ contributes to the CFFs $(\H_2, \E_2, \Ht_2, \Et_2)$.
The $\tau_3^{\mu\nu}$ and $\wt{\tau}_3^{\mu\nu}$ flip the photon helicities by one unit,
which cannot come from twist-2 GPDs.
Hence, the CFFs $(\H_3, \E_3, \Ht_3, \Et_3)$ require at least twist-3 GPDs or quark transversity GPD $F^q_T$, which is suppressed by the quark mass.
In the following, we confine our discussion within the leading perturbative order at twist-2, 
so we will only consider the first diagram on the right-hand side of \fig{fig:breit-dvcs-twist}.

Factorization requires one to specify the large momentum components of the partons 
by use of two auxiliary lightlike vectors $n$ and $\bar{n}$ satisfying $n \cdot \bar{n} = 1$.
Any vector $V$ can then be expanded as
\beq[eq:V-decomp]
	V^{\mu} = (V \cdot n) \, \bar{n}^{\mu} + (V \cdot \bar{n}) \, n^{\mu} + V_{T}^{\mu},
\eeq
with $V^2 = 2 (V \cdot n)(V \cdot \bar{n}) - \bm{V}_T^2$ and $\bm{V}_T^2 = -V_{T}^{\mu} V_{T \mu}$.
For each parton momentum $k$ that enters the hard part $C$, we approximate it by
\beq[eq:k-approx]
	k \to \hat{k} = (k \cdot n) \, \bar{n},
\eeq
which effectively also sets $\Delta$ to $\hat{\Delta} = (\Delta \cdot n) \, \bar{n}$ in $C$.
Also, we insert the projector $(\gamma \cdot \bar{n}) (\gamma \cdot n) / 2$ 
or $(\gamma \cdot n) (\gamma \cdot \bar{n}) / 2$ for each quark or antiquark line entering $C$.
This eventually gives a factorization formula for the virtual Compton amplitude,
\begin{align}
	T^{\mu\nu}
		= \sum_q \int_{-1}^1 dx \bb{ F^q(x, \xi, t) \, C_q^{\mu\nu}(x, \xi; n, \bar{n})
			+ \wt{F}^q(x, \xi, t) \, \wt{C}_q^{\mu\nu}(x, \xi; n, \bar{n}) }
			+ \order{\alpha_s, \frac{\sqrt{-t}}{Q} },
\label{eq:dvcs-T-factorize}
\end{align}
where the quark GPDs $F^q(x, \xi, t)$ and $\wt{F}^q(x, \xi, t)$ are defined as usual~\cite{Diehl:2003ny} using the vector $n$,
\bse\label{eq:GPD-def-q}\begin{align}
	F^q(x, \xi, t) & = 
		\int \frac{d \lambda}{4\pi} e^{-i \lambda x P \cdot n}
			\langle N(p', \lambda_N') | \,
			\bar{\psi}_{q}( \lambda n / 2)
			\, \gamma \cdot n \, 
			\psi_{q} ( - \lambda n / 2)
		\, | N(p, \lambda_N) \rangle	\nn\\
	& = \frac{1}{2 P \cdot n} \, \bar{u}(p', \lambda_N')
		\bb{ H^q(x, \xi, t) \, \gamma \cdot n 
			- E^q(x, \xi, t) \frac{i \sigma^{n \Delta}}{2m}
		} u(p, \lambda_N),	
	\label{eq:GPD-def-F-q} \\
	\wt{F}^q(x, \xi, t) & = 
		\int \frac{d \lambda}{4\pi} e^{-i \lambda x P \cdot n}
			\langle N(p', \lambda_N') | \,
			\bar{\psi}_{q}(\lambda n / 2)
			\, \gamma \cdot n \gamma_5 \, 
			\psi_{q}(- \lambda n / 2)
		\, | N(p, \lambda_N) \rangle	\nn\\
	& = \frac{1}{2 P \cdot n} \, \bar{u}(p', \lambda_N')
		\bb{ \wt{H}^q(x, \xi, t) \, \gamma \cdot n \gamma_5 
			- \wt{E}^q(x, \xi, t) \frac{\gamma_5 \Delta \cdot n}{2m}
		} u(p, \lambda_N),
\end{align}\ese
and we have suppressed the dependence on the factorization scale $\mu$.
Here the $\xi$ is defined as
\beq[eq:xi-def]
	\xi = \frac{\Delta \cdot n}{2 P \cdot n} = \frac{(p - p') \cdot n}{(p + p') \cdot n},
\eeq
which depends on the choice of $n$.
For the transversity GPDs, one needs to further specify the transverse indices, which also relies on the choice of $\bar{n}$.
GPDs defined using different choices of $n$ are equivalent up to higher twist.

The hard-part coefficient $C^{\mu\nu}$ or $\wt{C}^{\mu\nu}$ describes the interaction of the two photons with amputated partons
that are contracted with $\gamma \cdot \hat{P} / 2$ or $\gamma_5 \gamma \cdot \hat{P} / 2$, where 
$\hat{P} = (P \cdot n) \, \bar{n} = \hat{\Delta} / (2\xi)$.
Denoting the parton momenta as 
\beq
	\hat{k}_q = (\xi + x) \hat{P}, \quad
	\hat{k}_{\bar{q}} = (\xi - x) \hat{P}
\eeq
for the quark and antiquark entering the $C$, respectively, we have the hard coefficients,
\footnote{We take the convention $\epsilon_{0123} = -\epsilon^{0123} = 1$, 
which gives $\epsilon_{\perp}^{12} = -\epsilon_{\perp}^{21} = \epsilon^{+- 12} = 1$.}
\bse\label{eq:cffs-lo-exprs}\begin{align}
	C_q^{\mu\nu} &= e_q^2 \cdot \frac{1}{4\xi}
		\cc{ \frac{1}{(q' - \hat{k}_{\bar{q}})^2 + i \epsilon} \Tr\bb{ \gamma \cdot \hat{\Delta} \gamma^{\nu} \gamma\cdot (q' - \hat{k}_{\bar{q}}) \gamma^{\mu} }
			+ \frac{1}{(\hat{k}_{q} - q')^2 + i \epsilon} \Tr\bb{ \gamma \cdot \hat{\Delta} \gamma^{\mu} \gamma\cdot (\hat{k}_{q} - q') \gamma^{\nu} } 
		}	\nn\\
		&= e_q^2 \pp{ \frac{1}{\xi + x - i \epsilon} - \frac{1}{\xi - x - i \epsilon} } 
			\biggpp{ -g^{\mu\nu} + \frac{\hat{\Delta}^{\mu} q^{\prime \nu} + \hat{\Delta}^{\nu} q^{\prime \mu}}{\hat{\Delta} \cdot q'} }	\nn\\
		&\simeq e_q^2 \pp{ \frac{1}{\xi + x - i \epsilon} - \frac{1}{\xi - x - i \epsilon} } 
			\tau_1^{\mu\nu}, \\
	\wt{C}_q^{\mu\nu} &= e_q^2 \cdot \frac{1}{4\xi}
		\cc{ \frac{1}{(q' - \hat{k}_{\bar{q}})^2 + i \epsilon} \Tr\bb{ \gamma_5 \gamma \cdot \hat{\Delta} \gamma^{\nu} \gamma\cdot (q' - \hat{k}_{\bar{q}}) \gamma^{\mu} }
			+ \frac{1}{(\hat{k}_{q} - q')^2 + i \epsilon} \Tr\bb{ \gamma_5 \gamma \cdot \hat{\Delta} \gamma^{\mu} \gamma\cdot (\hat{k}_{q} - q') \gamma^{\nu} } 
		}	\nn\\
		&= e_q^2 \pp{ \frac{1}{\xi + x - i \epsilon} + \frac{1}{\xi - x - i \epsilon} } \frac{-i}{\hat{\Delta} \cdot q'} \epsilon^{\mu\nu \hat{\Delta} q'}	\nn\\
		&\simeq e_q^2 \pp{ \frac{1}{\xi + x - i \epsilon} + \frac{1}{\xi - x - i \epsilon} } 
			\wt{\tau}_1^{\mu\nu},
\end{align}\ese
where in the last steps of both coefficients, we approximated $\hat{\Delta}$ by $\Delta$, thereby dropping the dependence on $n$ and $\bar{n}$,
and neglected $t$ against $Q^2$.
Then to the LO at the LP, we have 
\begin{align}
	T^{\mu\nu}
		&= \frac{1}{2 P \cdot n} \, \bar{u}(p', \lambda_N') 
			\sum_q e_q^2 \cc{
				\int_{-1}^1 dx \pp{ \frac{1}{\xi + x - i \epsilon} - \frac{1}{\xi - x - i \epsilon} } 
				\bb{ H^q(x, \xi, t) \, \gamma \cdot n - E^q(x, \xi, t) \frac{i \sigma^{n \Delta} }{2m} } \tau_1^{\mu\nu}
				\right.\nn\\
				& \left. \hspace{6.5em}
				+ \int_{-1}^1 dx \pp{ \frac{1}{\xi + x - i \epsilon} + \frac{1}{\xi - x - i \epsilon} } 
				\bb{ \wt{H}^q(x, \xi, t) \, \gamma \cdot n \gamma_5 - \wt{E}^q(x, \xi, t) \frac{\gamma_5 \Delta \cdot n}{2m} } \wt{\tau}_1^{\mu\nu}
			}
			u(p, \lambda_N).
\label{eq:dvcs-T-LP}
\end{align}
Notice the explicit dependence on $n$ when compared with the decomposition in \eq{eq:CFF-decomp}.
 
The price of replacing $\hat{\Delta}$ by $\Delta$ in \eq{eq:cffs-lo-exprs} is power suppressed provided that 
the tensors to be contracted with $\tau_1^{\mu\nu}$ or $\wt{\tau}_1^{\mu\nu}$ in the DVCS amplitude
are power suppressed when contracting with $\bb{ \Delta - (\Delta \cdot n) \, \bar{n} }$, compared to when contracting with $\Delta$.
This anticipates those tensors to be constituted of such vectors $V$ that satisfy $|V\cdot \Delta| \gg |\Delta^2|$,
which is quite a loose constraint and is true as long as $V$ is not collinear to the target.
For the photon electroproduction, we can have $V \in \{\l, q, q'\}$ or their proper linear combinations.
Then we require that the choice of $n$ and $\bar{n}$ respect the power counting by maintaining 
\beq[eq:n-nb-condition]
	\left| 1 - \frac{(\Delta \cdot n) (\bar{n} \cdot V)}{\Delta \cdot V} \right| 
	\ll 1,
\eeq
which naturally picks $n$ to be a proper lightlike vector made of $V$'s, and $\bar{n}$ to be made of $\Delta$ like
\beq
	\bar{n} = \frac{1}{\Delta \cdot n} \pp{ \Delta - \frac{\Delta^2}{2 \Delta \cdot n} n }.
\eeq
It is convenient to picture this in the photon frame where $\Delta$ is highly boosted along the $z$ direction.

The remaining freedom is to fix $n$. Although it can have a variety of choices from $\l$ to $q'$, which differ not by a power suppressed amount, 
the condition in \eq{eq:n-nb-condition} implies that those choices yield only power suppressed differences in the GPDs, 
virtual Compton amplitudes, and observables.
To see this, we note that in a frame where $\Delta$ is boosted, $P$ is collinear to $\Delta$, in the sense that 
$P^2 \sim P \cdot \Delta \sim \order{\Delta^2}$ and $|P \cdot V| \gg |\Delta^2|$.
There exists a number $c = \O(1)$ such that $|(\Delta - c \, P) \cdot V| \ll |\Delta \cdot V|$
for all $V$'s, in particular for the one(s) that make up the $n$ vector.
So \eq{eq:n-nb-condition} also applies to $P$,
\beq
	\left| 1 - \frac{(P \cdot n) (\bar{n} \cdot V)}{P \cdot V} \right|
	= \left| 1 - \frac{\bb{ (c \, P - \Delta) \cdot n + \Delta \cdot n} (\bar{n} \cdot V)}{(c \, P - \Delta) \cdot V + \Delta \cdot V} \right|
	= \left| 1 - \frac{(\Delta \cdot n) (\bar{n} \cdot V)}{\Delta \cdot V} + \order{ \frac{(\Delta - c \, P) \cdot V}{\Delta \cdot V} } \right|
	\ll 1.
\eeq
As a result, for any such vector $V$, we have
\beq
	\frac{\Delta \cdot V}{P \cdot V} \simeq \frac{(\Delta \cdot n) (\bar{n} \cdot V)}{(P \cdot n) (\bar{n} \cdot V)} = \frac{\Delta \cdot n}{P \cdot n}.
\eeq
Relating back to \eq{eq:xi-def}, this shows that the value of $\xi$ does not depend on the choice of $n$ up to power correction.
Also, for the $\gamma$ matrices sandwiched between the spinors in \eq{eq:dvcs-T-LP}, we have the same approximation
\beq
	\frac{\gamma \cdot V}{P \cdot V} \simeq \frac{(\gamma \cdot n) (\bar{n} \cdot V)}{(P \cdot n) (\bar{n} \cdot V)} = \frac{\gamma \cdot n}{P \cdot n},
\eeq
because the vector $\bar{u}(p') \gamma^{\mu} u(p)$ scales in the same way as $\Delta^{\mu}$ and $P^{\mu}$.

The above discussion does not require $V^2 = 0$.
As a result, by replacing $n$ by $R$ in \eq{eq:dvcs-T-LP}, we recover the expression in \eq{eq:CFF-decomp}, and hence have the LP approximation
of the CFFs $(\H_1, \E_1, \Ht_1, \Et_1)$ at LO, 
\begin{align}
	\bigcc{\H_1, \E_1}(\eta, t / m^2, t / Q^2)
	&= \sum_q e_q^2 \int_{-1}^1 dx \pp{ \frac{1}{\xi + x - i \epsilon} - \frac{1}{\xi - x - i \epsilon} } 
			\bigcc{ H^q(x, \xi, t), E^q(x, \xi, t)}
			+ \order{\alpha_s, Q^{-1} }, \nn\\
	\bigcc{\Ht_1, \Et_1}(\eta, t / m^2, t / Q^2)
	&= \sum_q e_q^2 \int_{-1}^1 dx \pp{ \frac{1}{\xi + x - i \epsilon} + \frac{1}{\xi - x - i \epsilon} } 
			\bigcc{ \wt{H}^q(x, \xi, t), \wt{E}^q(x, \xi, t)}
			+ \order{\alpha_s, Q^{-1} },
\label{eq:cff-t2}
\end{align}
where we can use $\xi \simeq \eta$ on the right-hand side at LP accuracy.

In the photon frame (as defined in \sec{ssec:cffs-calc}), it is natural to choose $n \propto q'$.
In other frames, the choice of $n$ can vary in a variety of ways~\cite{Guo:2021gru}.
As shown above, all these choices are equivalent at the LP, but yield different remainders beyond.
This ambiguity is a special property of the off-forward kinematics of GPDs.
It does not apply to the inclusive DIS, where the nucleon momentum naturally defines the $\bar{n}$ and the factorization does not depend on the choice of $n$. 
In principle, for inclusively diffractive or TMD processes, one also has similar ambiguities, but this has not been studied in the literature.
One should expect that the different next-to-leading-power (NLP) remainders due to different choices of $n$ can be resolved in a complete NLP factorization~\cite{Guo:2021gru},
for which the ambiguity only causes difference at even higher powers. 
In this regard, one can fix the use of $n$ at all powers, expecting that the result is consistent and reliable at each fixed power.
This situation is similar to the renormalization or factorization scale dependence at each fixed perturbation order, which only gets canceled at the next order.
Only by calculating to all orders can one get rid of the scale dependence.
Similarly, only by including contributions from all powers can one have an $n$-independent result for the diffractive process,
although going to higher powers is technically much more challenging than going to higher perturbative orders.
We refer to \refcite{Guo:2021gru} for some detailed discussions and leave a more comprehensive study for future work.

\subsection{Breit frame and azimuthal modulations}
\label{ssec:breit}
It is not straightforward to disentangle the different CFFs in \eq{eq:CFF-decomp}, and thus the different GPDs in \eq{eq:cff-t2}.
This requires pinning down observables characterized by different combinations of the CFFs at the cross-section level, 
which requires a proper choice of frame.
The standard observables are azimuthal modulations, as the CFFs are associated with different spin structures of the nucleons or partons.
For this purpose, it is crucial that the CFFs in \eq{eq:cffs-exprs} not depend on any azimuthal angles.
This then fixes the frame uniquely to the Breit frame, up to a boost along the $z$ axis.

\begin{figure}[htbp]
\centering
	\begin{tabular}{cc}
	\includegraphics[scale=0.5]{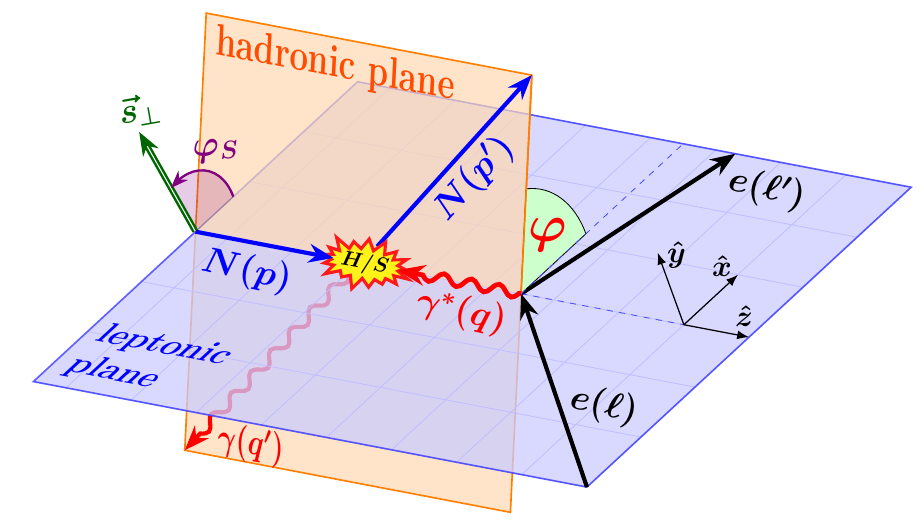} &
	\includegraphics[scale=0.5]{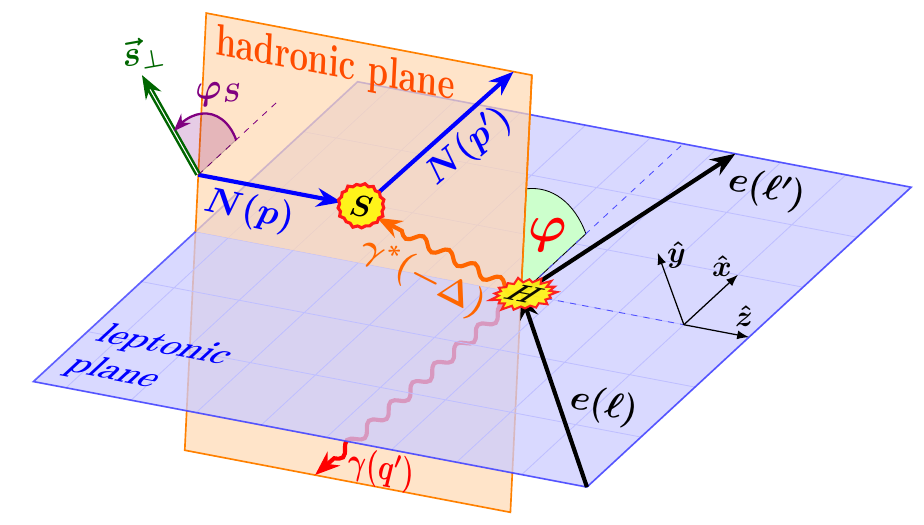} \\
	(a) & (b)
	\end{tabular}
	\caption{Breit frame for the photon electroproduction, specifically for the subprocesses (a) DVCS and (b) BH.}
\label{fig:breit-frame}
\end{figure}

The Breit frame is defined similarly to that in DIS or SIDIS, as shown in \fig{fig:breit-frame}(a).
The virtual photon $\gamma^*_{ee}(q = \l - \l')$ moves along the $-\hat{z}$ direction,
coming from the radiation of the electron, which scatters in the $\hat{x}$-$\hat{z}$ plane with $\l^x > 0$.
The nucleon target $N(p)$ is moving along the $\hat{z}$ direction, and scatters with $\gamma^*_{ee}$ to produce a nucleon $N(p')$ and photon $\gamma(q')$.
Each event is described by five independent variables $(Q^2, x_B, t, \varphi, \varphi_S)$, where
$Q^2$ and $t$ are defined in \eq{eq:breit-Q-t}, being Lorentz invariant together with $x_B = Q^2 / (2 p \cdot q)$.
The $\varphi$ and $\varphi_S$ describe the azimuthal angles of the scattered nucleon $N(p')$ 
and the transverse spin $\bm{s}_{\perp}$ of the initial-state nucleon $N(p)$, respectively;
both are with respect to the electron scattering plane.
The Breit frame observables are unchanged by an arbitrary boost along the $\hat{z}$ axis.
In particular, the BMK frame~\cite{Belitsky:2001ns} has fixed the initial-state nucleon to be at rest.

We note that the $\bm{s}_{\perp}$ is with respect to the Breit frame,
which differs from the nucleon transverse spin $\bm{S}_T$ in the lab frame by a rotation specified by the electron scattering.
The $\varphi_S$ is traded for the azimuthal angle $\varphi_\l$ of the scattered electron in the lab frame.
This transformation involves a nontrivial Jacobian factor as first pointed out by \refcite{Diehl:2005pc} in the context of SIDIS,
which has been omitted in the literature but can become important for the extraction of GPDs.
We review this problem in detail in Appendix~\ref{app:breit-jac} 
along with the kinematics and differential cross section formula in the Breit frame.
The distribution of $\varphi_S$ is nontrivial only when $S_T \equiv |\bm{S}_T| \neq 0$ in the lab frame.

The Breit frame is particularly suitable for describing the DVCS.
The amplitude has a simple dependence on the azimuthal angle $\varphi$ controlled by the helicity states of the colliding $N$ and $\gamma^*_{ee}$,
\beq
	\M^{\rm DVCS} \propto \, e^{i (\lambda_N - \lambda_{\gamma^*_{ee}}) \varphi},
\eeq
and thus, $|\M^{\rm DVCS}|^2$ contains various $\varphi$ modulations that are sensitive to the different spin structures in \eq{eq:CFF-decomp}.
For instance, to the LP in $ t / Q^2$ and $m^2 / Q^2$ in the unpolarized case, we have 
\begin{align}
	&|\M^{\rm DVCS}|^2
	= \frac{2e^6}{y^2 Q^2} \bb{
		\pp{ 1 + (1-y)^2 } A_0
		- 4 (1 - y) A_2 \cos2\varphi
	},
\end{align}
which contains only a constant and a $\cos2\varphi$ modulation. 
The coefficients are given by the CFFs $(\H, \E, \Ht, \Et)_{1,2}$ that receive contributions from twist-2 GPDs,
\begin{align}
	A_0 &= \sum_{i=1}^2 \bb{ 
		(1 - \eta^2) \bigpp{ | \H_i |^2 + | \Ht_i |^2 }
		- \pp{ \eta^2 + \frac{t}{4m^2} } | \E_i |^2
		- \frac{\eta^2 t}{4m^2} | \Et_i |^2
		- 2 \eta^2 \Re\bigpp{ \H_i \E_i^* + \Ht_i \Et_i^* }
	}, \nn\\
	A_2 &= (1 - \eta^2) \Re\bigpp{ \H_1 \H_2^* - \Ht_1 \Ht_2^* }
		- \pp{ \eta^2 + \frac{t}{4m^2} } \Re\bigpp{ \E_1 \E_2^* }
		+ \frac{\eta^2 t}{4m^2} \Re\bigpp{ \Et_1 \Et_2^* }
	\nn\\
	&\quad
		- \eta^2 \Re\bigpp{ \H_1 \E_2^* + \H_2 \E_1^* - \Ht_1 \Et_2^* - \Ht_2 \Et_1^* },
\end{align}
with $y = p \cdot q / p \cdot \l = Q^2 / \bb{ x_B (s - m^2) }$ given by $(x_B, Q^2)$ and the electron-nucleon c.m.\ energy squared $s$,
and $\eta \simeq x_B / (2 - x_B)$ at the LP.
The $\cos2\varphi$ modulation arises from the interference of $\lambda_{\gamma^*_{ee}} = 1$ and $\lambda_{\gamma^*_{ee}} = -1$ states,
which involves the mixing of the helicity-conserving CFFs $(\H_1, \E_1, \H_1, \Et_1)$ and double helicity-flipping CFFs $(\H_2, \E_2, \H_2, \Et_2)$.
The single helicity-flipping CFFs $(\H_3, \E_3, \H_3, \Et_3)$ are suppressed by not only high-twist partonic dynamics, but also by kinematics of the photon spin transfer.
Going beyond LP and taking the beam and target polarization effects into account will generate more azimuthal modulations
to set additional constraints on the CFFs, similarly to \refs{Belitsky:2001ns, Guo:2021gru, Shiells:2021xqo, Guo:2022cgq, Kriesten:2019jep, Kriesten:2020wcx, Kriesten:2020apm}.

However, the DVCS by itself does not account for all contributions to 
the physically measured cross section of the exclusive photon electroproduction in Eq.~(\ref{eq:dvcs-sdhep}).
The observables at the cross section level receive also contributions from the BH subprocess and its interference with the DVCS,
\beq[eq:dvcs-bh-xsec]
	|\M|^2 = |\M^{\rm DVCS}|^2 + |\M^{\rm BH}|^2 + 2 \Re\pp{ \M^{\rm DVCS} \M^{{\rm BH} *} }.
\eeq
While the Breit frame is convenient for describing the DVCS, it is rather inconvenient for the BH subprocess.
As shown in \fig{fig:breit-frame}(b), the BH diagram does not contain any intermediate state with momentum $q$.
Instead, the electron propagators that connect the two photon radiations have dependence on $\varphi$, cf.~\eq{eq:BH-lepton-tensor},
\begin{align}
	Q^2\mathcal{P}_1(\varphi) &= (\l + \Delta)^2 = t + 2 \l \cdot \Delta \supset - 2 \l^x \Delta^x = + 2 \l^x p^{\prime x} \propto \cos\varphi, \nn\\
	Q^2\mathcal{P}_2(\varphi) &= (\l - q')^2  = -2 \l \cdot q' \supset 2 \l^x q^{\prime x} \propto -\cos\varphi.
\label{eq:bh-p1-p2}
\end{align}
This means that
\footnote{The full explicit forms of $\mathcal{P}_1(\varphi)$ and $\mathcal{P}_2(\varphi)$
together with the amplitude squared in terms of Breit frame kinematics are given in Appendix~\ref{app:comp-ret}.}
\beq
	|\M^{\rm BH}|^2, \; 2 \Re\pp{ \M^{\rm DVCS} \M^{{\rm BH} *} }
	\propto 
	\frac{1}{\mathcal{P}_1(\varphi) \, \mathcal{P}_2(\varphi)}.
\eeq
Together with the $\varphi$ dependence from the BH numerators, this greatly contaminates the azimuthal modulations from the DVCS (see, e.g., \refcite{Belitsky:2001ns}).
Since the BH subprocess accounts for up to 90\% of the photon electroproduction rate~\cite{Kriesten:2020wcx, Kriesten:2020apm, Shiells:2021xqo}, 
its presence, including both its amplitude squared and its interference with the DVCS subprocess, 
obscures the clear physical understanding of the $\varphi$ modulations in terms of the $\gamma^*_{ee}$ polarization states,
and also complicates the extraction of the CFFs from them.

\subsection{Alternative perspective}
\label{ssec:change}
The reason behind this complication is that the Breit frame is constructed based on a state ($\gamma^*_{ee}$) that carries a hard virtuality.
Physically, a hard interaction happens ``instantaneously'' at a local spacetime point, where various microscopic subprocesses (i.e., Feynman diagrams) can occur.
One could get away with this in DIS or SIDIS because the conventional practice only works up to the LO in QED.
\footnote{Going to high orders of QED, one has similar problems in DIS and SIDIS~\cite{Liu:2020rvc, Liu:2021jfp, Qiu:2024arw, Cammarota:2025jyr} where extra care needs to be taken in defining the kinematic observables.}
But for the exclusive photon electroproduction process, one has additional independent hard microscopic subprocesses given by the BH 
even within the LO in QED.

It is the mixing of physical scales that makes the azimuthal modulations originate not just from dynamic origin 
but also messed up by arbitrary kinematic effects.
Specifically, as labeled in \fig{fig:breit-frame}, 
the ``DVCS blob'' in (a) (labeled by ``$H/S$'')
between the nucleon and $\gamma_{ee}^*$ involves both a hard scale $Q$ due to the virtual state $\gamma_{ee}^*$ 
and a soft scale $t$ from the nucleon diffraction,
while for the BH subprocess in (b), this blob separates into a soft vertex (labeled by ``$S$'') and a hard vertex (labeled by ``$H$'')
and the $\varphi$ angle is purely given at the hard vertex.

On the other hand, we notice that nucleon diffraction is unambiguously related to the soft scale $t$ defined by the momentum transfer $\Delta$.
This is characterized by the set of exchanged collinear partons in \fig{fig:breit-dvcs-twist} for the DVCS twist expansion
and by the $\gamma^*$ state in the BH subprocess [cf.~\fig{fig:breit}(b) or \fig{fig:breit-frame}(b)].
Therefore, we make the reorganization as

\beq
	\def\sc{1.5}
		\adjincludegraphics[valign=c, scale=0.6]{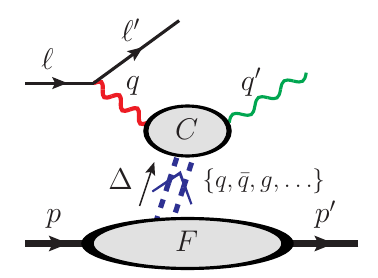}	
		\quad
		\Scale[\sc]{+}
		\quad
		\adjincludegraphics[valign=c, scale=0.6]{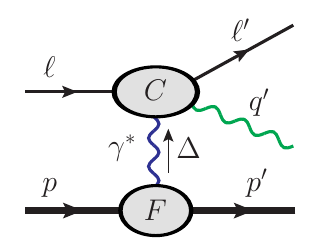} 
		\quad
		\Scale[\sc]{=}
		\quad
		\sum_{A^*}
		\adjincludegraphics[valign=c, scale=0.6]{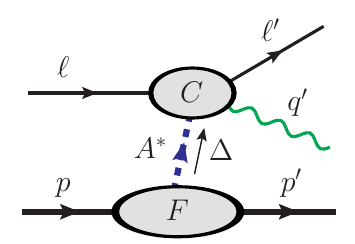} ,
\label{eq:breit-sdhep}
\eeq
where the collinear partons and the $\gamma^*$ are all collected by the same low-virtuality $A^*$ state, 
which carries the momentum $\Delta$.
\footnote{A tricky difference between this and the Breit frame description in \fig{fig:breit-frame}(b) is that 
now the $\gamma^*$ travels {\it from} the nucleon {\it to} the hard scattering,
while in the Breit frame it is radiated by the electron and absorbed by the nucleon. 
This causal flipping is allowed because $\Delta$ is a spacelike momentum and thus its temporal component can be reversed under a Lorentz transformation.
This explains why we define $\Delta$ oppositely to the normal convention, which introduces the extra minus signs in 
\eqss{eq:EM-form-factor}{eq:CFF-decomp}{eq:GPD-def-q}.}
This reorganization unifies the dynamics of DVCS and BH subprocesses on the same ground, 
simply as different channels in the $A^*$ state.
In so doing, we achieve a consistent separation of physical scales: 
the soft dynamics at scale $t$ in the diffraction part $F$
separated from the hard dynamics in $C$ by the low-virtuality $A^*$ state.
Then, instead of starting with the DVCS and BH separation in \eq{eq:dvcs-bh-amp} or \fig{fig:breit},
we concentrate on the division of the whole $2\to3$ physical process into
a single nucleon diffraction part and a hard $2\to2$ scattering part between the virtual $A^*$ and electron beam, 
along with a sum of all possible $A^*$ states.
In the next two Sections, we reformulate the photon electroproduction in this new perspective.

Before moving on, it is important to note that physics should not depend on what perspective or frame we pick to formulate the same process.
In Appendix~\ref{app:comp-ret}, we provide a covariant calculation that expresses the whole photon electroproduction amplitude squared in Lorentz invariants.
Then specifying the kinematic observables in the Breit frame reproduces the results in \refcite{Belitsky:2001ns},
whereas specifying the kinematics in the new SDHEP frame to be discussed below reproduces the results in \eq{eq:dvcs-LP-NLP-xsec}.
That said, the new perspective does provide fruitful physical insights about the dynamics, particularly the azimuthal modulations and their connections to GPDs.

\section{Reframing the exclusive photon electroproduction as an SDHEP}
\label{sec:dvcs-sdhep}

In fact, the organization of \eq{eq:breit-sdhep} in terms of the exchange state $A^*(\Delta)$ is not unique to the photon electroproduction process.
As pointed out in \refcite{Qiu:2022pla, Qiu:2024reu, Qiu:2024pvw}, 
all processes that are sensitive to GPDs share the same features of 
being diffractive for the hadron target $h$ while having a hard scattering scale at the same time.
They are at minimum $2\to3$ processes,
\beq[eq:sdhep]
	h(p) + B(p_2) \to h'(p') + C(q_1) + D(q_2),
\eeq
which we refer to as {\it single-diffractive hard exclusive processes (SDHEPs)},
and where the hadron $h$ is diffracted by the beam $B$ into $h'$,
with a production of two particles, $C$ and $D$, that carry large and balancing transverse momenta, $q_{1T}$ and $q_{2T}$, with respect to the $h$-$B$ collision axis.
The defining feature of SDHEPs is the presence of two distinct scales,
\beq[eq:hard-condition]
	q_T \sim q_{1T} \sim q_{2T} \gg \sqrt{-t} \sim \Delta_T,
\eeq
where $t = (p - p')^2$ and $\Delta_T = |\bm{p}_T - \bm{p}'_T|$ is the transverse momentum transfer of the hadron diffraction.

\begin{figure}[htbp]
	\def\sc{1.5}
	\centering
	\begin{align*}
		&\adjincludegraphics[valign=c, scale=0.55]{figures/sdhep-tot.pdf}	
			\Scale[\sc]{=}
			\adjincludegraphics[valign=c, scale=0.55]{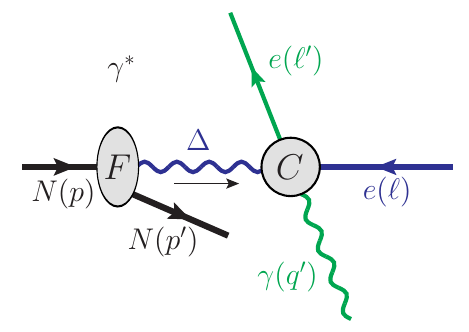} 
			\Scale[\sc]{+}
			\adjincludegraphics[valign=c, scale=0.55]{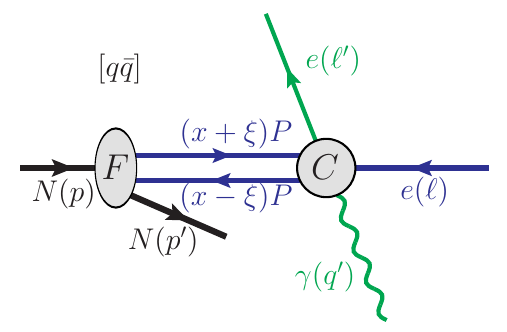} 
		\\
		&\hspace{4em}
			\Scale[\sc]{+}
			\adjincludegraphics[valign=c, scale=0.55]{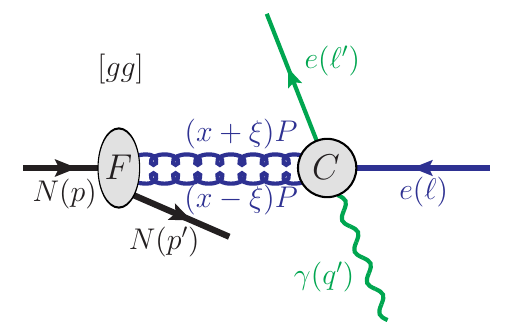}
			\Scale[\sc]{+}
			\adjincludegraphics[valign=c, scale=0.55]{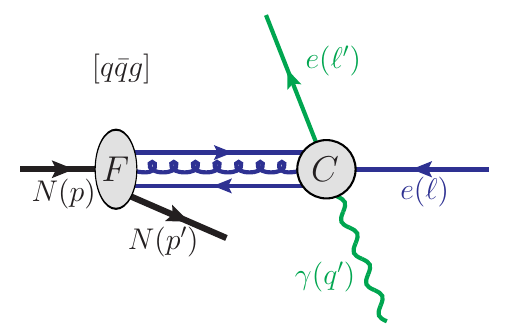} 
			\Scale[\sc]{+}
			\adjincludegraphics[valign=c, scale=0.55]{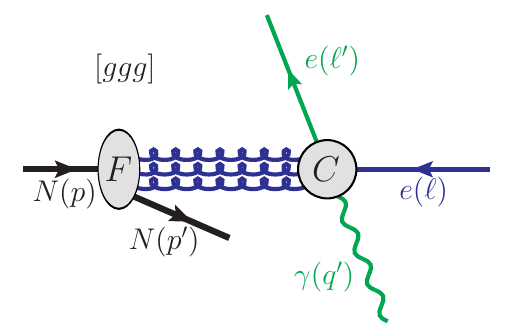}
			\Scale[\sc]{+ \cdots}
	\end{align*}
	\caption{Channel expansion of the photon electroproduction process in the light-cone gauge.
	The ``$\cdots$'' stands for channels with more than three partons. 
	}
	\label{fig:dvcs-channel}
\end{figure}

The exclusive photon electroproduction process in \eq{eq:dvcs-sdhep} is one specific SDHEP
with $h = h' = N$, $B = C = e$, and $D = \gamma$.
The momenta are correspondingly $(p_2, q_1, q_2) = (\l, \l', q')$.
As illustrated in \eq{eq:breit-sdhep}, the two-scale condition in \eq{eq:hard-condition} motivates a two-stage paradigm [\eq{eq:two-stage} or \fig{fig:dvcs-sdhep}].
The full process includes a complete sum of all independent particle states in the intermediate $A^*$, as shown in \fig{fig:dvcs-channel} in a physical gauge.
Due to the constraint in \eq{eq:hard-condition}, the intermediate lines in $A^*$ are perturbatively pinched on shell to the extent of order $\mathcal{O}(t)$~\cite{Kang:2014tta}. 
The contribution from the pinched region of a $j$-parton state is hence at most of order $(\sqrt{-t} / q_T)^{j-1}$ relative to the $A^* = \gamma^*$ channel at $j = 1$.
In this way, the factorization argument can be carried out in the SDHEP framework as simply as that in the Breit frame in \sec{ssec:cffs-calc}, 
with the hard scale $Q$ replaced by $q_T$.
The advantage of this formalism is that the $A^* = \gamma^*$ channel (for the BH subprocess) can be treated coherently with the parton channels (for the DVCS subprocess).

Similarly to the DIS, the argument from operator product expansion for the DVCS suggests that the photon electroproduction process might be factorizable 
at an arbitrary power of $\sqrt{-t} / q_T$. 
Detailing this full picture is beyond the scope of this paper. 
We will stop at the two-parton channel, i.e., the first three terms in the right-hand side expansion of \fig{fig:dvcs-channel}.
As we will see in \sec{sec:calc-sdhep}, the amplitude for the $\gamma^*$ channel scales as $1 / \sqrt{-t}$,
and the two-parton $[q\bar{q}]$ or $[gg]$ channel scales as $1 / q_T$.
Neglecting the three-parton states and beyond means that the error is of $\order{\sqrt{-t} / q_T^2}$,
and hence when squaring the whole amplitude, we shall only keep the square of the $\gamma^*$ channel, which is at order $1/t$,
and the interference of the $\gamma^*$ and the $[q\bar{q}]$ or $[gg]$ channels, which is at order $1 / (q_T \sqrt{-t})$.
The square of the two-parton channels contributes at order $1 / q_T^2$, 
which is the same as the interference between the $\gamma^*$ and the three-parton channels,
so should be neglected for a consistent power counting.

For convenience, in the following, 
we will refer to the $\gamma^*$ channel in the amplitude and its square in the cross section as the LP,
and the $[q\bar{q}]$ and $[gg]$ channels in the amplitude and their interference with the $\gamma^*$ channel in the cross section as the NLP.
We will defer further technical details to \sec{sec:calc-sdhep}. In this Section, let us comment on some general features of the SDHEP framework.

\subsection{Two-stage description of kinematics}
\label{ssec:sdhep}

\begin{figure}[htbp]
	\centering
		\includegraphics[scale=0.5]{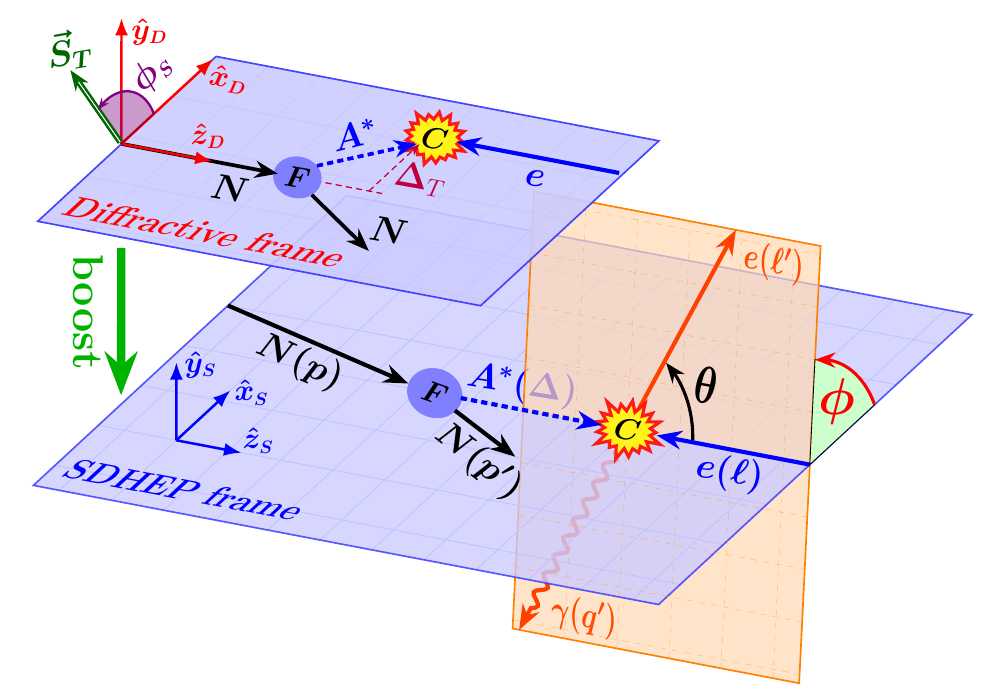}
	\caption{
		Diffractive (upper half) and SDHEP (lower half) frames for describing the photon electroproduction process.
		$F$ denotes the (nonperturbative) diffraction subprocess $N \to N + A^*$, which happens in the blue plane, 
		and $C$ denotes the hard interaction between $A^*$ and $e(\l)$ to produce $e(\l')$ and $\gamma$, 
		which happens in the orange plane. 
		The two planes intersect at the collision axis between $A^*$ and $e(\l)$ and form an angle of $\phi$. 
	}
\label{fig:sdhep-frame}
\end{figure}

Following the two-stage SDHEP paradigm, we describe the kinematics of the photon electroproduction process in a two-step manner,
as shown in \fig{fig:sdhep-frame}.

First, we describe the kinematics of the nucleon diffraction subprocess in \eq{eq:diffractive} in the {\it diffractive frame},
choosen as the c.m.\ of the target $N(p)$ and beam $e(\l)$, with 
$N$ along the $\hat{z}_{D}$ direction. 
In a fixed lab coordinate system, the diffractive kinematics in \eq{eq:diffractive} is fully captured by the momentum transfer vector
$\Delta = p - p'$ through its invariant mass $t = \Delta^2$, azimuthal angle $\phi_{\Delta}$, and skewness
\beq[eq:xi-sdhep]
	\xi = \frac{\Delta^+}{(p + p')^+} = \frac{\Delta \cdot n}{(p + p') \cdot n},
\eeq
which is similar to \eq{eq:xi-def} except that now we fix the lightlike vector $n^{\mu}$ to $\delta^{\mu-} = (1, 0, 0, -1) / \sqrt{2}$ in the diffractive frame.
This choice of $n$ is equivalent to the electron beam momentum $\l$.
Among the three variables $(t, \xi, \phi_{\Delta})$, only $t$ is Lorentz invariant. Both $\phi_{\Delta}$ and $\xi$ are frame dependent.

The $\phi_{\Delta}$ distribution is nontrivial only 
when the target has a transverse spin $\bm{S}_T$ to break the azimuthal rotation invariance.
To simplify the kinematic description, we choose the $\hat{x}_{D}$ axis of the diffractive frame to be along the 
diffraction direction, $\hat{x}_{D} = \bm{\Delta}_T / \Delta_T$, and $\hat{y}_{D} = \hat{z}_{D} \times \hat{x}_{D}$
is determined accordingly. 
This choice of coordinate system varies event by event, and trades the azimuthal angle $\phi_{\Delta}$ of the diffraction in a fixed lab
coordinate system for the azimuthal angle $\phi_S$ of the transverse spin $\bm{S}_T$ in the varying $\hat{x}_{D}$-$\hat{y}_{D}$ system,
as illustrated in the upper half of \fig{fig:sdhep-frame}.

Once the diffraction subprocess (and thereby the c.m.\ energy $\sqrt{\hat{s}}$ of the $2\to2$ hard scattering subprocess in \eq{eq:hard 2to2}) 
is determined, we perform a Lorentz transformation into the {\it SDHEP frame},
defined as the c.m.\ of $A^*(\Delta)$ and $e(\l)$ with $A^*$ moving along the $\hat{z}_S$ axis, 
to describe the $2\to2$ hard subprocess in terms of the polar and azimuthal angles $(\theta, \phi)$ of $e(\l')$ [or equivalently, $\gamma(q')$].
This is illustrated in the lower half of \fig{fig:sdhep-frame}, where the diffraction subprocess [\eq{eq:diffractive}] takes place in the diffractive plane (in blue), 
and the hard scattering subprocess [\eq{eq:hard 2to2}] in the scattering plane (in orange). 
The $\hat{x}_S$ axis lies on the diffractive plane, given by $\hat{x}_S = -\bm{p}_T / |\bm{p}_T|$, 
and $\hat{y}_S = \hat{z}_S \times \hat{x}_S$ is perpendicular to the diffractive plane. 
This $\hat{y}_S$ is the same as the $\hat{y}_{D}$ in the diffractive frame.

In this way, each event is described by five independent kinematic variables $(t, \xi, \phi_S, \theta, \phi)$,
where the first three describe the diffraction subprocess in the diffractive frame and the latter two describe the $2\to2$ hard collision in the SDHEP frame.
\footnote{In a sense, the SDHEP frame is analogous to the Breit frame by interchanging the roles of lepton and hadron, i.e., by the correspondence 
$[h(p), h'(p'), A^*(\Delta)] \leftrightarrow [e(\l), e(\l'), \gamma^*(q)]$. 
However, the hadron's transverse spin $\bm{S}_T$ is important to GPD studies. The azimuthal $\phi_S$ distribution is best described in the diffractive frame. 
The same quantity in the Breit frame is obtained by a Lorentz transformation from the lab frame, which involves a Jacobian factor that distorts the $\phi_S$ distribution, as detailed in Appendix~\ref{ssec:jacobian}.}
Among the five variables, only $t$ is Lorentz invariant.
However, as to be shown in \sec{ssec:trans-boost}, $\xi$ remains unchanged when going from the diffractive frame to the SDHEP frame
so is the same in both frames.
It is straightforward to work out the cross section formula,
\beq[eq:sdhep-xsec]
	\frac{d\sigma}{dt \, d\xi \, d\phi_S \, d\cos\theta \, d\phi}
	= \frac{\overline{|\M|^2}}{ (4\pi)^5 \, (1 + \xi)^2 \, s },
\eeq
where $s = (p + \l)^2$ is the c.m.\ energy squared.
This applies not only to the photon electroproduction, but also to any other SDHEP.

In Appendix~\ref{app:covariant}, we provide a set of definitions for the kinematic variables $(t, \xi, \phi_S, \theta, \phi)$ 
in terms of Lorentz-invariant expressions for convenience of experimental measurements.

\subsection{Lorentz transformation from the diffractive frame to SDHEP frame}
\label{ssec:trans-boost}

Both the diffractive and SDHEP frames have been chosen such that the nucleons and electron are highly boosted along the $\pm z$ directions, respectively.
So we naturally choose $n = (1, 0, 0, -1) / \sqrt{2}$ in both frames.
Going from the diffractive frame to the SDHEP frame is then straightforward by noting that
in both frames, the electron momentum $\l$ is along the minus light-cone direction, so that
the two frames can be connected by a Lorentz transformation $L$
that leaves the lightlike vector $n$ invariant, followed by a longitudinal boost along $\hat{z}_S$.
The special property of $L$ is such that any arbitrary vector $k$ transforms with its plus component unchanged, 
\beq
	(Lk)^+ = (Lk) \cdot n = (Lk) \cdot (Ln) = k \cdot n = k^+.
\eeq
So it is exactly a transverse boost~\cite{Diehl:2003ny}, which can be parametrized by 
a two-dimensional transverse vector $\bm{v}_T = (v_T^x, v_T^y)$,
\beq[eq:trans-boost]
	k' = L(\bm{v}_T) \cdot k
	= \pp{ k^{+}, \, k^- + \bm{k}_T \cdot \bm{v}_T + \frac{1}{2} k^+ v_T^2, \, \bm{k}_T + k^{+}\bm{v}_T },
\eeq
with $v_T^2 = |\bm{v}_T \cdot \bm{v}_T|$.

The $\bm{v}_T$ is determined by requiring the momentum $\Delta = (\Delta^+, (t + \Delta_T^2)/2\Delta^+, (\Delta_T, 0)_T)$ in the diffractive frame
to transform to $L(\bm{v}_T) \cdot \Delta = (\Delta^+, t/2\Delta^+, 0_T)$ in the SDHEP frame,
which gives
$\bm{v}_T = \pp{-\Delta_T/\Delta^+, 0}$, where we recall that the $\hat{x}_{D}$ is chosen along the 
direction of $\bm{\Delta}_T$.
Therefore, the transformation $L$ takes any vector $k = (k^+, k^-, k^x, k^y)$ to
\beq[eq:lorentz-trans-sdhep]
	L \cdot k = \pp{ k^+, \, k^- - k^x \frac{\Delta_T}{\Delta^+} + \frac{k^+}{2} \pp{ \frac{\Delta_T}{\Delta^+} }^2, \, k^x - \Delta_T \frac{k^+}{\Delta^+}, k^y},
\eeq
where we have written explicitly $\bm{k}_T = (k^x, k^y)$. 

Following $L$, one may perform a trivial boost along $\hat{z}_S$ to reach the c.m.\ of $A^*(\Delta)$ and $e(\l)$.
This boost does not change the $\xi$ and $\phi$.

Obviously, the diffraction kinematics $(t, \xi, \phi_S)$ does not require the diffractive frame to be the c.m.\ of the beam $e$ and target $N$.
In particular, one may just choose it as the target rest frame.
Any boost along $\hat{z}_{D}$ does not change $(t, \xi, \phi_S)$ or $\Delta_T$.
Consider then the full Lorentz transformation from the diffractive frame to the SDHEP frame as 
a boost $\Lambda(\beta_1)$ along the $\hat{z}_{D}$ in the diffractive frame,
followed by a transverse boost in \eq{eq:lorentz-trans-sdhep},
and then a boost $\Lambda(\beta_2)$ along the $\hat{z}_S$ into the SDHEP frame.
Any vector $k$ is transformed into
\begin{align}
	\Lambda(\beta_2) L(\bm{v}_T) \Lambda(\beta_1) k
	= \pp{ e^{\beta_1 + \beta_2} k^+, \, 
		e^{-(\beta_1 + \beta_2)} \pp{ k^- - k^x \frac{\Delta_T}{\Delta^+} + \frac{k^+}{2} \pp{ \frac{\Delta_T}{\Delta^+} }^2 }, 
		\, k^x - \Delta_T \frac{k^+}{\Delta^+}, k^y},
\label{eq:bbb}
\end{align}
where $\Delta^+$ is defined in the diffractive frame before the boost $\Lambda(\beta_1)$. 
\eq{eq:bbb} trivially combines the two boosts before and after the transverse boost.
\footnote{Note that this property is because the $\bm{v}_T$ in \eq{eq:trans-boost} is set by a physical momentum $\Delta$, 
which itself is also affected by the first boost $\Lambda(\beta_1)$. If $\bm{v}_T$ is an independent parameter, the simple combination in \eq{eq:bbb} will not hold.}
Hence, any choice of the diffractive frame with respect to the longitudinal boost $\Lambda(\beta_1)$ is compensated 
by the boost $\Lambda(\beta_2)$ to get to the SDHEP frame.

The boost $\Lambda(\beta_2)$ does not change the azimuthal angle $\phi$ or transverse momentum $q_T$ of $e(\l')$ [or $\gamma(q')$] 
but it changes the polar angle $\theta$.
This may motivate one to choose the independent kinematic variables as $(t, \xi, \phi_S, q_T, \phi)$ instead of $(t, \xi, \phi_S, \theta, \phi)$
to avoid necessity of staying in the $(A^*, \, e)$ c.m.\ frame, as done in \refs{Qiu:2022bpq, Qiu:2024mny}.
In that case, one may use $q_T = |\bm{\l}'_{T}| = |\bm{q}'_{T}| = (\sqrt{\hat{s}} / 2) \sin\theta$
to express the differential cross section in \eq{eq:sdhep-xsec} in terms of $q_T$ instead of $\theta$,
\begin{align}
	\frac{d\sigma}{dt \, d\xi \, d\phi_S \, dq_T^2 \, d\phi}
	&= \frac{2}{\hat{s} \sqrt{1 - 4 q_T^2 / \hat{s}}} \,
		\sum_{r = \pm 1}
		\frac{d\sigma}{dt \, d\xi \, d\phi_S \, d\cos\theta \, d\phi} \bigg|_{\cos\theta \to r \sqrt{1 - 4 q_T^2 / \hat{s}}}
		\nn\\
	&= \frac{2}{\hat{s} \sqrt{1 - 4 q_T^2 / \hat{s}}}
		\frac{1}{ (4\pi)^5 \, (1 + \xi)^2 \, s } \,
		\sum_{r = \pm 1} \overline{|\M|^2}\big|_{\cos\theta \to r \sqrt{1 - 4 q_T^2 / \hat{s}}},
\label{eq:diff-xsec-qt}
\end{align}
where the factor $1 / \sqrt{1 - 4 q_T^2 / \hat{s}}$ gives the Jacobian peak at $q_T = \sqrt{\hat{s}} / 2$.
The disadvantage of using $q_T$, however, is that it combines the two configurations of $r = \pm 1$ in \eq{eq:diff-xsec-qt}, 
which correspond to the final-state electron being in the forward or backward rapidity region,
while these two configurations are distinguishable in the electron-photon final states.
Hence, in the following, we will work with $\theta$ specifically in the c.m.\ frame of the hard collision, similarly to \refcite{Qiu:2023mrm},
with the conversion to $q_T$ to be easily carried out using \eq{eq:diff-xsec-qt}.

The fact that it is a transverse boost that connects the diffractive and SDHEP frames has the following advantages:
\begin{enumerate}
	\item The GPD definitions are the same in both frames since $n$ is invariant. 
		This is true for all unpolarized, polarized, and transversity GPDs.
		Although transversity GPDs contain both plus and transverse Lorentz indices in their operator definitions,
		namely the quark transversity involving $\bar{\psi} \gamma^+ \gamma^i \gamma_5 \psi$
		and the gluon transversity involving $F^{+i} F^{+j}$,
		the transverse indices $i$ and $j$ only mix with the plus index, which vanishes 
		for either $\gamma^+ \gamma^+$ or $F^{++}$ in the quark or gluon transversity.
		This property also extends to arbitrary-twist GPDs.
	\item The variable $\xi = [(p - p') \cdot n] / [(p + p') \cdot n]$ is the same in both frames.
		Therefore, unlike the Breit frame description where $\xi$ is only a parameter of the GPD,
		here it is also a useful kinematic variable.
	\item This transverse boost is part of the little group for a lightlike momentum along $-z$, so keeps both its momentum and spin state.
		As a result, the electron beam has the same spin vector $(\bm{S}_T^e, P_e)$ in both frames,
		although $\bm{S}_T^e = 0$ in most settings due to the chiral symmetry.
	\item Incidentally, if we describe the proton spin in terms of light-front helicity, then it will also stay invariant in both frames 
		and the $\bm{S}_T$ and $\phi_S$ in \fig{fig:sdhep-frame} will have the same definitions also in the SDHEP frame.
\end{enumerate}
These make the SDHEP frame rather convenient for describing the $2\to2$ hard scattering subprocess. 
We note that the transformation from the diffractive frame to the SDHEP frame can also be achieved by 
boosting along $-\bm{p}'$~\citep{Berger:2001xd} followed by a rotation around $\hat{y}$, 
but expressing in terms of the transverse boost elucidates the light-cone kinematics more clearly.

\subsection{Colliding kinematics}
\label{ssec:kins}

Since the longitudinal boosts in the diffractive and the SDHEP frames do not make significant roles, 
we shall simply parametrize the initial momenta of colliding hadron and lepton as
\beq[eq:lab-p]
	p^{\mu}_{D} = \pp{ p^+, \, \frac{m^2}{2 p^+}, \, \bm{0}_T }
	\qquad 
	\mbox{\rm and}
	\qquad 
    \ell^{\mu}_{D} = ( 0, \, \ell^-, \, \bm{0}_T)
\eeq
in the diffractive frame. 
The momentum transfer $\Delta$ and final-state nucleon momentum $p'$ can then be determined by $(t, \xi)$,
\beq[eq:lab-delta-p']
	\Delta^{\mu}_{D} = \pp{ \Delta^+, \, \frac{t + \Delta_T^2}{2 \Delta^+}, \, \bm{\Delta}_T }, \quad
	p^{\prime\mu}_{D} = \pp{ \frac{1 - \xi}{1 + \xi} p^+, \, \frac{1 + \xi}{1 - \xi} \frac{m^2 + \Delta_T^2}{2p^+}, \, - \bm{\Delta}_T },
\eeq
with $\Delta^+ = 2\xi p^+ / (1 + \xi)$, $\bm{\Delta}_T = (\Delta_T, 0)$, and $\Delta_T$ given by
\beq[eq:delta-T]
	\Delta_T^2 = \frac{(1 - \xi^2) (-t) - 4 \xi^2 m^2}{(1 + \xi)^2}.
\eeq
The spin vector of the proton can be obtained from $S_{\rm rest}^{\mu} = (0, \bm{S}_T, P_N)_{\rm c}$ in the proton rest frame
by boosting together with its momentum,
\beq[eq:lab-S]
	S^{\mu}_{D} = \pp{ P_N \frac{p^+}{m}, \, - P_N \frac{m}{2p^+}, \, \bm{S}_T },
\eeq
where $\bm{S}_T = (S^x_T, S^y_T) = S_T (\cos\phi_S, \sin\phi_S)$.
Setting $p^+ = m/\sqrt{2}$ in Eqs.~\eqref{eq:lab-p}--\eqref{eq:lab-S} brings us into the proton rest frame.

The c.m.\ energy squared of the $2\to2$ hard scattering is
\beq[eq:sdhep-shat]
	\hat{s} = (\Delta + \l)^2 = t + 2 \Delta^+ \l^-,
\eeq
which is written in terms of the kinematic variables in the diffractive frame.
This can be related to the overall c.m.\ energy squared,
\beq[eq:sdhep-s-approx]
	s = (p + \l)^2 = m^2 + 2p^+\l^-,
\eeq 
by
\beq[eq:sdhep-s-hat-approx]
	\hat{s} = t + \frac{2\xi (s - m^2)}{1 + \xi}
		\simeq \frac{2 \, \xi}{1 + \xi} s + \order{\frac{m^2}{s}, \frac{t}{q_T^2}}.
\eeq

Up to a longitudinal boost, the nucleon momenta and spin vector in the SDHEP frame
can be obtained by boosting Eqs.~\eqref{eq:lab-p}--\eqref{eq:lab-S} according to \eq{eq:lorentz-trans-sdhep},
\bse\label{eq:sdhep-p-p'-s}\begin{align}
	p^{\mu}_{S} & = 
		\pp{ p^+, 
			\, \frac{m^2}{2p^+} \bb{ 1 + \frac{(1+\xi)^2}{4\xi^2} \frac{\Delta_T^2}{m^2} }, 
			\, - \frac{1 + \xi}{2\xi}\bm{\Delta}_T },
	\\
	p^{\prime\mu}_{S} & =
		\pp{ \frac{1-\xi}{1+\xi} p^{+}, 
			\, \frac{1+\xi}{1-\xi} \frac{m^2}{2p^+} \bb{ 1 + \frac{(1 + \xi)^2}{4 \xi^2 } \frac{\Delta_T^2}{m^2} }, 
			\, - \frac{1 + \xi}{2\xi}\bm{\Delta}_T },
	\\
	S^{\mu}_{S} & =
		\pp{  P_N \frac{p^+}{m},  
			\, P_N \frac{m}{2p^+} \bb{ \pp{ \frac{1 + \xi}{2\xi} \frac{\Delta_T}{m} }^2 - 1} - \frac{1 + \xi}{2\xi} \frac{\bm{\Delta}_T\cdot \bm{S}_T}{p^+}, 
			\, \bm{S}_T - P_N \frac{\bm{\Delta}_T}{m} \frac{1 + \xi}{2\xi} } ,
	\label{eq:spin-v-sdhep}
	\\
	\Delta^{\mu}_{S} & =	
		\pp{  \Delta^+, 
		         \, \frac{t}{2\Delta^+}, 
		         \, \bm{0}_T } ,	
\end{align}\ese
whereas $\ell^{\mu}_{S} = \ell^{\mu}_{D}$ stays unchanged. 
The $\Delta_S^\mu$ and $\ell_S^\mu$ define the collision axis of SDHEP frame. 
For simplicity, we suppress the subscript ``$S$'' (indicating the SDHEP frame) for all momenta in the rest of this paper.
Evidently, in this frame, $p$ and $p'$ have the same transverse components $\bm{p}_T = -(1 + \xi) \bm{\Delta}_T / (2\xi)$ 
along the $-\hat{x}_S$ direction, just as shown in \fig{fig:sdhep-frame}.
The $\bm{\Delta}_T$ in \eq{eq:sdhep-p-p'-s} is the same as in \eq{eq:lab-delta-p'} and $\bm{\Delta}_T\cdot \bm{S}_T = \Delta_T S_T \cos\phi_S$.

In the SDHEP frame, i.e., the c.m.\ of $A^*(\Delta)$ and $e(\l)$, we have $\Delta^+ = \sqrt{\hat{s} / 2}$,
so setting the $p^+$ and $P^+$ to
\beq[eq:sdhep-p+]
	p^+ = (1 + \xi) P^+ = \frac{1 + \xi}{2\xi} \Delta^+ = \frac{1 + \xi}{2\xi} \sqrt{\frac{\hat{s}}{2}}
\eeq
will take \eq{eq:sdhep-p-p'-s} to the SDHEP frame.
In this frame, once we specify the observed final-state momentum $\ell'$ for the scattered electron or $q'$ for the produced photon, all momenta involved are uniquely fixed.  For example, if we choose $\ell'^{\mu}=\sqrt{\hat{s}}/2 \left(1, \sin\theta\cos\phi, \sin\theta\sin\phi,\cos\theta\right)_{\rm c}$, we obtain the virtual photon momentum $q$ of the DVCS subprocess and the produced photon momentum $q'$ in the SDHEP frame as
\bse\label{eq:sdhep-q-q'}\begin{align}
	q^{\mu} &= (\ell - \ell')^\mu 
	\approx \sqrt{\hat{s}}/2 \left(0, -\sin\theta\cos\phi, -\sin\theta\sin\phi, -(1+\cos\theta)\right)_{\rm c},
	\\
	q'^{\mu} &= (\Delta + \ell - \ell')^{\mu}
	= \sqrt{\hat{s}}/2 \left(1, -\sin\theta\cos\phi, -\sin\theta\sin\phi, -\cos\theta\right)_{\rm c},
\end{align}\ese
where ``$\approx$" is valid at the leading power in $|t|/\hat{s}$.

\subsection{Azimuthal dependence}
The two-stage description gives a clear physical understanding of the azimuthal distributions of both $\phi$ and $\phi_S$.
In both subprocesses, the initial states are along the $z$ axis such that the azimuthal distributions are solely determined by their spin states.
Schematically, the channel expansion in \fig{fig:dvcs-channel} of the full scattering amplitude can be written as
\beq[eq:sdhep-M-decomp-A-phi]
	\M_{\lambda_N \lambda_e}^{N e \to N e \gamma}(t, \xi, \phi_S, \theta, \phi) 
	= \sum_{A} \int \prod_i^{n_A-1} dx_i
	  	\bb{ e^{-i \lambda_N \phi_S} F^{A}_{\lambda_N}(t, \xi, \{x_i\} ) }
		\bb{ e^{i\pp{\lambda_A - \lambda_e} \phi} C^{A}_{\lambda_A \lambda_e}(\{x_i\} , \hat{s}, \theta) },
\eeq
where $x_i$ with $i=1,2, ..., n_A-1$ are independent parton momentum fractions,
with $n_A\geq 2$ referring to the number of exchanged active and physically polarized partons in the exchanged states $A^*$ in \fig{fig:dvcs-channel}. 
The $\{x_i\}$ reduces to the $x$ in Eq.~(\ref{eq:dvcs-T-LP}) at $n_A=2$,
while the $n_A = 1$ channel corresponds to the exchange of a virtual photon of the BH subprocess,
for which no convolution is involved.
In \eq{eq:sdhep-M-decomp-A-phi}, we have factored out the azimuthal dependence explicitly in both stages by rotation symmetry, 
$F^{A}$ is the hadron structure function associated with the diffraction $N \to N + A^*$, 
and $C^{A}$ is the corresponding amplitude of the hard scattering $A^*e \to e\gamma$.
The helicities $\lambda_N$ and $\lambda_e$ of the initial states are defined in the diffractive frame, with $\lambda_e$ being Lorentz invariant.
The helicity $\lambda_A$ of the intermediate state $A^*$ is given in the SDHEP frame.
It receives a precise definition only after we separate the two stages by factorization.

\begin{figure}[htbp]
	\centering
		\includegraphics[scale=0.55]{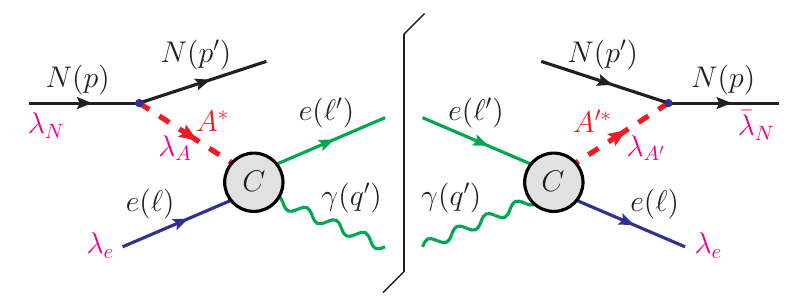}
	\caption{Interference of the $A^*$ channels in \eq{eq:M2-phi}. 
		Helicity indices are labeled using the same notations as \eq{eq:M2-phi}.}
\label{fig:sdhep-aa-interf}
\end{figure}

The cross section is obtained by squaring \eq{eq:sdhep-M-decomp-A-phi} and 
contracting with the beam and target spin density matrices,
\begin{align}
	\overline{|\M|^2}
	&= \sum_{\lambda_N, \bar{\lambda}_N, \lambda_e, \bar{\lambda}_e} 
	        \rho^{N}_{\lambda_N \bar{\lambda}_N} 
	        \rho^{e}_{\lambda_e\bar{\lambda}_e} 
		\M_{\lambda_N \lambda_e} \M_{\bar{\lambda}_N \bar{\lambda}_e}^*	
		\nn\\
	&=  \sum_{A, A'}  
		\bigg[ \sum_{\lambda_N, \bar{\lambda}_N}
		\rho^{N}_{\lambda_N \bar{\lambda}_N} \, e^{-i (\lambda_N - \bar{\lambda}_N) \phi_S}
			\, F^{A}_{\lambda_N} \bigpp{ F^{A'}_{\bar{\lambda}_N} }^*
		\bigg]
		\nn\\
	& {\hskip 0.3in}
		\otimes
		\bigg[ \sum_{\lambda_e,\bar{\lambda}_e} 
		\rho^{e}_{\lambda_e\bar{\lambda}_e}
		e^{i (\lambda_A - \lambda_{A'}) \phi} \, e^{-i (\lambda_e - \bar\lambda_{e}) \phi} \,
			\, C^{A}_{\lambda_A \lambda_e} \bigpp{ C^{A'}_{\lambda_{A'} \bar{\lambda}_e}  }^*
		\bigg],
		\nn\\
	&= \sum_{A, A'}  
		\bigg[ \sum_{\lambda_N, \bar{\lambda}_N}
		\rho^{N}_{\lambda_N \bar{\lambda}_N} \, e^{-i (\lambda_N - \bar{\lambda}_N) \phi_S}
			\, F^{A}_{\lambda_N} \bigpp{ F^{A'}_{\bar{\lambda}_N} }^*\bigg]
		\otimes
		\bigg[ \sum_{\lambda_e}
		\rho^{e}_{\lambda_e \lambda_e}
		e^{i (\lambda_A - \lambda_{A'}) \phi} \, 
			\, C^{A}_{\lambda_A \lambda_e} \bigpp{ C^{A'}_{\lambda_{A'} \lambda_e}  }^*
		\bigg], 
\label{eq:M2-phi}
\end{align}
where the symbol ``$\otimes$" represents the convolution of parton momentum fractions $ \prod_{i,j}^{n_A-1} dx_i\, dx'_j$. In \eq{eq:M2-phi}, 
the nucleon density matrix $\rho^{N}_{\lambda_N \bar{\lambda}_N}$
is given by the components of the spin vector in \eq{eq:lab-S},
and the electron density matrix $\rho^{e}_{\lambda_e\bar{\lambda}_e}$ is determined by the longitudinal polarization degree $P_e$,
\beq[eq:den-mat]
	\rho^{N}_{\lambda_N \bar{\lambda}_N} = \frac{1}{2}\begin{pmatrix}
		1 + P_N & S_T e^{-i \phi_S} \\
		S_T e^{i \phi_S} & 1 - P_N
	\end{pmatrix}_{\lambda_N \bar{\lambda}_N}, 
	\qquad
	\rho^{e}_{\lambda_e \bar{\lambda}_e} = \frac{1}{2}\begin{pmatrix}
		1 + P_e & 0 \\
		0 & 1 - P_e
	\end{pmatrix}_{\lambda_e \bar{\lambda}_e} ,
\eeq
where we take the electron density matrix to be diagonal, 
omitting the case with a transversely polarized electron beam (which is suppressed by the electron mass).
With the diagonal electron density matrix, we have organized \eq{eq:M2-phi} in the last step to make explicit the physical mechanisms for the azimuthal modulations of $\phi_S$ and $\phi$.
Both of them are due to interference of different helicity states, but the $\phi_S$ modulations come from that of the {\it same} particle $N$,
whereas the $\phi$ modulations involve interference of {\it different} particle states---including different particle species and numbers.
This is shown at the cut diagram level in \fig{fig:sdhep-aa-interf}.
It is this multi-particle interference that distinguishes the $\phi$ modulations in SDHEPs from those in inclusive processes~\cite{Bacchetta:2006tn, Bacchetta:2008xw, Lam:1978pu}.

For the nucleon target of spin-$1/2$, one can generate $\cos\phi_S$ and $\sin\phi_S$ modulations when there is a nonzero transverse spin $S_T$ 
which causes the two opposite helicity states to interfere.
For the $\phi$ modulations, 
the interference between $\lambda_A$ and $\lambda_{A^{\prime}}$ helicities leads to 
\beq[eq:A-A'-int]
	\cos(\lambda_A - \lambda_{A^{\prime}})\phi
	\; \mbox{ and/or } \;
	\sin(\lambda_A - \lambda_{A^{\prime}})\phi.
\eeq 
While the electron helicities do not interfere to enter \eq{eq:A-A'-int},
for a certain interference of $\lambda_A$ and $\lambda_{A^{\prime}}$ channels, 
the two terms $\lambda_e = \pm 1 / 2$ in \eq{eq:M2-phi} may not give the same contribution. 
In fact, the two modulations in \eq{eq:A-A'-int} have opposite parity properties.
Under a parity transformation followed by a rotation around the $\hat{y}_{S}$ axis by $\pi$ in \fig{fig:sdhep-frame},
the relevant variables change by 
\beq[eq:parity]
	(S_T, P_N, P_e, t, \xi, \theta, \phi_S, \phi)
	\to 
	(S_T, -P_N, -P_e, t, \xi, \theta, \pi - \phi_S, -\phi),
\eeq
so that $\sin\phi_S$ and $\cos(\lambda_A - \lambda_{A^{\prime}})\phi$ are parity even (P even),
while $\cos\phi_S$ and $\sin(\lambda_A - \lambda_{A^{\prime}})\phi$ are parity odd (P odd).
As we will see in \eq{eq:dvcs-LP-NLP-xsec}, these properties 
associate each polarization setting with specific azimuthal modulations.

In the photon electroproduction [\eq{eq:dvcs-sdhep}], we have three channels within the NLP,
with $A^*$ being a virtual photon $\gamma^*$, a pair of quark and antiquark $[q\bar{q}]$ or gluons $[gg]$.
For $A^* = \gamma^*$, we have three helicity states $\lambda_A^{\gamma} = (+1, 0, -1)$.
For $A^* = [q\bar{q}]$, only the unpolarized and polarized quark GPDs, $F^q$ and $\wt{F}^q$, contribute within NLP, 
both of which have $\lambda_A^{q\bar{q}} = 0$,
whereas for $A^* = [gg]$, 
all the gluon GPDs can contribute, of which $F^g$ and $\wt{F}^g$ have $\lambda_A^{gg} = 0$ 
while the transversity GPD $F^g_T$ has $\lambda_A^{gg} = \pm 2$.
As we will see in \sec{ssec:bh-amp}, among the three helicity states of the $\gamma^*$ channel, 
only the transverse polarizations $\lambda_A^{\gamma} = \pm 1$ belong to the LP, 
while the longitudinal polarization $\lambda_A^{\gamma} = 0$ is at NLP. 
So we regroup the different channels according to their powers,
\begin{enumerate}[leftmargin=4em, rightmargin=4em]
	\item [$\M_{\rm I}$:]	$A^* = \gamma_T^*$ with $\lambda_A^{\gamma} = \pm 1$;
	\item [$\M_{\rm II}$:] 	(1) $A^* = \gamma_L^*$ with $\lambda_A^{\gamma} = 0$; \;
					(2) $A^* = [q\bar{q}]$ with $\lambda_A^{q\bar{q}} = 0$; \;
					(3) $A^* = [gg]$ with $\lambda_A^{gg} = 0$; \\
					(4) $A^* = [gg]_T$ with $\lambda_A^{gg} = \pm 2$.
\end{enumerate}
The NLP approximation includes the square $|\M_{\rm I}|^2$ and the interference $\Re\bb{\M_{\rm I} \M_{\rm II}^*}$.
At LP, the two different $\gamma_T^*$ helicities do not interfere, so $|\M_{\rm I}|^2$ gives no $\phi$ dependence.
At NLP, the interference of $A^* = \gamma_T^*$ and $A^* = \gamma_L^*, [q\bar{q}]$, or $[gg]$ gives $\cos\phi$ and $\sin\phi$,
whereas the interference of $A^* = \gamma_T^*$ and $A^* = [gg]_T$ gives $\cos\phi$, $\sin\phi$, $\cos3\phi$, and $\sin3\phi$.

In this paper, we only work to the LO of QCD coupling $\alpha_s$, so will not include a detailed discussion of the gluon GPDs.
But it is already interesting to notice that the gluon transversity GPD gives new azimuthal correlations $\cos3\phi$ and $\sin3\phi$
which help to single them out.
Also, we note that in the kinematic region where the DVCS subprocess has an appreciable rate, the $q_T$ is in practice not too large,
so that $\alpha_s \sim \order{0.2}$ might be of the same order as the power expansion parameter $\sqrt{-t} / q_T$.
Therefore, it is likely to be numerically important to incorporate twist-3 GPDs together with the twist-2 gluon GPDs.
These involve three-parton channels $[q\bar{q}g]$ and $[ggg]$ in $A^*$, which carry helicities 
$\lambda_A^{q\bar{q}g} = \pm 1$ and $\lambda_A^{ggg} = \pm 1, \pm 3$.
Their interference with the $\gamma_T^*$ channel will further generate new $\cos2\phi$, $\sin2\phi$, $\cos4\phi$, and $\sin4\phi$
modulations, which again allow the twist-3 GPDs to be separated from the twist-2 ones.
In general, higher-twist GPDs can give higher-helicity parton states in $A^*$, which generate higher azimuthal modulation components
to enrich the observational signals and also in turn allow them to be singled out.
A detailed discussion of the contributions from high-twist GPDs are beyond the scope of this paper. 
They have, however, been studied extensively in the literature~\cite{Anikin:2000em, Penttinen:2000dg, Belitsky:2000vx, Kivel:2000cn, Radyushkin:2000ap, Diehl:2003ny, Belitsky:2001ns, Belitsky:2005qn, Kriesten:2019jep, Guo:2022cgq, Kivel:2000rb, Radyushkin:2000jy, Radyushkin:2001fc, Aslan:2018zzk, Blumlein:2006ia, Blumlein:2009hit, Braun:2011zr, Braun:2011dg, Braun:2012bg, Braun:2012hq, Braun:2014sta} in the Breit frame.
We hope to incorporate them in the SDHEP framework in a future work.

\section{Calculation in the SDHEP framework}
\label{sec:calc-sdhep}

The two-stage picture along with the SDHEP frame is suitable not only for organizing the factorization argument and expressing the kinematics,
but in fact also for calculating the amplitudes for the $2\to2$ hard scattering subprocesses 
because their initial states, especially the $A^*$, are exactly along the $\hat{z}_S$ axis,
making it convenient to use the helicity state of $A^*$. 
In this Section, we adopt the SDHEP framework in \sec{sec:dvcs-sdhep} and 
reformulate the calculation of the photon electroproduction in \eq{eq:dvcs-sdhep}.
We will first calculate the amplitudes of the one-particle $\gamma^*$ channel and two-parton $[q\bar{q}]$ channel,
and then combine them to get the cross section at NLP accuracy.
\footnote{This calculation was first done in \refcite{Yu:2023shd}. The presentation in this Section corrects one mistake thereof.
The result has been presented in \refcite{Qiu:2024reu}.}

\subsection{Amplitude of the $\gamma^*$ channel: Bethe-Heitler subprocess}
\label{ssec:bh-amp}

The amplitude of the $\gamma^*$ channel (i.e., the BH subprocess) counts as the leading power of $\sqrt{-t} / q_T$, scaling as $\O(1 / \sqrt{-t})$, to be shown shortly.
In the SDHEP frame, the $\gamma^*$ carries momentum $\Delta$
and collides with the electron in the c.m.\ frame,
\beq[eq:dvcs-bh]
	\gamma^*(\Delta, \lambda) + e(\l, \lambda_e) \to e(\l', \lambda'_e) + \gamma(q', \lambda_2),
\eeq
where $\lambda_e$, $\lambda'_e$, and $\lambda_2$ label the helicities of the three on-shell particles, and
the helicity index $\lambda$ of $\gamma^*$ is to be specified below. 
One difference from the DVCS amplitude calculation in \sec{ssec:dvcs-amp} is that 
here we need to keep $\Delta$ as exact because neglecting its virtuality may induce an error of order $1/q_T$ in the amplitude, 
which is suppressed with respect to the BH, but {\it not} to the DVCS. 
Denoting $\hat{s}$ as the {\it unapproximated} c.m.\ energy squared of the hard collision system, we have the kinematics,
\begin{align}
	\Delta &= \frac{\sqrt{\hat{s}}}{2} \pp{ \frac{\hat{s} + t}{\hat{s}}, 0, 0, \frac{\hat{s} - t}{\hat{s}} }_{\rm c},
	&\l &= \frac{\sqrt{\hat{s}}}{2} \pp{ \frac{\hat{s} - t}{\hat{s}}, 0, 0, - \frac{\hat{s} - t}{\hat{s}} }_{\rm c},	\nn\\
	\l' &= \frac{\sqrt{\hat{s}}}{2} \pp{ 1, \bm{q}_1 / |\bm{q}_1| }_{\rm c},
	&q' &= \frac{\sqrt{\hat{s}}}{2} \pp{ 1, -\bm{q}_1 / |\bm{q}_1| }_{\rm c},
\label{eq:bh-kins}
\end{align}
where $\bm{q}_1 = (\sqrt{\hat{s}} / 2) (\sin\theta \cos\phi, \sin\theta \sin\phi, \cos\theta)$ is determined by the angles $\theta$ and $\phi$.

The $\gamma^*$-channel amplitude trivially factorizes between the nucleon diffraction and $2\to2$ hard collision parts,
\begin{align}\label{eq:bh-FF}
	\M^{[1]}_{\lambda_e \lambda'_e \lambda_2} 
	&= -\frac{e}{t} \, \langle N(p', \lambda_N') | J^{\mu}(0) | N(p, \lambda_N) \rangle \cdot \langle e(\l', \lambda'_e) \gamma(q', \lambda_2) | \bb{ -i e J^e_{\mu}(0) } | e(\l, \lambda_e) \rangle	\nn \\
	&\equiv -\frac{e}{t} \, F^{\mu}(p, p') \, C^{\gamma}_{\mu}(\Delta, \l, \l', q'),
\end{align}
where $J^{\mu}$ and $F^{\mu}(p, p')$ are the same as in \eq{eq:EM-form-factor}, defining the form factors $F_{1,2}(t)$,
and $J^e_{\mu} = e_e \, \bar{\psi}_e \gamma_{\mu} \psi_e$ is the electron current with $e_e = -1$. 
In the second step of \eq{eq:bh-FF}, we defined the hard factor $C^{\gamma}_{\mu}$ to include the factor $-ie$.
As the nucleon system is highly boosted along the $z$ direction,
$F^{\mu}(p, p')$ has the leading component $F^+ \sim \order{\Delta^+}$.
In the SDHEP frame, the amplitude structure simplifies by use of Ward identities for both $F$ and $C^{\gamma}$,
\beq[eq:BH-ward]
	F \cdot \Delta = F^+ \Delta^- + F^- \Delta^+ = 0, 
	\quad
	C^{\gamma} \cdot \Delta = C^{\gamma +} \Delta^- + C^{\gamma -} \Delta^+ = 0, 
\eeq
such that 
\beq[eq:bh-aL-FH]
	F^+ C^{\gamma -} = F^- C^{\gamma +} 
	= - \frac{t}{2 (\Delta^+)^2} F^+ C^{\gamma +}.
\eeq
By the power counting $(F^+, F^-, \bm{F}_T) \sim (\Delta^+, t / \Delta^+, \sqrt{-t})$, we have
$(C^{\gamma +}, C^{\gamma -}, \bm{C}^{\gamma}_T) \sim \pp{1, t / (\Delta^+)^2, 1}$ in the SDHEP frame, instead of the superficial 
power counting $C^{\gamma}_{\mu} \sim \order{1}$.
Then the scalar product of $F$ and $C^{\gamma}_{\mu}$ is
\beq[eq:bh-F-H]
	F \cdot C^{\gamma} = 2 F^+ C^{\gamma -} - \bm{F}_T \cdot \bm{C}^{\gamma}_T
	= 2 (F\cdot n)(\bar{n} \cdot C^{\gamma}) - 
		\sum\nolimits_{\lambda = \pm} \bb{ F \cdot \epsilon_{\lambda}^*(\Delta)} \bb{\epsilon_{\lambda}(\Delta) \cdot C^{\gamma}}.
\eeq
Although this is only valid in the SDHEP frame, it equips the virtual photon with well-defined polarization states. 

The first term on the right-hand side of \eq{eq:bh-F-H} corresponds to a longitudinally polarized photon state $\gamma^*_L$, whose polarization vector 
\beq
	\epsilon_0^{\mu}(\Delta) \equiv \bar{n}^{\mu} = (1, 0, 0, 1)_{\rm c}  / \sqrt{2}
\eeq
is contracted with $C^{\gamma}$. 
Along with the $1/t$ factor from the photon propagator in \eq{eq:bh-FF}, it contributes to the amplitude at the power 
$1/t \times t/q_T = 1/q_T$, 
which is the same as the GPD channel. 
Within the NLP accuracy, we may therefore keep only the LP of $\sqrt{-t} / q_T$ 
when calculating the hard coefficient $\bar{n} \cdot C^{\gamma}$.

The second term in \eq{eq:bh-F-H} corresponds to two transverse polarization states $\gamma^*_T$, with the polarization vectors defined as
\beq[eq:ga*-pol-v]
	\epsilon_{\pm}^{\mu}(\Delta) = (0, \mp 1, - i, 0)_{\rm c} / \sqrt{2}.
\eeq
They contribute to the amplitude with a power counting $1/t \times \sqrt{-t} \sim 1 / \sqrt{-t}$, 
which is one power higher than the GPD channel.
Hence within the same NLP accuracy, 
we need to keep the hard coefficient $\epsilon_{\lambda} \cdot C^{\gamma}$ up to the subleading power of $\sqrt{-t} / q_T$.

The above discussion of the $\gamma^*$ channel is actually generic to all SDHEPs. 
For the photon electroproduction, the subprocess in \eq{eq:dvcs-bh} is an elementary scattering.
As we will see through explicit calculations, the $t$ dependence in 
$\epsilon_{\pm} \cdot C^{\gamma}$ arises from kinematic effects. 
Since there is no singularity associated with $t\to 0$ in $C^{\gamma}$, the subleading term starts at $t / q_T^2$ instead of $\sqrt{-t} / q_T$, 
so is power suppressed with respect to the GPD channel. 
Thus, it is a valid approximation to also neglect $t$ in the transverse amplitudes.

\begin{figure}[htbp]
\centering
	\includegraphics[clip, trim={-1em 0 -2em 0}, scale=0.75]{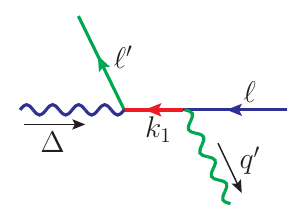}	
	\includegraphics[clip, trim={-2em 0 -1em 0}, scale=0.75]{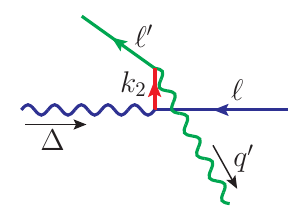}	
	\caption{LO diagrams for the BH subprocess, initialized by the virtual photon state $\gamma^*$.
	The red thick lines indicate propagators with high virtualities, which belong to the hard part.}
\label{fig:bh-lo}
\end{figure}

The hard scattering diagrams for the BH amplitude are shown in \fig{fig:bh-lo}. 
Denoting these hard coefficients as 
$C^{\gamma}_{\lambda, \lambda_e \lambda'_e \lambda_2} \equiv \epsilon_{\lambda}(\Delta) \cdot C^{\gamma}_{\lambda_e \lambda'_e \lambda_2}$ with 
$\lambda = \pm1$ or $0$ being the $\gamma^*$ helicity, 
we have
\begin{align}
	C^{\gamma}_{\lambda, \lambda_e \lambda'_e \lambda_2}
	& = -i e^2 \, \bar{u}_2 \bb{ \frac{1}{k_1^2} \slash{\epsilon}_{\lambda} \slash{k}_1 \slash{\epsilon}^*_{\lambda_2} 
		+ \frac{1}{k_2^2} \slash{\epsilon}^*_{\lambda_2} \slash{k}_2 \slash{\epsilon}_{\lambda} } u_1,
\label{eq:bh-amp-0}
\end{align}
where $k_1 = \l - q'$, $k_2 = \Delta + \l$,
$u_1 = u(\l, \lambda_e)$ and $u_2 = u(\l', \lambda'_e)$ are spinors of the initial- and final-state electrons, respectively, 
and $\epsilon_{\lambda} = \epsilon_{\lambda}(\Delta)$ and $\epsilon_{\lambda_2} = \epsilon_{\lambda_2}(q')$ 
are polarization vectors of the initial- and final-state photons, respectively.
The helicity amplitudes for $\gamma^*_L$ are
\beq[eq:bh-H-0]
	C^{\gamma}_{0, \pm\pm\mp}
	= -2e^2 \frac{t}{\hat{s}} \frac{\cos(\theta/2)}{\sqrt{1 - t/\hat{s}}} e^{\mp i \phi / 2}
	= -2e^2 \frac{t}{\hat{s}} \cos(\theta/2) e^{\mp i \phi / 2} + \order{t^2 / \hat{s}^2},
\eeq
which start at $\order{t/\hat{s}}$ as argued in \eq{eq:bh-aL-FH}. In the second step we only kept the leading terms.
The transverse photon polarizations give
\begin{align}
	C^{\gamma}_{\pm, \pm\pm\pm}
		& =  \mp \frac{2e^2}{\sin(\theta/2)} \sqrt{1 - \frac{t}{\hat{s}}} \, e^{\pm i \phi / 2}
		&& = \mp \frac{2e^2}{\sin(\theta/2)} \, e^{\pm i \phi / 2} + \order{t /\hat{s}}, \nn\\
	C^{\gamma}_{\pm, \pm\pm\mp}
		& =  \mp \frac{2 e^2 \cos^2(\theta/2)}{\sin(\theta/2)} \frac{t}{\hat{s}} \frac{1}{\sqrt{1 - t/\hat{s}}} \, e^{\pm i \phi / 2}
		&& = 0 + \order{t /\hat{s}}, \nn\\
	C^{\gamma}_{\pm, \mp\mp\pm}
		& =  \pm 2e^2 \sin(\theta/2) \frac{1}{\sqrt{1 - t/\hat{s}}} \, e^{\pm 3 i \phi / 2}
		&& = \pm 2e^2 \sin(\theta/2) \, e^{\pm 3 i \phi / 2} + \order{t /\hat{s}},
\label{eq:bh-H-t}
\end{align}
which scale as $\O(1)$, with subleading terms of $\order{t/\hat{s}}$.
All other helicity amplitudes are zero.

Note that the amplitudes $C^{\gamma}_{\pm, \pm\pm\mp}$ only contribute if we work up to next-to-next-to-leading-power (NNLP) precision. 
Then they would cause interference with $C^{\gamma}_{\pm, \mp\mp\pm}$ and give $\cos2\phi$ and $\sin2\phi$ azimuthal modulations.
To that precision, one shall not make the approximations in \eqs{eq:bh-H-0}{eq:bh-H-t}.

Together with Eqs.~\eqref{eq:bh-FF} and \eqref{eq:bh-F-H}, 
\eqs{eq:bh-H-0}{eq:bh-H-t} determine the helicity amplitudes for the $\gamma^*$ channel,
\bse\label{eq:BH-helicity-amplitudes}\begin{align}
	\M^{[1]}_{\pm\pm\pm} 
		& = \mp \frac{2e^3}{t} \, F_{\pm} \, \frac{e^{\pm i \phi/2}}{\sin(\theta/2)}  + \order{\sqrt{-t} / q_T^2 }, \\
	\M^{[1]}_{\pm\pm\mp} 
		& = \mp \frac{2e^3}{t} \, F_{\mp} \, \sin(\theta/2) e^{\mp 3 i \phi/2} 
		+ \frac{4e^3}{\hat{s}} \, F_0 \, \cos(\theta/2) e^{\mp i \phi/2}  + \order{\sqrt{-t} / q_T^2},
\end{align}\ese
which are the only four nonzero independent helicity amplitudes up to NLP.
Each of them are obtained by summing over the helicity of the intermediate virtual photon $\gamma^*$. 
The diffracted nucleon helicities are implicitly encoded in the EM form factors 
$F_0 = F \cdot n = F^+$ and $F_{\pm} = F \cdot \epsilon^*_{\pm}$. 
Note that $\M^{[1]}_{\pm\pm\pm}$ contains a singularity at $\theta = 0$,
where the photon $\gamma(q')$ is collinearly radiated by the electron $e(\l)$.

\subsection{Amplitude of the $[q\bar{q}]$ channel: Twist-2 DVCS at LO}
\label{ssec:dvcs-amp}

\begin{figure}[htbp]
\centering
	\includegraphics[trim={-1em 0 -2em 0}, clip, scale=0.75]{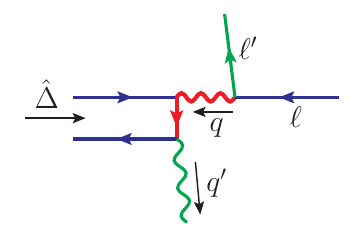}	
	\includegraphics[trim={-2em 0 -1em 0}, clip, scale=0.75]{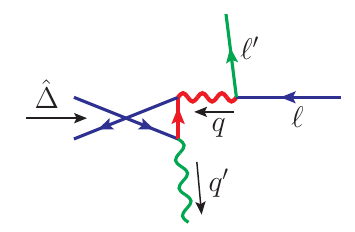}	
	\caption{LO hard scattering diagrams of the DVCS, initialized by the two-parton state $[q\bar{q}]$. 
	The red thick lines indicate propagators with high virtualities, which belong to the hard part.}
\label{fig:dvcs-lo}
\end{figure}

To the LO of $\alpha_s$, the two-parton channel is mediated by a $[q\bar{q}]$ state (considered in a physical gauge with $A\cdot n = 0$),
as shown in the second term of the right-hand side in \fig{fig:dvcs-channel}. 
In the limit $t / q_T^2 \to 0$, the two quark momenta $k_q$ and $\Delta - k_q$ are pinched on shell (see, for example, the argument in \refcite{Qiu:2022bpq}).
At a small but finite $t$, this pinch singularity is no longer exact, but only up to the extent $k_q^2 \sim (\Delta - k_q)^2 \sim \O(t)$.
Instead of giving a collinear divergence, the momentum region around that pinch surface gives an important contribution,
which can be extracted following the standard approximations in Eqs.~\eqref{eq:V-decomp}--\eqref{eq:dvcs-T-factorize} 
(one can find more details in, e.g., \refs{Qiu:2022bpq, Qiu:2022pla, Collins:1996fb, Collins:1998be, Collins:1989gx, Collins_2011, Yu:2023shd}),
with successive subtractions to avoid double counting of overlapped regions.
This results in a factorization formula of the two-parton channel amplitude, up to errors of $\O(\sqrt{-t} / q_T)$,
\begin{align}
	\M^{[2]}_{\lambda_e \lambda'_e \lambda_2}
		= \sum_q \int_{-1}^1 dx \bb{ F^q(x, \xi, t) \, C^q_{\lambda_e \lambda'_e \lambda_2}(x, \xi; \hat{s}, \theta, \phi)
			+ \wt{F}^q(x, \xi, t) \, \wt{C}^q_{\lambda_e \lambda'_e \lambda_2}(x, \xi; \hat{s}, \theta, \phi) },
\label{eq:dvcs-factorize-M}
\end{align}
where the quark GPDs $F^q$ and $\wt{F}^q$ are defined as \eq{eq:GPD-def-q}, 
with the $n \propto \l$ fixed by \eq{eq:bh-kins} in the SDHEP frame.
As commented in \sec{ssec:trans-boost}, since $n$ is invariant under a transverse boost or a longitudinal boost along the $z$ direction,
the GPDs $F^q$, $\wt{F}^q$, $H^q$, $E^q$, $\wt{H}^q$, and $\wt{E}^q$ are the same in both the diffractive and the SDHEP frames,
and so are $\xi$ and the parton momentum fraction $x$.
In \eq{eq:dvcs-factorize-M}, we have suppressed the factorization scale and omitted the gluon-channel terms within the LO accuracy.
Here $\hat{s}$, $\theta$, $\phi$, $\lambda_e$, $\lambda'_e$, and $\lambda_2$ are the same as in 
\eqs{eq:dvcs-bh}{eq:bh-kins} for the $\gamma^*$ channel.
The LP approximation of $\hat{s}$ is given in \eq{eq:sdhep-s-hat-approx}.

The hard coefficients $C^q$ and $\wt{C}^q$ in \eq{eq:dvcs-factorize-M} are given by the scattering of a collinear $[q\bar{q}]$ pair of zero net helicity
with an electron, 
\beq[eq:dvcs-hard]
	[q\bar{q}](\hat{\Delta}, 0) + e(\l, \lambda_e) \to e(\l', \lambda'_e) + \gamma(q', \lambda_2),
\eeq 
as given by the two diagrams in \fig{fig:dvcs-lo}.
To the LP of the DVCS amplitude, we have projected $\Delta$ on shell, $\hat{\Delta} = (\Delta \cdot n) \bar{n}$.
The other momenta $\l$, $\l'$, and $q'$ are the same as \eq{eq:bh-kins}, where $t$ can also be neglected with respect to $\hat{s}$.
The SDHEP frame is the c.m.\ of the hard scattering in \eq{eq:dvcs-hard}.
By introducing two auxiliary lightlike vectors $w$ and $\bar{w}$,
\beq[eq:dvcs-w-wbar]
	\bar{w} = \frac{1}{\sqrt{2}} \pp{ 1, \bm{q}_1 / |\bm{q}_1|}_{\rm c}, \quad
	w = \frac{1}{\sqrt{2}} \pp{ 1, -\bm{q}_1 / |\bm{q}_1|}_{\rm c},
	\quad
	w \cdot \bar{w} = 1,
\eeq
the kinematics can be described to the LP of DVCS as
\beq
	\hat{\Delta} = \sqrt{\hat{s}/2} \,\bar{n}, \quad
	\l = \sqrt{\hat{s}/2} \,n, \quad
	\l' = \sqrt{\hat{s}/2} \,\bar{w}, \quad
	q' = \sqrt{\hat{s}/2} \,w,
\eeq
with the simple scalar products,
\beq
	w \cdot n = \bar{w} \cdot \bar{n} = (1 - \cos\theta) / 2, \quad
	w \cdot \bar{n} = \bar{w} \cdot n = (1 + \cos\theta) / 2.
\eeq

As indicated by \eq{eq:dvcs-hard} and \fig{fig:dvcs-lo}, 
we calculate the hard coefficients by viewing both the quark $q$ and antiquark $\bar{q}$ as {\it entering} the hard interaction, 
carrying momenta $(\xi + x) \hat{P}$ and $(\xi - x) \hat{P}$, respectively, with $\hat{P} \equiv (P \cdot n) \bar{n}$. The results can be directly extrapolated to the full
$x$ range $[-1, 1]$ by keeping all relevant $i\epsilon$ prescriptions explicitly. 
Following the two-stage paradigm, it is convenient to introduce a variable change, as in \refs{Qiu:2022pla, Qiu:2023mrm, Qiu:2022bpq, Qiu:2024mny},
\beq[eq:x2z]
	x \to x_{\xi} = (x + \xi) / (2 \xi),
\eeq 
and then the two quarks carry momenta $x_{\xi} \, \hat{\Delta}$ and $(1 - x_{\xi}) \hat{\Delta}$, respectively. 
The hard coefficients $C^q$ and $\wt{C}^q$ are obtained by contracting the amputated parton lines with 
$\gamma \cdot \hat{\Delta} / 2$ and $\gamma_5\gamma \cdot \hat{\Delta} / 2$ (with an additional $1/2\xi$ factor), 
for $F^q$ and $\wt{F}^q$, respectively.

The P-even hard coefficient $C^q$ for the quark flavor $q$ is
\begin{align}
	2 \xi C^q_{\lambda_e \lambda'_e \lambda_2}
	= -\frac{i e_q^2 e^3}{q^2} \pp{ \bar{u}_2 \gamma_{\mu} u_1 \, \epsilon^*_{\lambda_2,\nu} }
		\bb{ \frac{1}{k_1^2 + i \epsilon} 
			\Tr\pp{\frac{\Slash{\hat{\Delta}}}{2} \gamma^{\nu} \slash{k}_1 \gamma^{\mu} }
		+ \frac{1}{k_2^2 + i \epsilon} 
			\Tr\pp{\frac{\Slash{\hat{\Delta}}}{2} \gamma^{\mu} \slash{k}_2  \gamma^{\nu}  }
		},
\label{eq:dvcs-amp-0}
\end{align}
where $u_1$, $u_2$, and $\epsilon_{\lambda_2}$ are the same as in \eq{eq:bh-amp-0},
and we have the following internal momentum definitions,
\begin{align}
	q & = \l - \l', &
	q^2 & = -2 \l \cdot \l' = - \hat{s} \, (1 + \cos\theta) / 2, \nn\\
	k_1 & = q' - (1-x_{\xi}) \hat{\Delta}, &
	k_1^2 &= -2(1-x_{\xi}) \hat{\Delta} \cdot q' = - (1-x_{\xi}) \, \hat{s} \, (1 + \cos\theta) / 2, \nn\\
	k_2 & = x_{\xi} \hat{\Delta} - q', &
	k_2^2 &= -2 x_{\xi} \hat{\Delta} \cdot q' = - x_{\xi} \, \hat{s} \, (1 + \cos\theta) / 2.
\end{align}
One can immediately notice that the two terms in the curly bracket in \eq{eq:dvcs-amp-0} are related to each other by a 
charge conjugation and $x_{\xi} \to 1 - x_{\xi}$, up to an overall minus sign, so we have
$C^q_{\lambda_e \lambda'_e \lambda_2}(x_{\xi}) = -C^q_{\lambda_e \lambda'_e \lambda_2}(1 - x_{\xi})$,
or 
\beq
	C^q_{\lambda_e \lambda'_e \lambda_2}(x, \xi) = -C^q_{\lambda_e \lambda'_e \lambda_2}(-x, \xi),
\eeq
in terms of the original GPD variable $x$.
This property goes beyond LO and ensures that only the charge-conjugation-even (C-even) unpolarized GPD component
\beq[eq:C-even-F]
	F^{q+}(x, \xi, t) \equiv F^q(x, \xi, t) - F^q(-x, \xi, t)
\eeq 
is probed by the DVCS.
Evaluating the fermion traces, we have
\begin{align}
	2 \xi C^q_{\lambda_e \lambda'_e \lambda_2}
		= -\frac{i e_q^2 e^3}{q^2} \pp{ \bar{u}_2 \gamma_{\mu} u_1 \, \epsilon^*_{\lambda_2,\nu} }
			\bb{ -g^{\mu\nu} + \frac{\bar{n}^{\mu} w^{\nu} + \bar{n}^{\nu} w^{\mu}}{\bar{n} \cdot w} }
			\bb{ \frac{1}{x_{\xi} - 1 + i \epsilon} + \frac{1}{x_{\xi} - i \epsilon} }.
\label{eq:dvcs-ce-tensor}
\end{align}

Similarly, replacing $\Slash{\hat{\Delta}}$ in \eq{eq:dvcs-amp-0} by $\gamma_5\Slash{\hat{\Delta}}$ gives the P-odd hard coefficient,
\begin{align}
	2 \xi \wt{C}^q_{\lambda_e \lambda'_e \lambda_2}
		= - \frac{i e_q^2 e^3}{q^2} \pp{ \bar{u}_2 \gamma_{\mu} u_1 \, \epsilon^*_{\lambda_2,\nu} }
			\frac{i \, \epsilon^{\mu \nu \bar{n} w}}{\bar{n} \cdot w}
			\bb{ \frac{1}{x_{\xi} - 1 + i \epsilon} - \frac{1}{x_{\xi} - i \epsilon} },
\label{eq:dvcs-co-tensor}
\end{align}
which is invariant under $x_{\xi} \to 1 - x_{\xi}$ (or equivalently, $x \to -x$) and thus probes the C-even polarized GPD component,
\beq[eq:C-even-Ft]
	\wt{F}^{q+}(x, \xi, t) \equiv \wt{F}^q(x, \xi, t) + \wt{F}^q(-x, \xi, t),
\eeq 
a property that also goes beyond LO.

Substituting their explicit expressions for the spinors and polarization vector in Eqs.~\eqref{eq:dvcs-ce-tensor} and \eqref{eq:dvcs-co-tensor},
we have the nonzero helicity amplitudes in the SDHEP frame,
\bse\label{eq:DVCS-hel-amps-hard}\begin{align}
	C^q_{\pm\pm\pm} & = +e_q^2 e^3 \sqrt{\frac{2}{\hat{s}}} 
		\cdot \frac{1}{2 \xi} \bb{ \frac{1}{x_{\xi} - 1 + i \epsilon} + \frac{1}{x_{\xi} - i \epsilon} } 
		\cdot \frac{1}{\cos^3(\theta/2)} \, e^{\mp i \phi / 2}, 
	\\
	C^q_{\pm\pm\mp} & = -e_q^2 e^3 \sqrt{\frac{2}{\hat{s}}} 
		\cdot \frac{1}{2 \xi} \bb{ \frac{1}{x_{\xi} - 1 + i \epsilon} + \frac{1}{x_{\xi} - i \epsilon} }  
		\cdot \frac{\sin^2(\theta/2)}{\cos^3(\theta/2)} \, e^{\mp i \phi / 2},
	\\
	\wt{C}^q_{\pm\pm\pm} & = \pm e_q^2 e^3 \sqrt{\frac{2}{\hat{s}}} 
		\cdot \frac{1}{2 \xi} \bb{ \frac{1}{x_{\xi} - 1 + i \epsilon} - \frac{1}{x_{\xi} - i \epsilon} } 
		\cdot \frac{1}{\cos^3(\theta/2)} \, e^{\mp i \phi / 2},
	\\
	\wt{C}^q_{\pm\pm\mp} & = \pm e_q^2 e^3 \sqrt{\frac{2}{\hat{s}}} 
		\cdot \frac{1}{2 \xi} \bb{ \frac{1}{x_{\xi} - 1 + i \epsilon} - \frac{1}{x_{\xi} - i \epsilon} } 
		\cdot \frac{ \sin^2(\theta/2) }{\cos^3(\theta/2)} \, e^{\mp i \phi / 2}.
\end{align}\ese
We see that the $x_{\xi}$ dependence factors out of the $\theta$ and $\phi$ dependence, while the latter are experimental observables.
In terms of $x$, the $x_{\xi}$ factor becomes
\beq[eq:dvcs-coefs-x]
	\frac{1}{2\xi} \bb{ \frac{1}{x_{\xi} - 1 + i \epsilon} \pm \frac{1}{x_{\xi} - i \epsilon} } 
	= \frac{1}{x - \xi + i \epsilon} \pm \frac{1}{x + \xi - i \epsilon} ,
\eeq
which gives the ``zeroth GPD moments'' when convoluted with $F^q$ and $\wt{F}^q$, respectively,
\beq[eq:dvcs-GPD-moments]
	F^{q, +}_0(\xi, t) \equiv \int_{-1}^1 dx \, \frac{F^{q, +}(x, \xi, t)}{x - \xi + i \epsilon}, \quad
	\wt{F}^{q, +}_0(\xi, t) \equiv \int_{-1}^1 dx \, \frac{\wt{F}^{q, +}(x, \xi, t)}{x - \xi + i \epsilon},
\eeq
Then the DVCS amplitudes in \eq{eq:dvcs-factorize-M} are
\bse\label{eq:dvcs-helicity-amplitudes}\begin{align}
	\M^{[2]}_{\pm\pm\pm}
	& = e^3 \sqrt{\frac{2}{\hat{s}}} \bb{ \frac{1}{\cos^3(\theta/2)} e^{\mp i \phi / 2} } \cdot \sum_q e_q^2 
			\bb{ F^{q, +}_0(\xi, t) \pm \wt{F}^{q, +}_0(\xi, t) }
		+ \order{\sqrt{-t} / q_T },	\\
	\M^{[2]}_{\pm\pm\mp}
	& = e^3 \sqrt{\frac{2}{\hat{s}}} \bb{ \frac{ \sin^2(\theta/2) }{\cos^3(\theta/2)} e^{\mp i \phi / 2} } \cdot \sum_q e_q^2 
			\bb{ -F^{q, +}_0(\xi, t) \pm \wt{F}^{q, +}_0(\xi, t) }
		+ \order{\sqrt{-t} / q_T },
\end{align}\ese
which are written as helicity amplitudes for the electrons and photon, 
constituting four nonzero independent helicity structures up to NLP, similarly to \eq{eq:BH-helicity-amplitudes}.
The nucleon helicities are implicitly encoded in the GPDs.

Contrary to \eq{eq:BH-helicity-amplitudes}, the singularity now occurs at $\theta = \pi$,
where the photon $\gamma(q')$ is radiated collinearly by the quarks.
By cutting $q_T$ below, $q_T \geq q_{T{\rm min}} \gg \sqrt{-t} $, we not only ensure the hard collision in the factorization formula [\eq{eq:dvcs-factorize-M}],
but also avoid collinear photon radiations in both the BH and DVCS subprocesses.

\subsection{Combining the BH and DVCS}
\label{ssec:bh-dvcs}

Up to the NLP accuracy, the full amplitude of the photon electroproduction is given by
the sum of \eqs{eq:BH-helicity-amplitudes}{eq:dvcs-helicity-amplitudes}.
The LP contribution starts at $\O(1/\sqrt{-t})$ and is given by the $\gamma^*_T$ channel (in the BH subprocess),
\begin{align}\label{eq:LP-amplitudes}
	\M^{\rm LP}_{\lambda_N \lambda_N' \pm \pm \pm}
		= \pp{ F_{\lambda_N\lambda_N'} \cdot \epsilon^*_{\pm} } \A^T_{\pm \pm \pm}, 
	\quad
	\M^{\rm LP}_{\lambda_N \lambda_N' \pm \pm \mp}
		= \pp{ F_{\lambda_N\lambda_N'} \cdot \epsilon^*_{\mp} } \A^T_{\pm \pm \mp},
\end{align}
where we have made explicit the dependence on the nucleon helicities $\lambda_N$ and $\lambda_N'$ 
encoded in the EM form factor $F_{\lambda_N\lambda_N'}^{\mu}(p, p') = \langle N(p', \lambda_N') | J^{\mu}(0) | N(p, \lambda_N) \rangle$,
and the reduced hard scattering amplitudes $\A^T_{\lambda_e \lambda'_e \lambda_2}$ can be easily obtained by matching to \eq{eq:BH-helicity-amplitudes},
\beq[eq:bh-AT]
	\A^T_{\pm\pm\pm} = \mp \frac{2e^3}{t} \frac{e^{\pm i \phi / 2}}{\sin(\theta/2)}, \quad
	\A^T_{\pm \pm \mp} = \mp \frac{2e^3}{t} \sin(\theta/2) \, e^{\mp 3 i \phi / 2}.
\eeq
The NLP contribution is at $\O(1/q_T)$, given by the DVCS subprocess and the $\gamma^*_L$ channel (in the BH subprocess),
\bse\label{eq:NLP-amplitudes}\begin{align}
	\M^{\rm NLP}_{\lambda_N \lambda_N' \pm\pm\pm}
	& = ( \F_{\lambda_N\lambda_N'} \pm \Ft_{\lambda_N\lambda_N'} ) \, \G_{\pm \pm \pm} ,	\\
	\M^{\rm NLP}_{\lambda_N \lambda_N' \pm\pm\mp}
	& = ( -\F_{\lambda_N\lambda_N'} \pm \Ft_{\lambda_N\lambda_N'} ) \, \G_{\pm \pm \mp} 
		+ \pp{ F_{\lambda_N\lambda_N'} \cdot n } \A^L_{\pm \pm \mp},
\end{align}\ese
where we defined the shorthand notations for the GPD convolutions,
\begin{align}
	\F_{\lambda_N\lambda_N'} &= \F_{\lambda_N\lambda_N'}(\xi, t) 
		\equiv \sum\nolimits_q e_q^2 \, F^{q, +}_{0, \lambda_N \lambda_N'}(\xi, t)
		= \frac{1}{2P \cdot n} \bar{u}(p', \lambda_N') \bb{ \H(\xi, t) \, \gamma \cdot n - \E(\xi, t) \, \frac{i \sigma^{n \Delta}}{2m} } u(p, \lambda_N), 
	\nn\\
	\Ft_{\lambda_N\lambda_N'}  &= \Ft_{\lambda_N\lambda_N'}(\xi, t) 
		\equiv \sum\nolimits_q e_q^2 \, \wt{F}^{q, +}_{0, \lambda_N \lambda_N'}(\xi, t)
		= \frac{1}{2P \cdot n} \bar{u}(p', \lambda_N') \bb{ \Ht(\xi, t) \, \gamma \cdot n \gamma_5 - \Et(\xi, t) \, \frac{\gamma_5 \Delta  \cdot n}{2m} } u(p, \lambda_N),
\label{eq:F-Ft-aa'}
\end{align}
with similar definitions for the complex-valued GPD moments $\H$, $\E$, $\Ht$, and $\Et$,
\footnote{It is the real and imaginary parts of these moments $\H$, $\E$, $\Ht$, and $\Et$ 
that are referred to as ``Compton form factors'' in some literature (see, e.g., \refcite{Kumericki:2016ehc}),
but we reserve this terminology for the nonperturbative definitions given in \eq{eq:CFF-decomp}.}
\beq[eq:dvcs-GPD-form-factors]
	\bigcc{ \H, \E, \Ht, \Et }(\xi, t)
		\equiv 
		\sum\nolimits_q e_q^2 \, \bigcc{ H^{q, +}_0, E^{q, +}_0, \wt{H}^{q, +}_0, \wt{E}^{q, +}_0 }(\xi, t).
\eeq 
These moments agree with the LO CFFs in \eq{eq:cff-t2} up to a minus sign for $(\Ht, \Et)$.
The associated hard coefficients $\G_{\lambda_e \lambda'_e \lambda_2}$ in \eq{eq:NLP-amplitudes} 
can be obtained by matching with \eq{eq:dvcs-helicity-amplitudes},
\begin{gather}
	\cc{ \G_{\pm\pm\pm}, \, \G_{\pm\pm\mp} } 
	= e^3 \sqrt{\frac{2}{\hat{s}}} \, e^{\mp i \phi / 2} 
		\cc{ \frac{1}{\cos^3(\theta/2)}, \, \frac{ \sin^2(\theta/2) }{\cos^3(\theta/2)} }.
\end{gather}

The BH $\gamma_L^*$ channel only contributes to the helicity amplitudes $\M^{\rm NLP}_{\lambda_N \lambda_N' \pm\pm\mp}$
and its hard coefficients $\A^L_{\pm \pm \mp}$ can be easily obtained from \eq{eq:BH-helicity-amplitudes},
\beq
	\A^L_{\pm\pm\mp} = \frac{4e^3}{\hat{s}} \cos(\theta/2) e^{\mp i \phi/2}.
\eeq

Now we derive the most general expression for the amplitude squared, allowing arbitrary polarizations of the initial-state nucleon and electron in the diffractive frame. 
These are introduced through the density matrices in \eqs{eq:M2-phi}{eq:den-mat} with the polarization parameters $(S_T, P_N, P_e)$ defined in the lab frame.
The nucleon spin average can be performed in a Lorentz covariant way by using 
\beq
	\sum_{\lambda_N, \bar{\lambda}_N} u(p, \lambda_N) \, \rho^{N}_{\lambda_N \bar{\lambda}_N} \bar{u}(p, \bar{\lambda}_N)
	=\frac{1}{2}(\slash{p} + m) \pp{1 + \gamma_5 \Slash{S}},
\eeq
with the covariant spin four-vector $S^{\mu}$ that takes the explicit forms 
\eqs{eq:lab-S}{eq:spin-v-sdhep} in the diffractive and the SDHEP frames, respectively.

For the $\gamma^*$ channel,
the transverse polarization vectors for $\gamma^*_T$ in \eq{eq:ga*-pol-v} are written in the SDHEP frame.
They transform into
\beq[eq:lab-e]
	\epsilon^{\mu}_{{D}, \pm} = \epsilon^{\mu}_{\pm} + \frac{\bm{\Delta}_T \cdot \bm{\epsilon}_{\pm,T}}{\Delta^+} n^{\mu}
\eeq
in the diffractive frame by the inversion of \eq{eq:lorentz-trans-sdhep},
whereas the $F \cdot n$ for $\gamma^*_L$ is the same in both the diffractive and the SDHEP frames. 

\subsubsection{Leading power}

The LP of the amplitude square starts at $\O(1/t)$, given by the square of the $\gamma^*_T$ channel in \eq{eq:LP-amplitudes},
\begin{align}
	\overline{|\M|_{\rm LP}^2} &\, = \rho^{N}_{\lambda_N\bar{\lambda}_N} \, \rho^e_{\lambda_e \bar{\lambda}_e} \,
		\M^{\rm LP}_{\lambda_N \lambda_N' \lambda_e \lambda'_e \lambda_2} \,
		\M^{{\rm LP} *}_{\bar{\lambda}_N \lambda_N' \bar{\lambda}_e \lambda'_e \lambda_2} 	\nn\\
	& \, = \rho^e_{++} \pp{ | \A^T_{+++} |^2 \, \vv{ F_+ F^*_+ }
									+ | \A^T_{++-} |^2 \, \vv{ F_- F^*_- } }
			+ \rho^e_{--} \pp{ | \A^T_{---} |^2 \, \vv{ F_- F^*_- }
									+ | \A^T_{--+} |^2 \, \vv{ F_+ F^*_+ } },
\end{align}
where the sum over repeated indices is implied and we introduced the notation 
\beq[eq:vFF]
	\vv{ F_{\lambda} F^*_{\lambda'} }
	\equiv 
	\rho^N_{\lambda_N\bar{\lambda}_N} \bigpp{ F_{\lambda_N \lambda_N'} \cdot \epsilon^*_{\lambda} } 
		\bigpp{ F_{\bar{\lambda}_N \lambda_N'}^* \cdot \epsilon_{\lambda'} }.
\eeq
From \eq{eq:bh-AT}, we see that
\begin{equation*}
	| \A^T_{+++} |^2 = | \A^T_{---} |^2 = \pp{ \frac{2e^3}{t} }^2 \frac{1}{\sin^2(\theta/2)}, \quad
	| \A^T_{++-} |^2 = | \A^T_{--+} |^2 = \pp{ \frac{2e^3}{t} }^2 \sin^2(\theta/2),
\end{equation*}
such that
\begin{align}\label{eq:dvcs-M2-LP-0}
	\overline{|\M|_{\rm LP}^2} &\, = 
		\pp{ \frac{2e^3}{t} }^2 
		\bb{ \frac{1}{\sin^2(\theta/2)} \pp{ B_0 + P_e B_1 }
			+ \sin^2(\theta/2) \pp{ B_0 - P_e B_1 }
		},
\end{align}
where $B_0$ and $B_1$ are computed explicitly,
representing the unpolarized rate and (unnormalized) net helicity of the transverse photon $\gamma^*_T$,
\begin{align}
	B_0 &= \frac{\vv{ F_+ F^*_+ } + \vv{ F_- F^*_- }}{2} 
		= -\pp{ 2m^2 + \frac{1 - \xi^2}{2 \xi^2} t } \pp{ F_1^2 - \frac{t}{4m^2} F_2^2 } 
			- t (F_1 + F_2)^2, 
	\nn\\
	B_1 &= \frac{\vv{ F_+ F^*_+ } - \vv{ F_- F^*_- }}{2} 
		= -(F_1 + F_2) \cc{ P_N \bb{ F_1 \pp{ \frac{4 \xi m^2}{1 + \xi} + \frac{t}{\xi} } + t \, F_2}
			+ \frac{\bm{S}_T \cdot \bm{\Delta}_T }{2m} \bb{ 4m^2 F_1 + \frac{1 + \xi}{\xi} \, t \, F_2 } },
\end{align}
with $\bm{S}_T \cdot \bm{\Delta}_T = S_T \Delta_T \cos\phi_S$ evaluated in the diffractive frame.
Organizing \eq{eq:dvcs-M2-LP-0} in terms of the polarization parameters, we have
\begin{align}\label{eq:dvcs-M2-LP}
	\overline{|\M|_{\rm LP}^2} &\, = \pp{ \frac{2e^3 m }{t} }^2 
		\Sigma^{\rm LP}_{UU} \cdot \pp{ 1 + P_e P_N A^{\rm LP}_{LL} + P_e S_T A^{\rm LP}_{TL} \cos\phi_S },
\end{align}
with the dimensionless polarization parameters,
\bse\label{eq:dvcs-pol-LP}\begin{align}
	\Sigma^{\rm LP}_{UU} &
		= \bb{ \frac{1}{\sin^2(\theta/2)} + \sin^2(\theta/2) } 
			\bb{ \pp{ \frac{1 - \xi^2}{2 \xi^2} \frac{-t}{m^2} - 2 } \pp{ F_1^2 - \frac{t}{4m^2} F_2^2 } 
				- \frac{t}{m^2} (F_1 + F_2)^2 },	\\
	A^{\rm LP}_{LL} &
		= \bigpp{ \Sigma_{UU}^{\rm LP} }^{-1} \cdot \bb{ \frac{1}{\sin^2(\theta/2)} - \sin^2(\theta/2) } 
			(F_1 + F_2) \bb{ F_1 \pp{ \frac{-t}{\xi m^2} - \frac{4 \xi}{1 + \xi} } - \frac{t}{m^2} \, F_2},	\\
	A^{\rm LP}_{TL} &
		= \bigpp{ \Sigma_{UU}^{\rm LP} }^{-1} \cdot \frac{\Delta_T}{2m} \bb{ \frac{1}{\sin^2(\theta/2)} - \sin^2(\theta/2) } 
			(F_1 + F_2) \bb{ -4 F_1 + \frac{1 + \xi}{\xi} \, \frac{-t}{m^2} \, F_2 }.
\end{align}\ese
Note that the unpolarized part $\Sigma^{\rm LP}_{UU}$ is positive definite by the kinematic constraint $-t \geq 4 \xi^2 m^2 / (1 - \xi^2)$.
Here and below, we mark the polarization asymmetry quantities with
superscripts that refer to the power of $\sqrt{-t}/q_T$ at which they contribute 
and subscripts that refer sequentially to the nucleon and electron polarizations, 
with ``$U$'', ``$L$'', and ``$T$'' standing for ``unpolarized'', ``longitudinally polarized'', and ``transversely polarized'', respectively.
In case a single polarization configuration contains more than one azimuthal modulations, 
we further distinguish the asymmetries with additional subscripts $1$, $2$, etc.

Evidently from \eq{eq:BH-helicity-amplitudes}, at the LP, 
the two $\gamma^*_T$ helicities are associated with different helicity structures of the other three on-shell particles in the hard scattering amplitudes. 
There is no interference between the right-handed and left-handed $\gamma^*$ states,
and hence no nontrivial $\phi$ correlation between the diffraction and scattering planes. 

\subsubsection{Next-to-leading power}

The NLP of the amplitude squared is of $\O(1/q_T \sqrt{-t})$, given by the interference of the $\gamma_T^*$ channel in \eq{eq:LP-amplitudes}
with the $\gamma^*_L$ and DVCS channels in \eq{eq:NLP-amplitudes},
\begin{align}
	\overline{|\M|_{\rm NLP}^2} 
	&= 2 \Re\bb{ \rho^N_{\lambda_N\bar{\lambda}_N} \, \rho^e_{\lambda_e \bar{\lambda}_e} \,
		\M^{\rm NLP}_{\lambda_N \lambda_N' \lambda_e \lambda'_e \lambda_2} \,
		\M^{{\rm LP} *}_{\bar{\lambda}_N \lambda_N' \bar{\lambda}_e \lambda'_e \lambda_2} 
		}	\nn\\
	&= \frac{4 e^6}{-t} \sqrt{\frac{2}{\hat{s}}}
		\bigg\{ 
			\frac{4}{\sin\theta (1 + \cos\theta)}
			\Re\bb{ 
				e^{-i \phi} \rho^e_{++} \bigvv{ (\F + \Ft) \, F^*_+ } + e^{i \phi} \rho^e_{--} \bigvv{ (-\F + \Ft) \, F^*_- }
			}	\nn\\
	&\hspace{5em}
			+ \frac{\sin\theta (1-\cos\theta)}{(1+\cos\theta)^2}
			\Re\bb{ e^{i \phi} \rho^e_{++} \bigvv{ (-\F + \Ft) \, F^*_- } + e^{-i \phi} \rho^e_{--} \bigvv{ (\F + \Ft) \, F^*_+ } } \nn\\
	&\hspace{5em}
			+ \frac{\sin\theta}{2\xi} \frac{1}{P^+}
			\Re\bb{ e^{i \phi} \rho^e_{++} \vv{ F_0 \, F^*_- } - e^{-i \phi} \rho^e_{--} \vv{ F_0\, F^*_+ } }
		\bigg\},
\label{eq:dvcs-M2-NLP-0}
\end{align}
where we introduced notations similar to \eq{eq:vFF},
\begin{align}
	\vv{ \F F^*_{\pm} } 
	&\equiv 
	\rho^N_{\lambda_N\bar{\lambda}_N} \, \F_{\lambda_N \lambda_N'} 
		\bigpp{ F_{\bar{\lambda}_N \lambda_N'}^* \cdot \epsilon_{\pm} },
	\nn\\
	\bigvv{ \Ft F^*_{\pm} }
	&\equiv 
	\rho^N_{\lambda_N\bar{\lambda}_N} \, \Ft_{\lambda_N \lambda_N'}
		\bigpp{ F_{\bar{\lambda}_N \lambda_N'}^* \cdot \epsilon_{\pm} }, 
	\nn\\
	\vv{ F_0 F^*_{\pm} }
	&\equiv
	\rho^N_{\lambda_N\bar{\lambda}_N} 
		\bigpp{ F_{\lambda_N \lambda_N'} \cdot n }
		\bigpp{ F_{\bar{\lambda}_N \lambda_N'}^* \cdot \epsilon_{\pm} }, 
\end{align}
with $\F_{\lambda_N \lambda_N'}$ and $\Ft_{\lambda_N \lambda_N'}$ given by \eq{eq:F-Ft-aa'}.
In the second step of \eq{eq:dvcs-M2-NLP-0}, we used the explicit forms of the hard scattering amplitudes.
Clearly, the NLP contributions are proportional to $\cos\phi$ or $\sin\phi$, coinciding with the helicity structures in the interference. 
Furthermore, the nucleon spin average introduces dependence on $\bm{S}_T$, which enters linearly through
$\bm{S}_T \cdot \bm{\Delta}_T \propto \cos\phi_S$ or $\bm{S}_T \times \bm{\Delta}_T \propto \sin\phi_S$ 
and contributes nontrivial diffractive azimuthal $\phi_S$ distributions. 
In this way, we organize \eq{eq:dvcs-M2-NLP-0} according to the spin and azimuthal dependence,
\begin{align}
	\overline{|\M|_{\rm NLP}^2} &= 
		\frac{4 e^6}{-t} \frac{m}{\sqrt{\hat{s}}} \,
		\bb{
			\pp{ \Sigma_{UU}^{\rm NLP} + P_e P_N \Sigma_{LL}^{\rm NLP} } \cos\phi 
			+ \pp{ P_e \Sigma_{UL}^{\rm NLP} + P_N \Sigma_{LU}^{\rm NLP} }\sin\phi 	\right.\nn\\
	&\hspace{2em}\left.
			+ S_T \pp{ \Sigma_{TU, 1}^{\rm NLP} \cos\phi_S \sin\phi + \Sigma_{TU, 2}^{\rm NLP} \sin\phi_S \cos\phi }
			+ P_e S_T \pp{ \Sigma_{TL, 1}^{\rm NLP} \cos\phi_S \cos\phi + \Sigma_{TL, 2}^{\rm NLP} \sin\phi_S \sin\phi }
		},
\label{eq:dvcs-M2-NLP}
\end{align}
with the polarization coefficients,
\bse\label{eq:dvcs-pol-NLP}\begin{align}
	\Sigma_{UU}^{\rm NLP} &=
		\frac{\Delta_T}{2m} \frac{1 + \xi}{\xi} 
		\cc{ \frac{2\sin\theta}{\xi} \pp{ F_1^2 - \frac{t}{4m^2} F_2^2 }
			- \frac{4 + (1 - \cos\theta)^2}{\sin\theta \cos^2(\theta/2)} 
				\bb{ F_1 \Re\H - \frac{t}{4m^2} F_2 \Re\E 
					+ \xi  (F_1 + F_2) \Re\Ht 
				}	
		},	\\
	\Sigma_{LL}^{\rm NLP} &=
		- \frac{\Delta_T}{m}
		\cc{ (F_1 + F_2) 
			\bb{ \sin\theta \pp{ \frac{1 + \xi}{\xi} F_1 + F_2 }
				+ \frac{3 - \cos\theta}{\sin\theta} \pp{ (1 + \xi) \Re\H + \xi \Re\E }
			}
			\right.\nn\\
		& \hspace{4.5em} \left.
			+ \frac{3 - \cos\theta}{\sin\theta} 
				\bb{ \frac{1 + \xi}{\xi} F_1 \Re\Ht - 
					\pp{ \xi F_1 + (1 + \xi) \frac{t}{4m^2} F_2 } \Re\Et 
				}
		},	\\
	\Sigma_{TL, 1}^{\rm NLP} &=
		2(F_1 + F_2)
		\cc{ \sin\theta \bb{ F_1 + \pp{ \frac{\xi}{1 + \xi} + \frac{t}{4\xi m^2} } F_2 }
			+ \frac{3 - \cos\theta}{\sin\theta} 
			\bb{ \xi \Re\H + \pp{ \frac{\xi^2}{1 + \xi} + \frac{t}{4m^2} } \Re\E } 
		} 
		\nn\\
		& \hspace{1em}
			- 2 \frac{3 - \cos\theta}{\sin\theta} 
			\cc{ 
				\bb{ \xi F_1 - \frac{t}{4m^2} \frac{1 - \xi^2}{\xi} F_2 } \Re\Ht
				+ \bb{ \pp{ \frac{\xi^2}{1 + \xi} + \frac{t}{4m^2} } F_1
					+ \frac{\xi t}{4 m^2} F_2 } 
					\Re\Et
			},	\\
	\Sigma_{TL, 2}^{\rm NLP} &=
	 	2 (F_1 + F_2)
		\bb{ \sin\theta \pp{ F_1 + \frac{t}{4m^2} F_2 }
			- \xi \frac{3 - \cos\theta}{\sin\theta} 
			\pp{ \Re\Ht + \frac{t}{4m^2} \Re\Et }
		}	\nn\\
		& \hspace{1em}
			+ 2 \frac{3 - \cos\theta}{\sin\theta}
			\cc{ 
				\bb{ \xi F_1 - \frac{1 - \xi^2}{\xi} \frac{t}{4m^2} F_2 } \Re\H
				+ \bb{ \pp{ \xi + \frac{t}{4 \xi m^2} } F_1 + \frac{\xi t}{4m^2} F_2 } \Re\E
			},	\\
	\Sigma_{UL}^{\rm NLP} &=
		- \frac{\Delta_T}{m} \frac{1 + \xi}{\xi} \frac{3 - \cos\theta}{\sin\theta}
		\bb{ F_1 \Im\H - \frac{t}{4m^2} F_2 \Im\E + \xi \, (F_1 + F_2) \Im\Ht }, 
	\\
	\Sigma_{LU}^{\rm NLP} &=
		- \frac{\Delta_T}{2m} \frac{4 + (1 - \cos\theta)^2}{\sin\theta \cos^2(\theta/2)}
		\bigg[ (F_1 + F_2) \pp{ (1 + \xi) \Im\H + \xi \Im\E }
			+ \frac{1 + \xi}{\xi} F_1 \Im\Ht	 - \pp{ \xi F_1 + \frac{t}{4m^2} (1 + \xi) F_2 } \Im\Et
		\bigg],	\\
	\Sigma_{TU, 1}^{\rm NLP} &=
		\frac{4 + (1 - \cos\theta)^2}{\sin\theta \cos^2(\theta/2)}
		\cc{ (F_1 + F_2)
			\bb{ \xi \Im\H + \pp{ \frac{\xi^2}{1 + \xi} + \frac{t}{4m^2} } \Im\E }
			\right.\nn\\
		& \hspace{9em} \left.
			- \pp{ \xi F_1 + \frac{1 - \xi^2}{\xi} \frac{-t}{4m^2} F_2 } \Im\Ht
			- \bb{ \frac{\xi^2}{1 + \xi} F_1 + \frac{t}{4m^2} (F_1 + \xi F_2) } \Im\Et
		},	\\
	\Sigma_{TU, 2}^{\rm NLP} &= 
		\frac{4 + (1 - \cos\theta)^2}{\sin\theta \cos^2(\theta/2)}
		\cc{ \xi (F_1 + F_2) \pp{ \Im\Ht + \frac{t}{4m^2} \Im\Et }
			\right.\nn\\
		& \hspace{9em} \left.
			- \pp{\xi F_1 + \frac{1 - \xi^2}{\xi} \frac{-t}{4m^2} F_2 } \Im\H
			- \bb{ \pp{ \xi + \frac{t}{4\xi m^2} } F_1 + \frac{\xi t}{4m^2} F_2 } \Im\E
		}.
\end{align}\ese

\subsubsection{Combination}

Combining the LP in \eq{eq:dvcs-M2-LP} and NLP in \eq{eq:dvcs-M2-NLP} 
contribution and substituting them for the $\overline{|\M|^2}$ in \eq{eq:sdhep-xsec}, 
we have the differential cross section for the photon electroproduction process,
\begin{align}
	\frac{d\sigma}{dt \, d\xi \, d\phi_S \, d\cos\theta \, d\phi}
	& = \frac{1}{(2\pi)^2} \frac{d\sigma^{\rm unpol}}{dt \, d\xi \, d\cos\theta}
		\cdot \Big[ 1 + P_e P_N A^{\rm LP}_{LL} + P_e S_T A^{\rm LP}_{LT}  \cos\phi_S
		\nn\\
		&\hspace{11em}
			+ \pp{ A_{UU}^{\rm NLP} + P_e P_N A_{LL}^{\rm NLP} } \cos\phi 
			+ \pp{ P_e A_{UL}^{\rm NLP} + P_N A_{LU}^{\rm NLP} }\sin\phi 	\nn\\
		&\hspace{11em}
			+ S_T \pp{ A_{TU, 1}^{\rm NLP} \cos\phi_S \sin\phi + A_{TU, 2}^{\rm NLP} \sin\phi_S \cos\phi }
				\nn\\
		&\hspace{11em}
			+ P_e S_T \pp{ A_{TL, 1}^{\rm NLP} \cos\phi_S \cos\phi + A_{TL, 2}^{\rm NLP} \sin\phi_S \sin\phi }
		\Big],
\label{eq:dvcs-LP-NLP-xsec}
\end{align}
where 
\beq
	\frac{d\sigma^{\rm unpol}}{dt \, d\xi \, d\cos\theta}
	= \frac{\alpha_e^3}{(1 + \xi)^2} \frac{m^2}{s \, t^2} \Sigma^{\rm LP}_{UU}
\eeq
is the unpolarized differential cross section with the azimuthal dependence integrated out, 
the LP polarization parameters $\Sigma^{\rm LP}_{UU}$, $A^{\rm LP}_{LL}$, and $A^{\rm LP}_{LT}$ are given in \eq{eq:dvcs-pol-LP},
and the NLP ones are obtained from \eq{eq:dvcs-pol-NLP} by
\begin{align}
	A_X^{\rm NLP}	
	= \frac{-t}{m \sqrt{\hat{s}}} \frac{1}{\Sigma^{\rm LP}_{UU}}
		\Sigma_X^{\rm NLP},
\label{eq:pol-asys}
\end{align}
with $\hat{s}$ given by \eq{eq:sdhep-s-hat-approx} and $X$ denoting various polarization subscripts, $X \in \{ UU, LL, \cdots \}$. 
Evidently, the NLP does not change the event rate, but only introduce azimuthal $\cos\phi$ and $\sin\phi$ modulations. 
Measuring the latter offers an efficient way to determine the GPDs, 
with knowledge of the EM form factors $F_{1, 2}$ from elastic electron-nucleon scattering experiments.
The combinations of various polarizations and azimuthal modulations in \eq{eq:dvcs-LP-NLP-xsec} agree with the parity constraint in \eq{eq:parity}.

Notably, GPDs enter in a linear way, and we have in total eight NLP polarization parameters together with 8 real degrees of freedom of the GPD moments.
The $A_{UU}^{\rm NLP}$, $A_{LL}^{\rm NLP}$, $A_{TL, 1}^{\rm NLP}$, and $A_{TL, 2}^{\rm NLP}$ depend on the real parts, 
and $A_{UL}^{\rm NLP}$, $A_{LU}^{\rm NLP}$, $A_{TU, 1}^{\rm NLP}$, and $A_{TU, 2}^{\rm NLP}$ depend on the imaginary parts.
Therefore, the DVCS can in principle fully determine the GPD moments $\H$, $\E$, $\Ht$, and $\Et$, for both real and imaginary parts. 
While the real parts constrain the principle values of the GPD integrals in \eq{eq:dvcs-GPD-moments}, 
the imaginary parts can directly constrain the GPD values at $x = \pm\xi$.

To understand the uniqueness of the GPD solutions from the eight polarization parameters, 
we note that the GPD dependence in \eq{eq:dvcs-pol-NLP} is controlled by the matrix $M$,
\begin{align}
	M = \begin{bmatrix}
		F_1	& -\dfrac{t}{4m^2} F_2		& \xi (F_1 + F_2)	& 0  \\
		(1 + \xi) (F_1 + F_2)	& \xi (F_1 + F_2) 	& \dfrac{1 + \xi}{\xi} F_1	& - \xi F_1 - (1+\xi) \dfrac{t}{4m^2} F_2	\\
		\xi (F_1 + F_2) 	& \pp{ \dfrac{\xi^2}{1 + \xi} + \dfrac{t}{4m^2} } (F_1 + F_2)	
								& - \xi F_1 + \dfrac{t}{4m^2} \dfrac{1 - \xi^2}{\xi} F_2	& - \pp{ \dfrac{\xi^2}{1 + \xi} + \dfrac{t}{4m^2} } F_1 - \dfrac{\xi \, t}{4m^2} F_2 \\
		\xi F_1 - \dfrac{t}{4m^2} \dfrac{1 - \xi^2}{\xi} F_2 	&  \pp{ \xi + \dfrac{t}{4\xi m^2} } F_1 + \dfrac{\xi \, t}{4m^2} F_2
								& -\xi (F_1 + F_2) 	& -\dfrac{\xi \, t}{4m^2} (F_1 + F_2)
	\end{bmatrix}.
\label{eq:coef-matrix}
\end{align}
Denoting $M_i$ as the $i$-th row vector of $M$ and $V_\F = (\H, \E, \Ht, \Et)^T$, we can rewrite \eq{eq:dvcs-pol-NLP} as
\bse\label{eq:dvcs-pol-M}\begin{align}
	\Sigma_{UU}^{\rm NLP} &=
		\frac{\Delta_T}{2m} \frac{1 + \xi}{\xi} 
		\bb{ \frac{2\sin\theta}{\xi} \pp{ F_1^2 - \frac{t}{4m^2} F_2^2 }
			- \frac{4 + (1 - \cos\theta)^2}{\sin\theta \cos^2(\theta/2)} 
				\pp{ M_1 \cdot \Re V_\F }	
		},	\\
	\Sigma_{LL}^{\rm NLP} &=
		- \frac{\Delta_T}{m}
		\bb{ \sin\theta (F_1 + F_2) \pp{ \frac{1 + \xi}{\xi} F_1 + F_2 }
			+ \frac{3 - \cos\theta}{\sin\theta} \pp{ M_2 \cdot \Re V_\F }
		},	\\
	\Sigma_{TL, 1}^{\rm NLP} &=
		2 \sin\theta \, (F_1 + F_2)\bb{ F_1 + \pp{ \frac{\xi}{1 + \xi} + \frac{t}{4\xi m^2} } F_2 }
		+ \frac{2(3 - \cos\theta)}{\sin\theta} \pp{ M_3 \cdot \Re V_\F },	\\
	\Sigma_{TL, 2}^{\rm NLP} &=
	 	2 \sin\theta \, (F_1 + F_2) \pp{ F_1 + \frac{t}{4m^2} F_2 } 
		+ \frac{2(3 - \cos\theta)}{\sin\theta} \pp{ M_4 \cdot \Re V_\F }, \\
	\Sigma_{UL}^{\rm NLP} &=
		- \frac{\Delta_T}{m} \frac{1 + \xi}{\xi} \frac{3 - \cos\theta}{\sin\theta}
		\pp{ M_1 \cdot \Im V_\F }, 
	\\
	\Sigma_{LU}^{\rm NLP} &=
		- \frac{\Delta_T}{2m} \frac{4 + (1 - \cos\theta)^2}{\sin\theta \cos^2(\theta/2)}
		\pp{ M_2 \cdot \Im V_\F },	\\
	\Sigma_{TU, 1}^{\rm NLP} &=
		\frac{4 + (1 - \cos\theta)^2}{\sin\theta \cos^2(\theta/2)}
		\pp{ M_3 \cdot \Im V_\F },	\\
	\Sigma_{TU, 2}^{\rm NLP} &= 
		\frac{4 + (1 - \cos\theta)^2}{\sin\theta \cos^2(\theta/2)}
		\pp{ M_4 \cdot \Im V_\F }.
\end{align}\ese
It is interesting that the real and imaginary parts of the GPD moments are controlled by the same matrix $M$.
After measuring the asymmetries, subtracting the BH contributions using the known information of $F_1$ and $F_2$, and factoring out the $\theta$ dependence,
we end up with a system of linear equations,
\beq[eq:F-eqs]
	M \cdot V_\F = \hat{V}_{\rm exp},
\eeq
where $\hat{V}_{\rm exp} = (\hat{V}_{\rm exp}^1, \hat{V}_{\rm exp}^2, \hat{V}_{\rm exp}^3, \hat{V}_{\rm exp}^4)^T$ are the experimentally reconstructed (complex) values of the left-hand sides.
Since the determinant of the coefficient matrix $M$ is nonzero,
\beq
	\det M = -\xi^2 (1 + \xi) \pp{ 1 + \frac{1 - \xi^2}{\xi^2} \frac{t}{4m^2} }^2 \pp{ F_1^2 - \frac{t}{4m^2} F_2^2 }^2,
\eeq
\eq{eq:F-eqs} can be easily solved with a unique set of solutions for the GPD moments, $V_\F = M^{-1} \hat{V}_{\rm exp}$.

We remark that, as mentioned in \sec{ssec:sdhep}, $\phi_S$ is defined only when $S_T \neq 0$. 
In case of $S_T = 0$, we shall simply integrate over $\phi_S$ (which is equivalently to integrate over $\phi_{\Delta}$ in a fixed lab coordinate setting) 
and use in place of \eq{eq:dvcs-LP-NLP-xsec},
\begin{align}
	\frac{d\sigma}{dt \, d\xi \, d\cos\theta \, d\phi}
	& = \frac{1}{2\pi} \frac{d\sigma^{\rm unpol}}{dt \, d\xi \, d\cos\theta}
		\cdot \Big[ 1 + P_e P_N A^{\rm LP}_{LL}
			+ \pp{ A_{UU}^{\rm NLP} + P_e P_N A_{LL}^{\rm NLP} } \cos\phi 
			+ \pp{ P_e A_{UL}^{\rm NLP} + P_N A_{LU}^{\rm NLP} }\sin\phi
		\Big],
\label{eq:dvcs-LP-NLP-xsec-no-st}
\end{align}
with the meaning of each quantity the same as in \eq{eq:dvcs-LP-NLP-xsec}.

Beyond the NLP, the cross section receives contributions from three sources: 
(1) the NNLP part from the BH amplitude squared,
(2) the twist-2 DVCS amplitude squared,
and
(3) the interference between the BH and twist-3 DVCS amplitudes.
They will modify the unpolarized amplitude square $\Sigma_{UU}^{\rm LP}$ and add new azimuthal modulations $(\cos2\phi, \sin2\phi)$, 
both at $\order{1 / q_T^2}$.
However, it is important to notice that they will not contribute to the $\cos\phi$ and $\sin\phi$ modulations, 
and hence will only modify the associated asymmetries in \eq{eq:pol-asys} by
\begin{align}
	A_X^{\rm NLP}	 
	\to \frac{-t}{m \sqrt{\hat{s}}} \frac{\Sigma_X^{\rm NLP}}{\Sigma^{\rm LP}_{UU} + \Sigma^{\rm NNLP}_{UU}}
	= A_X^{\rm NLP}
		\cdot \bb{1 + \order{t / q_T^2} }.
\end{align}
Therefore, the azimuthal asymmetries are accurate up to corrections of $\order{t / q_T^2}$,
which is of the same order as the contributions from twist-4 GPDs.

Within the NLP, however, one does receive higher order corrections at $\order{\alpha_s}$, including especially the gluon GPDs.
The gluon GPD $F^g$ and $\wt{F}^g$ contribute to the $\cos\phi$ and $\sin\phi$ modulations,
whereas its transversity GPD $F^g_T$ also contribute to new $\cos3\phi$ and $\sin3\phi$ modulations.
This introduces new asymmetries as well as modifying the ones in \eq{eq:pol-asys} by
\begin{align}
	A_X^{\rm NLP}	 
	\to \frac{-t}{m \sqrt{\hat{s}}} \frac{1}{\Sigma^{\rm LP}_{UU}}
		\Sigma_X^{\rm NLP}
		\cdot \bb{1 + \order{\alpha_s} }	
	= A_X^{\rm NLP}
		\cdot \bb{1 + \order{\alpha_s} }.
\end{align}
The high-order corrections have been considered in \refs{Ji:1997nk, Ji:1998xh, Belitsky:1997rh, Mankiewicz:1997bk, Hoodbhoy:1998vm, Belitsky:1999sg, Belitsky:2000jk, Freund:2001rk, Freund:2001hm, Freund:2001hd, Pire:2011st, Moutarde:2013qs, Braun:2020yib, Braun:2022bpn, Ji:2023xzk, Belitsky:2001ns, Kriesten:2019jep} in the Breit frame, 
yet are interesting to be adapted to the SDHEP frame. 
We leave the detailed discussion to a future work.

\section{Comparison between the Breit frame and SDHEP frame}
\label{sec:unique}

We have seen how the SDHEP frame unifies the BH and DVCS subprocesses and simplifies the azimuthal structure.
It naturally describes the photon electroproduction process using a consistent factorization approach.
Yet, it is still intriguing to parametrize the full process by the 2 EM form factors plus 18 CFFs.
As we saw in \sec{ssec:cffs-calc}, the CFFs have close relations to the GPDs.
In this Section, we give a further discussion as to how CFFs are represented in the SDHEP frame.
This will give a better understanding about the azimuthal structures in the photon electroproduction,
and in the end uncover the uniqueness of the Breit frame in terms of the CFF description 
and the SDHEP frame in terms of the coherent azimuthal modulations.

\subsection{Compton form factors in the SDHEP frame}
\label{ssec:cffs-obs}

If we take the CFF decompositions in \eq{eq:CFF-decomp} and attempt to formulate the observables in the SDHEP frame,
one immediately notices that the scalar arguments of the CFFs contain both $\theta$ and $\phi$ dependence in terms of the SDHEP observables,
\begin{align}
	P \cdot R = \frac{\hat{s}}{8\xi} \bb{ 1 + \pp{ 1 + \frac{t}{\hat{s}} } \cos\theta - (1 + \xi) \frac{2 \Delta_T}{\sqrt{\hat{s}}} \sin\theta \cos\phi },
\label{eq:cff-pk-sdhep}
\end{align}
with $\hat{s}$ given in \eq{eq:sdhep-s-hat-approx}.
[The $\theta$ dependence also comes from $Q^2 = (\hat{s} - t) (1 + \cos\theta) / 2$ from the $R^2$ in \eq{eq:cff-args}.]
Hence the CFFs shall be expressed as
\beq
	\bigcc{ \H_i, \E_i, \wt{\H}_i, \wt{\E}_i } = \bigcc{ \H_i, \E_i, \wt{\H}_i, \wt{\E}_i }(\xi, t / \hat{s}, \theta, \phi),
\eeq
with dependence on both the hadronic observables $(\xi, t)$ and hard scattering variables $(\hat{s}, \theta, \phi)$.
This feature is also true for the expressions in \eq{eq:cffs-exprs} in the Breit frame, 
where one should note that $Q$ belongs to the hard scattering. 

The $\phi$ dependence of the CFFs is because in the SDHEP frame, 
the hard-scattering plane spanned by the two photons $\gamma^*_{ee}$ and $\gamma$ 
differs from the hadronic plane by a rotation $\phi$ around the $\hat{z}_S$ axis.
This ``geometric twist'' between the two planes 
causes the internal propagators in the virtual Compton amplitude to carry explicit $\phi$ dependence,
similar to $\varphi$ dependence in \eq{eq:bh-p1-p2} for the BH.
Hence the CFFs generally contain arbitrary azimuthal dependence in the SDHEP frame.

\begin{figure}[htbp]
	\def\sc{1.5}
	\centering
	\begin{align*}
			\adjincludegraphics[valign=c, scale=0.55]{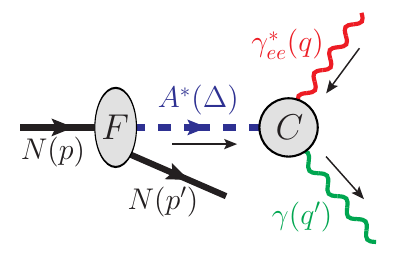} 
		&\Scale[\sc]{=}
			\adjincludegraphics[valign=c, scale=0.55]{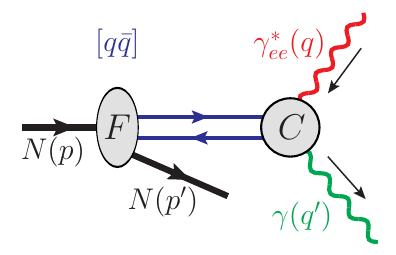} 
			\Scale[\sc]{+}
			\adjincludegraphics[valign=c, scale=0.55]{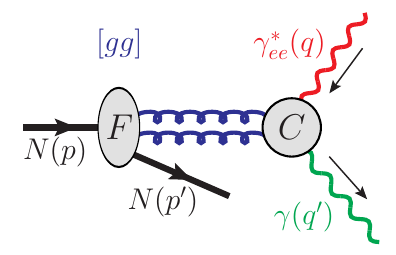} 
		\\
		&\Scale[\sc]{+}
			\adjincludegraphics[valign=c, scale=0.55]{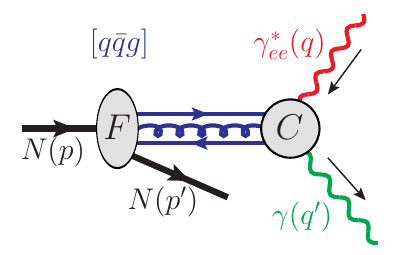} 
			\Scale[\sc]{+}
			\adjincludegraphics[valign=c, scale=0.55]{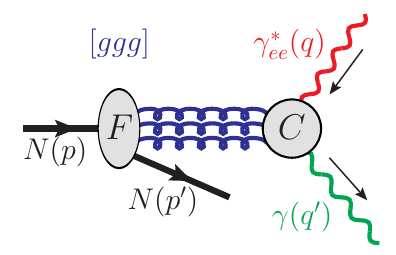}
			\Scale[\sc]{+ \cdots}
	\end{align*}
	\caption{Channel expansion of the virtual Compton amplitude in the light-cone gauge in the SDHEP frame.}
	\label{fig:cff-sdhep}
\end{figure}

However, contrary to the BH, the ``geometric twist'' here is connected by a set of pinched parton lines, 
which takes a systematic expansion in terms of the parton channels, as shown in \fig{fig:cff-sdhep},
and can be factorized into a hadronic part and a hard scattering at each power of $\sqrt{-t} / Q$.
Then the ``geometric twist'' at the $j$-th power appears as the spin of the $(j+2)$-parton state,
and the hard scattering happens on a single plane that is rotated with respect to the parton state by an angle $\phi$.
It is their contraction that yields a $\cos(j \phi)$ dependence within the CFFs at the $j$-th power of $\sqrt{-t} / Q$.
This picture is confirmed by the fact that the $\phi$ dependence in \eq{eq:cff-pk-sdhep} is suppressed by $\Delta / \sqrt{\hat{s}} = \order{\sqrt{-t} / Q}$.
Hence, one only generates a $\cos^j (\phi) \sim \cos(j\phi)$ component at $\order{\sqrt{-t} / Q}^j$.

Due to such $\phi$ dependence, the SDHEP picture is not particularly useful for formulating the nonperturbative extraction of the CFFs, 
whereas it is very convenient for a factorization treatment of the CFFs.
At each power of  in terms of $\sqrt{-t} / Q$, the CFFs can be factorized into new (twist-2 or higher) GPDs associated with new azimuthal modulations
to allow the contraction of GPDs. 

In contrast, we notice in \sec{ssec:breit} that in the Breit frame, aside from the $\P_1(\varphi)\P_2(\varphi)$ factor, 
there is a finite set of azimuthal modulations due to the finite number of photon spin configurations in \eq{eq:cff-helicity-structure}.
This apparent contradiction is because going from the photon frame in \eq{eq:cff-helicity-structure} to the SDHEP frame in \fig{fig:cff-sdhep}
involves a nontrivial Lorentz transformation that mixes the helicity structures of GPDs of various twists.
Even though the parton helicities of the GPDs in the photon frame are restricted to $(0, \pm1, \pm 2)$ due to helicity conservation,
they can mix with a tower of multiple-twist GPDs with various helicity states under this transformation.

\subsection{Uniqueness of the frames}
\label{ssec:unique}

\begin{figure}[htbp]
	\centering
		\begin{tabular}{cc}
			\includegraphics[scale=0.5]{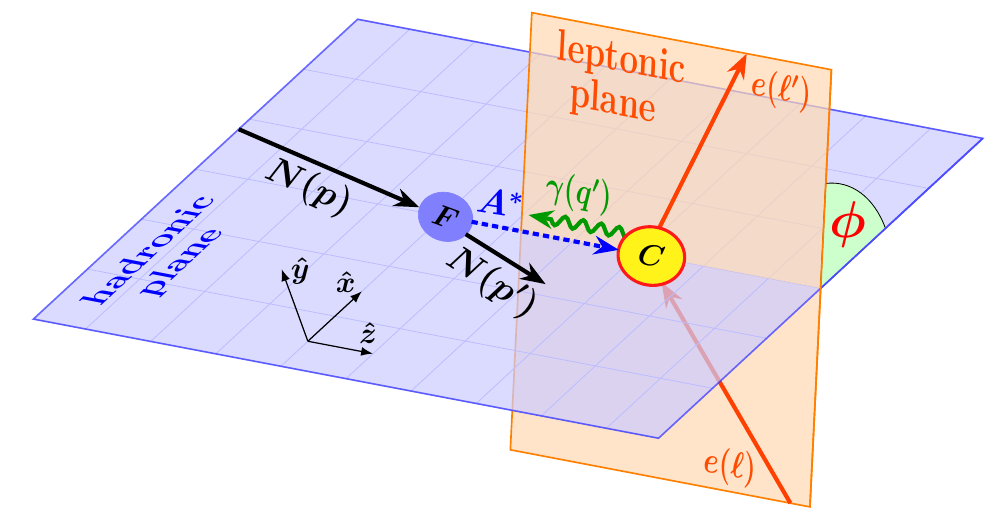} &
			\includegraphics[scale=0.5]{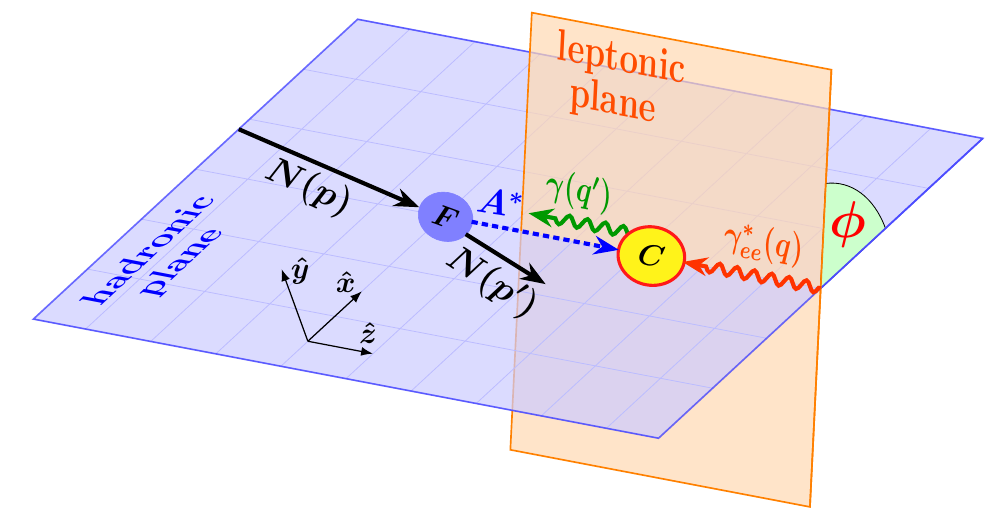}\\
			(a) & (b)
		\end{tabular} 
	\caption{
		(a) Photon frame for describing the photon electroproduction process in \eq{eq:dvcs-sdhep}.
		(b) is the DVCS subprocess, omitting the electron lines on the leptonic plane.
		Replacing the $A^*$ by $\gamma^*$ in (a) gives the BH subprocess.
	}
\label{fig:mbreit-frame}
\end{figure}

Having seen the pros and cons of both the Breit and SDHEP frames,
one might be tempted to combine the advantages of two frames by using the ``SDHEP-modified'' Breit frame, as shown in \fig{fig:mbreit-frame},
which is in fact the photon frame in \eq{eq:cff-helicity-structure}.
Now we still sit on the plane of the hadronic diffraction, which emits the same $A^*(\Delta)$ state that goes into the hard collision.
However, we describe the hard collision in a way similar to the Breit frame, 
with the incoming and outgoing electrons determining a leptonic plane, and their momentum difference $q = \l - \l'$ aligned with the $-\hat{z}$ direction.
Then by momentum conservation, the real photon must be produced to move along the $-\hat{z}$ direction as well.
This photon frame differs from the Breit frame by a slight Lorentz transformation of order $\sqrt{-t} / Q$.
The azimuthal angle $\phi$ is now defined as the relative angle between the hadronic and the leptonic planes.
Each event can be described by $(t, \xi, Q^2, \phi_S, \phi)$ or $(t, x_B, Q^2, \phi_S, \phi)$ 
with $\xi$ defined similarly using the new $z$ axis, $(t, x_B, Q^2)$ the same as the Breit frame, and $\phi_S$ the same as the SDHEP observable.
\footnote{The photon frame is similar to the $RX$ frame used in \refcite{Arens:1996xw} (in the Fig.~5 there) for the inclusively diffractive $ep$ collision,
where the role of $A^*$ is played by an exchanged Pomeron and the final state $X$ has a nonzero invariant mass so is at rest. 
The same two-stage SDHEP formalism shall also be applicable there, with the azimuthal $\phi$ distribution used to separate Pomeron helicity structures.}

The photon frame apparently unifies the SDHEP and the Breit frames. It describes both the BH and DVCS subprocesses with the same kinematic setting.
For the BH, all the hard scattering momenta lie on the same leptonic plane, so one should not expect a $\phi$ dependence in the electron propagators.
For the DVCS, the hadronic part sits right in the transverse basis spanned by the two photon momenta $q$ and $q'$, 
so can be well described by the nonperturbative CFFs.

However, one issue undermines all these advantages. The general use of an azimuthal angle in describing an event distribution is due to the fact that 
the scattering configuration is not completely fixed on a certain plane, leaving a free azimuthal degree of freedom.
That is, the absolute values of momenta should not change with the azimuthal angle.
This is true for both the BH and DVCS in the SDHEP frame.
It is also true for the DVCS in the Breit frame, but for the BH, the rotation around the $z$ direction causes the internal electron momentum to change,
and hence induces $\varphi$ dependence in the denominators.
For the photon frame in \fig{fig:mbreit-frame}(a), it is important to notice that once the nucleon momentum $p$ is fixed under a certain c.m.\ energy $\sqrt{s}$,
the electron momentum $\l$ depends on $\phi$, in both its longitudinal and transverse components
\footnote{The full expression of $\l$ in terms of $\phi$ at a given $p$ and $s$ is too unwieldy to show.};
otherwise $s$ would not be a fixed quantity,
\beq
	s = (p + \l)^2 = m^2 + 2 p \cdot \l \supset -2 p^x \l^x = -2 p^x \l_{T} \cos\phi.
\eeq
The $\phi$ dependence in $\l$ then carries over to the electron propagators in the BH and turns the result even more complicated than in the Breit frame.

Therefore, to make the azimuthal description useful at all, one needs to have at least one of the initial-state particles along the $z$ axis.
The Breit frame chooses it as the nucleon, while the SDHEP frames goes for the electron.
In both cases, the azimuthal angle is defined as that between the hadronic and leptonic planes.

Then if the aim is to avoid azimuthal dependence in the denominators, one has to guarantee that 
at least the four momenta $(\Delta, \l, \l', q')$ in the BH subprocess stay on the same plane as $\phi$ changes. 
Since $\Delta$ is necessarily on the hadronic plane, 
the only way to guarantee that is to make both $\l$ and $\Delta$ on the $z$ axis, and define the azimuthal rotation with respect to that axis.
This uniquely leads to the SDHEP frame.

On the other hand, if one wants to stick to the nonperturbative description and extraction of the CFFs from the DVCS, 
it is necessary that the CFFs are expressed only of scalar quantities, without dependence on the azimuthal angle.
Then one must ensure that the scalar products $P \cdot R$ and $\Delta \cdot R$ remain unchanged under azimuthal rotation [cf.~\eq{eq:cff-pk-sdhep}].
This simply requires $(P, \Delta, R)$ to be on the same plane and the azimuthal angle to be defined as between this plane and the leptonic plane.
These two planes are connected by the $\gamma^*_{ee}$, whose direction is naturally chosen as the $z$ axis,
and thus we are led uniquely to the Breit frame.

In either case, the frame can differ by a boost along the $z$ axis, which does not affect the azimuthal description.

This whole argument leads to the conclusion that one cannot have a single frame that is suitable for describing the photon electroproduction 
in terms of the all-order CFFs 
while maintaining a good physical understanding of the azimuthal modulations.
Either virtue picks a unique frame.
The practical choice of frame hence depends on the priority of the analysis.
One has to recognize that both the CFFs in the Breit frame and the azimuthal asymmetries in the SDHEP frame
depend on multiple variables like $(\eta, t, Q^2)$ or $(\xi, t, \theta)$.
However clear the theoretical formalism is, experimental measurements are going to be challenging.
In this regard, we remind ourselves that it is the GPDs that are of physical interest, while a full extraction of the all-order CFFs is not necessary.
So we propose that one should make the experimental analysis as easy and unambiguous as possible
and hence should choose the SDHEP frame, using the azimuthal asymmetries to extract the GPD information.
Not to use CFFs will not cause a loss of information. 
As argued in \sec{ssec:cffs-obs}, the CFFs contain GPDs of all twists, each of which is associated with a unique azimuthal component in the SDHEP frame.
The full information of CFFs would be recovered by a fine enough resolution of the azimuthal structures 
as well as a factorization treatment at high enough twists.

\section{Conclusion}
\label{sec:conclusion}

Extracting the GPD information encoded in the deeply virtual Compton scattering (DVCS) 
might be the top priority for most of the GPD programs in the current and future experiments.
Even though the DVCS depends on the GPDs via simple moments [\eq{eq:dvcs-GPD-moments}], 
fully extracting these moments is obscured in the conventional analysis in the Breit frame 
by the accompanying Bethe-Heitler (BH) process
that interferes with the DVCS and has a much larger rate.
The main reason is that the Breit frame focuses on the virtual photon $\gamma^*_{ee}$ in the DVCS that does not exist in the BH.

In this paper, we proposed a new perspective to analyze the whole process.
Instead of sticking to the DVCS part of the full process to focus on the exclusive scattering of the $\gamma^*_{ee}$ and the nucleon,
we analyze the whole photon electroproduction process as a single-diffractive hard exclusive process (SDHEP),
which orients around the quasireal state $A^*$ that connects the collinear diffraction part and the hard scattering part.
This new viewpoint unifies the BH and DVCS in a coherent framework with a clear factorization structure,
where the BH happens at the leading power, corresponding to a photon-mediated channel ($A^* = \gamma^*$)
while the DVCS is simply a two-parton channel ($A^* = [q\bar{q}]$ or $[gg]$) or higher so it contributes at subleading powers.
Within the new SDHEP picture and frame,
the disadvantage of the Breit frame due to the presence of the BH is now turned into a major advantage that
the interference of the photon channel with the parton channels gives rise to rich and well defined azimuthal correlations
between the diffraction and the hard scattering planes, 
which gives a clear prescription to extract the GPD moments and allows to distinguish GPDs of different parton spin structures.
At the leading power, we found eight polarization asymmetries associated with unique azimuthal correlations 
that correspond exactly to the eight real degrees of freedom of the GPD moments.

The advantage of the Breit frame compared to the SDHEP frame is that 
it has a close analogy to that for the inclusive DIS and thereby provides a suitable frame to 
analyze the DVCS amplitude in terms of the independent Compton form factors (CFFs)
which we have defined in a nonperturbative way.
We have argued that the Breit frame is the unique frame to suitably analyze the CFFs.
The SDHEP frame is not particularly useful for measuring the CFFs 
because it expresses them in terms of variables that depend on the azimuthal angle.
Rather, following the SDHEP picture, the same CFFs shall be analyzed in the same way as the whole photon electroproduction amplitude 
in terms of the channel expansion, receiving contributions from GPDs of all twists with each twist generating new azimuthal modulations.
We also argued that the SDHEP frame is the unique one to give a coherent azimuthal correlation description.
This then forces one to trade off between a nonperturbative CFF description and the azimuthal modulations.

In the hard exclusive limit, the CFFs, like the structure functions in DIS, receive contributions from all twist levels, with the leading one arising from twist-2 GPDs.
While all of these contain valuable information, we emphasize that neither CFFs nor GPDs are direct experimental observables. 
Having access to only a finite number of experimental observables permits the extraction of only a limited set of GPDs encoded within CFFs. 
Since it is GPDs that are of our ultimate interest, we should choose an analysis procedure that facilitates their extraction as directly as possible, 
without the intermediate step of first determining CFFs. 
To this end, we propose analyzing or reanalyzing the photon electroproduction process within the SDHEP framework, utilizing the full azimuthal modulation information.

\section*{Acknowledgements}
We thank M.\ Diehl, Y.\ Guo, C.\ E.\ Hyde, S.\ Lee, E.\ Moffat, M.\ G.\ Santiago, Y.\ Wang, and F.\ Yuan for the helpful discussions and communications. 
This work was supported in part by the U.S.\ Department of Energy (DOE) Contract No.\ DE-AC05-06OR23177, 
under which Jefferson Science Associates, LLC operates Jefferson Lab, 
and the Jefferson Lab LDRD program Projects No.\ LD2312 and No.\ LD2406. 
The work of N.\ S.\ and Z.\ Y.\ was also supported by the DOE, Office of Science, Office of Nuclear Physics in the Early Career Program. 
This work has benefited from interactions within the Quark-Gluon Tomography (QGT) Topical Collaboration funded by the DOE, Office of Science, Office of Nuclear Physics with Award No.\ DE-SC0023646.

\appendix

\section{Decomposition of the virtual Compton amplitude into Compton form factors}
\label{app:cffs}

Here we provide the derivation for the CFF decomposition of the Compton amplitude in \eq{eq:compton-amp}
\beq
	T^{\mu\nu}(P, \Delta, R)
	= \bar{u}(p', \lambda_N') \, \Gamma^{\mu\nu}(P, \Delta, R) \, u(p, \lambda_N),
\eeq
where parity symmetry constrains
\beq[eq:T-parity]
	\Gamma^{\mu\nu}(P, \Delta, R) 
	= \gamma^0 \, \Gamma_{\mu\nu}(\bar{P}, \bar{\Delta}, \bar{R}) \, \gamma^0,
\eeq
with $\bar{V}^{\mu} = V_{\mu}$.
It is well known that $\Gamma^{\mu\nu}(P, \Delta, R)$ can be written as a linear combination of the 16 Dirac matrices,
\beq[eq:gamma-decomp]
	\Gamma^{\mu\nu}
	= S^{\mu\nu} \mathbf{1} + P^{\mu\nu} \gamma_5 + V^{\mu\nu}{}_{\alpha} \, \gamma^{\alpha} + A^{\mu\nu}{}_{\alpha} \, \gamma^{\alpha} \gamma_5
		+ \frac{1}{2} T^{\mu\nu}{}_{\alpha\beta} \, \sigma^{\alpha\beta}.
\eeq
When sandwiched between the on-shell spinors $\bar{u}(p', \lambda_N')$ and $u(p, \lambda_N)$, they are subject to more constraints from the Gordon identities,
\bse\label{eq:gordon}\begin{align}
	\bar{u}(p')\gamma^{\mu}\,u(p) 
		& = \bar{u}(p') \bb{\frac{P^{\mu}}{m} - \frac{i\sigma^{\mu\Delta}}{2m}} u(p), 		
		\label{eq:gordon-V}	\\
	0	& = \bar{u}(p') \bb{\frac{\Delta^{\mu}}{2m} - \frac{i\sigma^{\mu P}}{m}} u(p), 		
		\label{eq:gordon-S}	\\
	\bar{u}(p') \gamma^{\mu}\gamma_5 u(p) 
		& = \bar{u}(p') \bb{- \frac{\Delta^{\mu}}{2m} + \frac{i\sigma^{\mu P}}{m}} \gamma_5 \, u(p), 		
		\label{eq:gordon-A}\\
	0 	& = \bar{u}(p') \bb{\frac{P^{\mu}}{m} - \frac{i\sigma^{\mu \Delta}}{2m}} \gamma_5 \, u(p).
		\label{eq:gordon-P}
\end{align}\ese
The last two involve the pseudotensor Dirac matrix, which can be related back to the tensor Dirac matrix by 
\beq[eq:Tt]
	i \sigma^{\mu\nu} \gamma_5 = -\frac{1}{2} \epsilon^{\mu\nu\alpha\beta} \sigma_{\alpha\beta}.
\eeq
	
The reduction of the Dirac structures in \eq{eq:gamma-decomp} follows the procedure in \refs{Diehl:2001pm, Meissner:2009ww}:
\begin{enumerate}
	\item \eq{eq:gordon-V} converts the vector structure into scalar and tensor structures.
	\item \eq{eq:gordon-A} converts the axial-vector structure into pseudoscalar and pseudotensor structures.
	\item \eq{eq:gordon-P} converts the pseudoscalar structure into a pseudotensor,
		\beq
			\bar{u}(p') \gamma_5 \, u(p) 
				= \bar{u}(p') \frac{i \sigma^{R \Delta} \gamma_5}{2 P \cdot R} \, u(p)
				= \bar{u}(p') \frac{i \sigma^{P \Delta} \gamma_5}{2 P^2} \, u(p).
		\eeq
	\item \eq{eq:Tt} converts any pseudotensor structure into a tensor structure.
	\item Now we are reduced to only scalar and tensor structures. 
		\eq{eq:gordon-S} further converts any tensor structure contracted with $P$ into a scalar structure. 
\end{enumerate}
This gives the complete Dirac basis composed of
\begin{itemize}
\item 10 scalar structures: $S^{\mu\nu} \mathbf{1}$, where 
	\beq[eq:S-10]
		S^{\mu\nu} = g^{\mu\nu} \mbox{ or any one of } (P^{\mu}, \Delta^{\mu}, R^{\mu}) \otimes (P^{\nu}, \Delta^{\nu}, R^{\nu});
	\eeq
\item and 23 tensor structures:
	\begin{enumerate}
		\item[(1)] $\sigma^{\mu\nu}$;
		\item[(2)] $\sigma^{\mu a} b^{\nu}$ and $\sigma^{\nu a} b^{\mu}$, where $a \in \{ \Delta, R \}$ and $b \in \{ P, \Delta, R \}$;
		\item[(3)] $\sigma^{R \Delta} S^{\mu\nu}$, where $S^{\mu\nu}$ is the same as the scalar structure in \eq{eq:S-10}.
	\end{enumerate}
\end{itemize}

We would like to convert the uncontracted tensor structures into contracted ones.
In the general case, the three momenta $(P, \Delta, R)$ are linearly independent, so that 
$\epsilon^{\mu P \Delta R}$ is the remaining degree of freedom in the four-dimensional vector space.
Following the same strategy to be detailed in Appendix~\ref{app:covariant}, one can construct a covariant set of coordinate axes,
\begin{align}
	T^{\mu} &= \frac{ P^{\mu} }{ m \sqrt{1 - t / 4m^2} }, \nn\\
	Z^{\mu} &= \frac{ \Delta^{\mu} }{ \sqrt{ -t } },	\nn\\
	X^{\mu} &= \frac{4 \, \eta \, m}{Q^2} \sqrt{ \frac{ 1 - t / 4m^2 }{ 1 - \eta^2 / \eta_0^2 } }
		\bb{ R^{\mu} - \frac{Q^2}{\eta (4 m^2 - t)} P^{\mu} - \frac{Q^2}{2t} \Delta^{\mu} }, \nn\\
	Y^{\mu} & = \epsilon^{\mu T Z X} \propto \epsilon^{\mu P \Delta R},
\label{eq:XYZT-cffs}
\end{align}
where $\eta$ is defined before \eq{eq:cffs-exprs} and has an upper bound $\eta_0$,
\beq
	\eta_0 = \frac{1}{1 + t / Q^2} \sqrt{\frac{-t}{4m^2 - t}}.
\eeq
These coordinate axes are orthogonal to each other and normalized such that $T^2 = -X^2 = -Y^2 = -Z^2 = 1$ and 
\beq[eq:cff-g-txyz]
	g^{\mu\nu} = T^{\mu} T^{\nu} - X^{\mu} X^{\nu} - Y^{\mu} Y^{\nu} - Z^{\mu} Z^{\nu}.
\eeq
They correspond to the rest frame of $P$, where $p$ and $p'$ are along the $+z$ and $-z$ directions, respectively,
and $q$ and $q'$ define the $x$ direction.
The only kinematic singularity happens at the boundary $\eta = \eta_0$ when the two photons are also along the $z$ axis.
Since it has no associated divergence in the whole virtual Compton tensor [\eq{eq:compton-amp}], 
we can omit it in the derivation and eventually approach it by taking the $\eta \to \eta_0$ limit.
We will come back to this after presenting the final result in \eqs{eq:p-even-tau}{eq:p-odd-tau}.

Now let us first consider the tensor $\sigma^{\mu a}$ with one open index. 
The $\mu$ index can be expanded using \eq{eq:cff-g-txyz},
\begin{align}
	\sigma^{\mu a}
	= g^{\mu}{}_{\nu} \sigma^{\nu a}
	& = T^{\mu} \sigma^{T a} - X^{\mu} \sigma^{X a} - Y^{\mu} \sigma^{Y a} - Z^{\mu} \sigma^{Z a}.
\end{align}	
Except for the third term, the other three are all converted to have no open indices, being $\sigma^{P a}$, $\sigma^{\Delta a}$, or $\sigma^{R a}$.
The term involving $Y$ is proportional to 
\begin{align}
	\epsilon^{\mu P \Delta R} \epsilon^{\nu P \Delta R} \sigma_{\nu a} 
	& = \frac{1}{2} \epsilon^{\mu P \Delta R} \epsilon^{\nu P \Delta R} \epsilon_{\nu a \alpha\beta} (i \sigma^{\alpha\beta} \gamma_5)	\nn\\
	& = - \frac{1}{2} \epsilon^{\mu P \Delta R} 
		\cc{ (P \cdot a) \Delta_{[\alpha} R_{\beta]} - (\Delta \cdot a)P_{[\alpha} R_{\beta]} + (R \cdot a) P_{[\alpha} \Delta_{\beta]} } 
		(i \sigma^{\alpha\beta} \gamma_5)	\nn\\
	& = -\epsilon^{\mu P \Delta R} 
		\cc{ (P \cdot a) (i \sigma^{\Delta R} \gamma_5) - (\Delta \cdot a) (i \sigma^{P R} \gamma_5)+ (R \cdot a) (i \sigma^{P \Delta} \gamma_5) },
\label{eq:cff-Ycont}
\end{align}	
where in the second line we used the notation $a_{[\mu} b_{\nu]} \equiv a_{\mu} b_{\nu} - a_{\nu} b_{\mu}$.
So we see that while the $T$, $Z$, and $X$ terms convert the tensor $\sigma^{\mu a}$ to fully contracted ones, 
the $Y$ term converts it to a fully contracted pseudotensor with the coefficient proportional to $\epsilon^{\mu P \Delta R}$.

Similarly, for the fully uncontracted tensor $\sigma^{\mu\nu}$, we have
\begin{align}
	\sigma^{\mu\nu} &= g^{\mu\alpha} g^{\nu\beta} \sigma_{\alpha\beta}
	= - T^{[\mu} X^{\nu]} \sigma^{TX} - T^{[\mu} Y^{\nu]} \sigma^{TY} - T^{[\mu} Z^{\nu]} \sigma^{TZ}
		+ X^{[\mu} Y^{\nu]} \sigma^{XY} + X^{[\mu} Z^{\nu]} \sigma^{XZ} + Y^{[\mu} Z^{\nu]} \sigma^{YZ}.
\end{align}
Here, the terms not involving $Y$ only give regular tensors with regular coefficients carrying the $(\mu, \nu)$ indices.
The terms involving one single contraction with $Y$ give similar results as \eq{eq:cff-Ycont}.

Therefore, besides the scalar structures, we are left with only fully contracted tensors or pseudotensors. The coefficients of the latter can only take the forms of
$a^{\mu} \epsilon^{\nu P \Delta R}$ or $a^{\nu} \epsilon^{\mu P \Delta R}$.
Furthermore, the tensors $\sigma^{P \Delta}$ and $\sigma^{P R}$ can be converted to scalars using \eq{eq:gordon-S}, 
and the two pseudotensors $\sigma^{P \Delta} \gamma_5$ and $\sigma^{\Delta R} \gamma_5$ can be converted to pseudoscalars using \eq{eq:gordon-P}.
This gives the complete list as
\begin{itemize}
\item 10 scalars $S^{\mu\nu} \mathbf{1}$ and 10 tensors $\sigma^{R \Delta} S^{\mu\nu}$, with $S^{\mu\nu}$ the same as \eq{eq:S-10};
\item 6 pseudotensors $i \sigma^{R P} \gamma_5 \wt{S}^{\mu\nu}$ and 6 pseudoscalars $\gamma_5 \wt{S}^{\mu\nu}$, 
	with 
	\beq[eq:cff-Vt]
		\wt{S}^{\mu\nu} = \epsilon^{\mu P \Delta R} (P, \Delta, R)^{\nu}, \; (P, \Delta, R)^{\mu} \epsilon^{\nu P \Delta R},
	\eeq
\end{itemize}
where the pseudoscalar structures come from the conversion of pseudotensors 
$\sigma^{P \Delta} \gamma_5$ and $\sigma^{\Delta R} \gamma_5$ by \eq{eq:gordon-P}.
In this way, we effectively showed that the $13$ tensors with open indices are not completely independent. 
One of them can be reduced to the other twelve.
The resultant 32 structures all satisfy the parity symmetry in \eq{eq:T-parity}.
A further observation is that the scalar structure $\mathbf{1}$ can be replaced by 
a fully contracted vector $\gamma \cdot R$ and tensor $\sigma^{R \Delta}$ by \eq{eq:gordon-V},
and the pseudotensor structure $i \sigma^{R P} \gamma_5$ can be replaced by 
a pseudoscalar $\Delta \cdot R \gamma_5$ and contracted axial-vector $\gamma \cdot R \gamma_5$ by \eq{eq:gordon-A}.
So we can write 
\beq[eq:Tmn-decomp-1]
	T^{\mu\nu}(P, \Delta, R)
	= \frac{1}{2 P \cdot R} \bar{u}(p', \lambda_N') \bb{ \sum_{i = 1}^{10} \pp{ A_i \gamma \cdot R + B_i \frac{i \sigma^{R \Delta}}{2 m} } S_i^{\mu\nu}
		+ \sum_{i = 1}^{6} \pp{ \wt{A}_i \gamma \cdot R \gamma_5 + \wt{B}_i \frac{\Delta \cdot R \gamma_5}{2 m} } \wt{S}_i^{\mu\nu} } u(p, \lambda_N).
\eeq
The reason for using this form is to mimic the form factor decomposition of GPDs [\eq{eq:GPD-def-q}].

The remaining constraint for the coefficients $S^{\mu\nu}$ and $\wt{S}^{\mu\nu}$ comes from gauge invariance,
\beq
	q_{\mu} T^{\mu\nu} = T^{\mu\nu} q'_{\nu} = 0.
\eeq
This can be easily achieved by using
\beq[eq:gauge-proj]
	T^{\mu\nu} 
	= \wt{g}^{\mu\alpha} T_{\alpha\beta} \wt{g}^{\beta \nu}, \quad
	\wt{g}^{\mu\nu} = g^{\mu\nu} - \frac{q^{\prime \mu} q^{\nu}}{q \cdot q'},
\eeq
which projects all tensors in $T$ into gauge-invariant ones.
At this stage, we can see the advantage of using fully contracted Dirac matrices, 
because then the gauge projector only acts on regular tensors made of momentum vectors.
Now the explicit use of $q$ and $q'$ in the projector $\wt{g}^{\mu\nu}$ makes it not convenient to use 
$\{ P, \Delta, R \}$ as the independent momentum basis in $S_i^{\mu\nu}$ and $\wt{S}_i^{\mu\nu}$, 
so we change for $\{ P, q, q' \}$.
Since $\wt{g}^{\mu\nu} q'_{\nu} = q_{\mu} \wt{g}^{\mu\nu} = 0$, the allowed tensors in $S_i^{\mu\nu}$ and $\wt{S}_i^{\mu\nu}$
are simply constructed from $\{ g^{\mu\nu}$, $P^{\mu} P^{\nu}$, $q^{\mu} P^{\nu}$, $P^{\mu} q^{\prime \nu}$, $q^{\mu}q^{\prime \nu} \}$
and $\{ P^{\mu} \epsilon^{\nu P q q'}$, $q^{\mu} \epsilon^{\nu P q q'}$, $\epsilon^{\mu P q q'} P^{\nu}$, $\epsilon^{\mu P q q'} q^{\prime \nu} \}$, 
respectively.
This reduces \eq{eq:Tmn-decomp-1} into $2 \times 5 + 2 \times 4 = 18$ independent form factors.

The independence of the spinor structures and momentum tensors in \eq{eq:Tmn-decomp-1} allows us to 
rewrite the decomposition in any suitable linear combinations of $S_i$ and $\wt{S}_i$.
So we reorganize them by defining
\begin{align}
	g_{\perp}^{\mu\nu} & = 
		g^{\mu\nu} - \frac{q^{\prime \mu} q^{\nu}}{q \cdot q'} - \frac{q^{\mu} q^{\prime \nu}}{q \cdot q'} +  \frac{q^2 q^{\prime \mu} q^{\prime \nu}}{(q \cdot q')^2}, \nn\\
	P_{\perp}^{\mu} & = g_{\perp}^{\mu\nu} P_{\nu}
		= P^{\mu} - \frac{P \cdot q}{q \cdot q'} q^{\prime \mu} - \frac{P \cdot q'}{q \cdot q'} q^{\mu} + \frac{q^2 P \cdot q'}{(q \cdot q')^2} q^{\prime \mu},
	\quad
	(P_{\perp}^2 \equiv P_{\perp}^{\mu} P_{\perp, \mu} )
\label{eq:trans-q-q2}
\end{align} 
which automatically satisfy 
\beq
	g_{\perp}^{\mu\nu} = g_{\perp}^{\nu\mu}, \quad
	g_{\perp}^{\mu\nu} q_{\nu} = g_{\perp}^{\mu\nu} q'_{\nu} = 0, \quad
	g_{\perp \mu}^{\mu} = 2, \quad
	P_{\perp} \cdot q = P_{\perp} \cdot q' = 0.	
\eeq
This is as if we expand in the ``light-cone'' coordinate frame spanned by $q$ and $q'$, except that $q$ is an off-shell photon momentum.
The same frame has also been used in \refs{Braun:2012bg, Braun:2012hq, Braun:2014sta} to write the helicity amplitudes.
Given the three independent momenta $\{ P, q, q' \}$, $P_{\perp}$ is the only available transverse vector.
The remaining transverse direction is simply given by
$\epsilon^{\mu P_{\perp} q q'} = \epsilon^{\mu P q q'}$.
Then we choose the P-even tensor structures as
\bse\label{eq:p-even-tau}\begin{align}
	\tau_1^{\mu\nu} & = -g_{\perp}^{\mu\nu}, \label{eq:p-even-tau1} \\
	\tau_2^{\mu\nu} & = -g_{\perp}^{\mu\nu} + 2 \frac{P_{\perp}^{\mu} P_{\perp}^{\nu}}{P_{\perp}^2},\\
	\tau_3^{\mu\nu} & = \frac{1}{q \cdot q'} \pp{ q^{\mu} - \frac{q^2}{q \cdot q'} q^{\prime \mu}} P_{\perp}^{\nu}, \\
	\tau_4^{\mu\nu} & = \frac{1}{q \cdot q'} P_{\perp}^{\mu} q^{\prime \nu}, \\
	\tau_5^{\mu\nu} & = \frac{1}{q \cdot q'} \pp{ q^{\mu} - \frac{q^2}{q \cdot q'} q^{\prime \mu}} q^{\prime \nu},
\end{align}\ese
and the P-odd ones as
\bse\label{eq:p-odd-tau}\begin{align}
	\wt{\tau}_1^{\mu\nu} & = i \frac{ P_{\perp}^{\mu} \epsilon^{\nu P q q'} - P_{\perp}^{\nu} \epsilon^{\mu P q q'} }{(q \cdot q') P_{\perp}^2}
		= -i \frac{\epsilon^{\mu\nu q q'}}{q \cdot q'}, \label{eq:p-odd-tau1}\\
	\wt{\tau}_2^{\mu\nu} & = i \frac{ P_{\perp}^{\mu} \epsilon^{\nu P q q'} + P_{\perp}^{\nu} \epsilon^{\mu P q q'} }{(q \cdot q') P_{\perp}^2}, \\
	\wt{\tau}_3^{\mu\nu} & = \frac{i}{(q \cdot q')^2} \pp{ q^{\mu} - \frac{q^2}{q \cdot q'} q^{\prime \mu}} \epsilon^{\nu P q q'}, \\
	\wt{\tau}_4^{\mu\nu} & = \frac{i}{(q \cdot q')^2} \epsilon^{\mu P q q'} q^{\prime \nu},
\end{align}\ese
which result in the decomposition in \eq{eq:CFF-decomp} and thereby define the CFFs.

The tensors in \eqs{eq:p-even-tau}{eq:p-odd-tau}
have direct correspondence to the photon helicity structures if we work in the photon frame where both photons move along the $-z$ direction,
as discussed in \eq{eq:cff-helicity-structure}.
There are in total $3 \times 2 = 6$ helicity structures associated with the photons, 
where the transverse polarizations are described by $\tau_1, \tau_2, \wt{\tau}_1$, and $\wt{\tau}_2$,
and the longitudinal polarization of the incoming virtual photon $\gamma^*$ is described by $\tau_3$ and $\wt{\tau}_3$.
The other three, $\tau_4$, $\tau_5$, and $\wt{\tau}_4$, 
correspond to the longitudinal polarization of the outgoing photon, which give zero contribution to the helicity amplitude.

Finally, we note that the tensors in \eqs{eq:p-even-tau}{eq:p-odd-tau} contain the same kinematic singularity as \eq{eq:XYZT-cffs}
due to the explicit use of the transverse vector $P_{\perp}$, which vanishes at the collinear boundary $\eta = \eta_0$,
\beq
	P_{\perp}^2 = \frac{t}{4 \, (1 + t / Q^2)^2} \pp{ \frac{1}{\eta^2} - \frac{1}{\eta_0^2} }.
\eeq
At this point, there is no reference vector to define a transverse direction, 
so rotational invariance dictates the CFFs associated with $\tau_2^{\mu\nu}$ and $\wt{\tau}_2^{\mu\nu}$ to vanish;
that is, we expect $\H_2(\eta, t/m^2, t/Q^2)$ and $\Ht_2(\eta, t/m^2, t/Q^2)$ to vanish as powers of $(\eta - \eta_0)$ as $\eta \to \eta_0$.
The structures $\tau_3$ and $\wt{\tau}_3$ vanish by themselves, so $\H_3$ and $\Ht_3$ do not have such constraints.

\section{Kinematics and cross section formula in the Breit frame}
\label{app:breit-jac}

Since the kinematic observables $(Q^2, x_B, t, \varphi, \varphi_S)$ in the Breit frame 
of the photon electroproduction [in Eq.~\eqref{eq:one-photon} or \eqref{eq:dvcs-sdhep}] are unaffected by a longitudinal boost, 
we present the formulae in the BMK frame~\cite{Belitsky:2001ns}, 
which is a special case of the Breit frame with the nucleon target set at rest. 
The virtual photon $\gamma^*_{ee}(q = \ell - \ell')$ in the DVCS subprocess moves along the $-z$ direction.
The $x$ axis lies on the electron scattering plane, with $\ell$ having positive $x$ component. 

\subsection{Kinematics}
\label{ssec:breit-kins}

We neglect the electron mass but keep the proton mass $m$. Defining the Lorentz invariants as usual,
\beq
	Q^2 = -q^2, \quad
	x_B = \frac{Q^2}{2p \cdot q}, \quad
	y = \frac{p \cdot q}{p \cdot \ell}, \quad
	\gamma = \frac{2 x_B m}{Q}, \quad
	t = (p - p')^2,
\eeq
we have the explicit expressions of the momenta,
\begin{align}
	p &= m(1, 0, 0, 0)_{\rm c}, \quad
	q = \frac{Q}{\gamma} \pp{ 1, 0, 0, -\sqrt{1 + \gamma^2} }_{\rm c}, \nn\\
	\ell &= \frac{Q}{y \gamma} \pp{ 1, \gamma \sqrt{ \frac{1 - y - y^2 \gamma^2 / 4}{1 + \gamma^2} }, 
		0, -\frac{1 + y \gamma^2 / 2}{\sqrt{1 + \gamma^2}} }_{\rm c}, \nn\\
	\ell' &= \ell - q = \frac{Q}{y \gamma} \pp{ 1 - y, \gamma \sqrt{ \frac{1 - y - y^2 \gamma^2 / 4}{1 + \gamma^2} }, 
		0, -\frac{1 - y - y \gamma^2 / 2}{\sqrt{1 + \gamma^2}} }_{\rm c}, \nn\\
	p' &= \pp{ m - \frac{t}{2m}, p_T \cos\varphi, p_T \sin\varphi, \frac{m}{\sqrt{1 + \gamma^2}}\bb{ \frac{t}{2m^2} - x_B \pp{ 1 - \frac{t}{Q^2}} } }_{\rm c}, \nn\\
	q' &= \pp{ \frac{Q}{\gamma} + \frac{t}{2m}, -p_T \cos\varphi, -p_T \sin\varphi, 
		-\frac{m}{\sqrt{1 + \gamma^2}}\bb{ \frac{t}{2m^2} + \frac{Q}{\gamma m} + x_B \pp{ 1 + \frac{t}{Q^2}} } }_{\rm c},	\nn\\
	\Delta &= p - p'
		= \pp{ \frac{t}{2m}, -p_T \cos\varphi, -p_T \sin\varphi, -\frac{m}{\sqrt{1 + \gamma^2}}\bb{ \frac{t}{2m^2} - x_B \pp{ 1 - \frac{t}{Q^2}} } }_{\rm c}.
\label{eq:bmk-kins}
\end{align}
Here $p_T$ is the transverse momentum of the final-state nucleon and photon,
\beq[eq:breit-pT]
	p_T = \frac{1}{\sqrt{1 + \gamma^2}} \sqrt{(1-x_B) (-t) \pp{ 1 + x_B \frac{t}{Q^2}} - x_B^2 m^2 \pp{ 1 + \frac{t}{Q^2}}^2 }.
\eeq
It scales as $p_T \sim \sqrt{(1-x_B)(-t) - x_B^2 m^2}$ at large $Q$, 
which is the same as the $\Delta_T$ in \eq{eq:delta-T} if we use $x_B \simeq 2 \xi / (1 + \xi)$.
The target spin vector in the BMK frame is simply
\beq[eq:spin-BMK]
	S_{\rm BMK}^{\mu} = (0, s_{\perp} \cos\varphi_S, s_{\perp} \sin\varphi_S, s_{\parallel})_{\rm c}
		\equiv (0, \bm{s}_{\perp}, s_{\parallel})_{\rm c},
\eeq
where $s_{\perp} > 0$ and $s_{\parallel}$ are, respectively, the transverse and longitudinal spin components in the BMK frame.
\footnote{To avoid confusion, we use lower-case symbols $(s_{\perp}, s_{\parallel})$ in the BMK frame
and capital symbols $(S_T, P_N)$ in the lab frame
to refer to the transverse and longitudinal polarization degrees of the target.
}

Defining $\theta_q = \vv{ \bm{\ell}, \bm{q} }$ as the angle between the initial-state electron $e(\ell)$ and $\gamma_{ee}^*(q)$ in the BMK frame, we have
\beq[eq:theta-l]
	\cos\theta_q = \frac{1 + y \gamma^2 / 2}{\sqrt{1 + \gamma^2}}, \quad
	\sin\theta_q = \gamma \sqrt{ \frac{1 - y - y^2 \gamma^2 / 4}{1 + \gamma^2} }.
\eeq
In the large-$Q$ limit, $\theta_q$ is a very small angle,
\beq
	\cos\theta_q \simeq 1 + \order{\gamma^2}, \quad
	\sin\theta_q \simeq \gamma \sqrt{1 - y} + \order{\gamma^3}.
\eeq
This fact is only true in the BMK frame,
and will no longer hold when the proton is boosted along the $z$ direction.
But since we have expressed $\theta_q$ in terms of Lorentz invariants, 
the results in the following will be true in any frame.

\subsection{Transformation of the target spin vector}
\label{ssec:jacobian}
When the nucleon target is polarized, 
the transformation from the lab frame to the BMK frame induces a nontrivial rotation of the target spin.
In this subsection, we derive this and the associated Jacobian factor,
generalizing the discussion in \refcite{Diehl:2005pc} to allow the nucleon to have both longitudinal and transverse spins.

Let us first consider the lab frame, where the target is at rest and has a fixed spin vector,
with its transverse component along a (fixed) azimuthal angle $\psi_0$ (normally chosen as $0$).
We denote the target spin vector as
\beq
	S_{\rm lab}^{\mu}	= (0, S_T \cos\psi_0, S_T \sin\psi_0, P_N)_{\rm c}.
\eeq
The kinematics of the final-state electron is described by $(Q^2, x_B)$ and an azimuthal angle $\varphi_\l$ in the lab frame.
The transformation from the lab frame to BMK frame is given by 
a rotation $R_z(-\varphi_\l)$ around the $z$ axis that sets the electrons on the $x$-$z$ plane,
followed by a rotation $R_y(-\theta_q)$ around the $y$ axis to set $\gamma_{ee}^*(q)$ along the $-z$ direction,
where $\theta_q$ is the angle between $e(\ell)$ and $\gamma_{ee}^*(q)$ and is given in \eq{eq:theta-l}.
This transforms the spin vector into 
\begin{align}
	S_{\rm BMK}^{\mu} 
	= R_y(-\theta_q) R_z(-\varphi_\l) S_{\rm lab}^{\mu}
	= (0, S_T \cos\theta_q \cos\psi - P_N \sin\theta_q, \, S_T \sin\psi, \, S_T \sin\theta_q \cos\psi + P_N \cos\theta_q)_{\rm c},
\label{eq:spin-rotation}
\end{align}
with $\psi = \psi_0 - \varphi_\l$ being the azimuthal angle of the spin vector after the first rotation.
Compared with \eq{eq:spin-BMK}, this defines the spin parameters $(s_{\perp}, s_{\parallel}, \varphi_S)$ in the BMK frame.

The purpose of this transformation is to use $\varphi_S$ as a kinematic variable in the BMK frame in the place of $\varphi_\l$, 
so we want to express $(s_{\perp}, s_{\parallel}, \varphi_\l, \psi)$ in terms of $(S_T, P_N, \varphi_S)$.

Combining \eqs{eq:spin-BMK}{eq:spin-rotation} gives us the three equations,
\bse\begin{align}
		s_{\perp} \cos\varphi_S &= S_T \cos\theta_q \cos\psi - P_N \sin\theta_q, \label{eq:Sa}\\
		s_{\perp} \sin\varphi_S &= S_T \sin\psi, \label{eq:Sb}\\
		s_{\parallel} &= S_T \sin\theta_q \cos\psi + P_N \cos\theta_q. \label{eq:Sc}
\end{align}\ese
Using the first two equations to get rid of $\psi$, we obtain
\begin{align}
	\pp{ 1 - \sin^2\theta_q \sin^2\varphi_S} s_{\perp}^2 
		+ 2 \pp{ P_N \sin\theta_q \cos\varphi_S } s_{\perp}
		+ P_N^2 \sin^2\theta_q - S_T^2 \cos^2\theta_q
	= 0.
\label{eq:quadra-sperp}
\end{align}
In order for $s_{\perp}$ to have a real solution, $\varphi_S$ needs to satisfy the constraint,
\beq[eq:constraint-phis]
	\sin^2\varphi_S \leq \frac{S_T^2}{\Lambda^2 \sin^2\theta_q },
\eeq
where $\Lambda^2 = S_T^2 + P_N^2 = s_{\perp}^2 + s_{\parallel}^2$ is the total polarization degree squared of the target.
This constraint is quite easily satisfied in practice because
in usual experimental setup, we either have $P_N = 0$ or $S_T = 0$, so that we have
either $\sin^2\varphi_S \leq 1 / \sin^2\theta_q$ or $\sin^2\varphi_S \leq 0$.
In the former case of a purely transversely polarized target, because $\sin\theta_q \ll 1$, this simply gives $\varphi_S \in [0, 2\pi)$;
in the latter case of a purely longitudinally polarized target, we simply have $\varphi_S = 0$ or $\pi$.
However, in the most general case with both $P_N$ and $S_T$ nonzero, especially when $S_T \ll P_N$,
\eq{eq:constraint-phis} constrains $\varphi_S$ to be around $0$ and $\pi$.
With $\varphi_S$ constrained by \eq{eq:constraint-phis}, we can solve \eq{eq:quadra-sperp} to get $s_{\perp}$,
\beq[eq:ST-sol]
	s_{\perp} = \frac{1}{1 - \sin^2\theta_q \sin^2\varphi_S} 
		\bb{ \cos\theta_q \sqrt{ S_T^2 - \Lambda^2 \sin^2\theta_q \sin^2\varphi_S } - P_N \sin\theta_q \cos\varphi_S },
\eeq
which is picked from the two solutions to the quadratic equation \eqref{eq:quadra-sperp} by requiring $s_{\perp} > 0$ for the special case $P_N = 0$.

At the kinematic boundary (if it can be reached), $S_T^2 - \Lambda^2 \sin^2\theta_q \sin^2\varphi_S = 0$, we have 
\beq
	\sin^2\theta_q \sin^2\varphi_S = \frac{S_T^2}{\Lambda^2}, \quad
	\sin^2\theta_q \cos^2\varphi_S = \sin^2\theta_q - \frac{S_T^2}{\Lambda^2},
\eeq
the second of which gives
\beq[eq:bound-phiS-sols]
	\sin\theta_q \cos\varphi_S = \pm \frac{1}{\Lambda} \sqrt{\Lambda^2\sin^2\theta_q - S_T^2}.
\eeq
Then we have
\beq
	s_{\perp} = - \frac{P_N \sin\theta_q \cos\varphi_S}{1 - \sin^2\theta_q \sin^2\varphi_S}
	= \mp \frac{\Lambda}{P_N} \sqrt{\Lambda^2\sin^2\theta_q - S_T^2}.
\eeq
Clearly, when $P_N > 0$, we need to choose the second ($-$) solution of \eq{eq:bound-phiS-sols}, 
while for $P_N < 0$, we need to choose the first ($+$) solution.
Therefore, we have
\beq
	\sin^2\varphi_S = \frac{S_T^2}{\Lambda^2 \sin^2\theta_q }, \quad
	\cos\varphi_S = -\sgn{P_N} \sqrt{1 - \frac{S_T^2}{\Lambda^2 \sin^2\theta_q} }, \quad
	s_{\perp} = \frac{\Lambda}{|P_N|} \sqrt{\Lambda^2\sin^2\theta_q - S_T^2}.
\eeq
The special case is when $S_T = 0$, where we have $\varphi_S = \pi$ when $P_N > 0$ and $\varphi_S = 0$ when $P_N < 0$.

Substituting \eq{eq:ST-sol} into \eqs{eq:Sa}{eq:Sb}, we get $\psi$,
\begin{align}
	\cos\psi &= \frac{1}{S_T \pp{1 - \sin^2\theta_q \sin^2\varphi_S} }
		\bb{ P_N \sin\theta_q \cos\theta_q \sin^2\varphi_S + \cos\varphi_S \sqrt{ S_T^2 - \Lambda^2 \sin^2\theta_q \sin^2\varphi_S } }, \nn\\
	\sin\psi &= \frac{ \sin\varphi_S }{S_T \pp{ 1 - \sin^2\theta_q \sin^2\varphi_S } }
		\bb{ \cos\theta_q \sqrt{ S_T^2 - \Lambda^2 \sin^2\theta_q \sin^2\varphi_S } - P_N \sin\theta_q \cos\varphi_S }.
\end{align}
Using these, we can get $s_{\parallel}$ from \eq{eq:Sc}, 
\beq[eq:SL-sol]
	s_{\parallel} = \frac{P_N \cos\theta_q + \sin\theta_q \cos\varphi_S \sqrt{ S_T^2 - P^2 \sin^2\theta_q \sin^2\varphi_S }}{1 - \sin^2\theta_q \sin^2\varphi_S}.
\eeq

Finally, let us calculate the Jacobian of the conversion from $\psi$ (or $\varphi_\l$) to $\varphi_S$. Using \eqs{eq:Sa}{eq:Sb}, we have
\beq[eq:tan-phis-phil]
	\tan\varphi_S = \frac{S_T \sin\psi}{S_T \cos\theta_q \cos\psi - P_N \sin\theta_q}.
\eeq
Taking derivative with respect to $\varphi_S$ on both sides gives
\begin{align}
	\frac{d \psi}{d \varphi_S}
	= \frac{1}{\cos^2\varphi_S} \frac{\pp{ S_T \cos\theta_q \cos\psi - P_N \sin\theta_q }^2}{S_T \pp{ S_T \cos\theta_q - P_N \sin\theta_q \cos\psi }}.
\end{align}
Using the explicit results of $\psi$, we have
\beq[eq:jac-phi]
	J = \frac{d \psi}{d \varphi_S} = \left| \frac{d \varphi_\l}{d \varphi_S} \right|
	= \frac{\cos\theta_q - \alpha_L}{1 - \sin^2\theta_q \sin^2\varphi_S},
\eeq
where we defined 
\beq[eq:alpha-L]
	\alpha_L = \frac{P_N \sin\theta_q \cos\varphi_S}{\sqrt{ S_T^2 - \Lambda^2 \sin^2\theta_q \sin^2\varphi_S }}.
\eeq
Note that we use $\varphi_S$ only when $S_T \neq 0$, and in actual cases of such, $S_T^2 \gg \Lambda^2 \sin^2\theta_q$,
so we do not worry about the singularity when the denominator of $\alpha_L$ vanishes.

Taking $P_N = 0$, \eq{eq:jac-phi} reduces to the Jacobian factor in the Eq.~(9) of \refcite{Diehl:2005pc}.
Here \eq{eq:jac-phi} extends this to a general situation for a more comprehensive theory description.

Before moving on, we note that in the massless target limit, by setting $\gamma \to 0$, the Jacobian becomes trivial.
This conclusion is nontrivial from our derivation because we have been working in the target rest frame which does not exist for massless particles.
However, since the result is not affected by a longitudinal boost in the Breit frame, the Jacobian obtained is boost invariant.
This trivial limit is actually due to a new symmetry for massless particles.
In both the lab and Breit frames, the nucleon carries a lightlike momentum along $z$,
such that the Lorentz transformation connecting the two frames is simply a little group transformation of the massless nucleon,
which keeps its helicity unchanged; the only effect of the transformation is due to the rotation $R_z(-\phi_\l)$.

\subsection{Cross section formula in the Breit frame}

Because of the extra Jacobian factor, we provide for completeness a detailed derivation of the differential cross section formula in the BMK frame.

We start in the lab frame by writing the final-state phase space in $\ell'$, $p'$, and $q'$,
\begin{align}
	\sigma &= \frac{1}{2(s-m^2)} 
		\int \frac{d^3\bm{p'}}{(2\pi)^3 2 E_{p'}} \frac{d^3\bm{\ell'}}{(2\pi)^3 2 E_{\ell'}} \frac{d^3\bm{q'}}{(2\pi)^3 2 E_{q'}}
		\, \overline{|\M|^2}
		\, (2\pi)^4 \delta^{(4)}(p + \ell - p' - \ell' - q') \nn\\
	&= \frac{1}{2(s-m^2)} 
		\int \frac{d^3\bm{p'}}{(2\pi)^3 2 E_{p'}} \frac{d^3\bm{\ell'}}{(2\pi)^3 2 E_{\ell'}}
		\, \overline{|\M|^2}
		\, (2\pi) \, \delta^+\pp{ (p + q - p')^2 },
\end{align}
where $\overline{|\M|^2}$ is the square of the total amplitude, with initial-state spins averaged with proper density matrices.
In the second step, we introduced $q = \ell - \ell'$ and used the $\delta$-function to integrate over $q'$.
The $\delta^+(k^2)$ function is a shorthand notation for $\delta(k^2) \theta(k^0)$, 
which is a normal $\delta$-function constrained with a positive energy component. 
Similarly to DIS or SIDIS, it is straightforward to express the $\l'$ phase space in $(x_B, Q, \varphi_\l)$,
\begin{align}
	\sigma &= \int \frac{y^2}{Q^2} \frac{d x_B \, dQ^2 \, d \varphi_{\l}}{(4\pi)^3}
		\int \frac{d^3\bm{p'}}{(2\pi)^3 2 E_{p'}}
		\, \overline{|\M|^2}
		\, (2\pi) \, \delta^+\pp{ (p + q - p')^2 }.
\label{eq:phase-space-xQ}
\end{align}

Now because both the $p'$ integration measure and the amplitude squared are Lorentz invariants,
we can express the $p'$ phase space in the BMK frame in $(t, p^{\prime z}, \varphi)$, 
for which the Jacobian can be worked out easily to give
\beq
	\frac{d^3\bm{p'}}{2 E_{p'}}
	= \frac{d |t| d p^{\prime z} d \varphi }{4 m}.
\eeq
Then we can use the kinematics in \eq{eq:bmk-kins} to write the $\delta$-function in \eq{eq:phase-space-xQ} 
as a constraint on $p^{\prime z}$,
\begin{align}
	\delta^+\pp{ (p + q - p')^2 }
	&= \delta\pp{\frac{Q^2}{x_B} - Q^2 + t - \frac{2Q}{\gamma} \pp{ \frac{2 m^2 - t}{2 m} - p^{\prime z} \sqrt{1 + \gamma^2} } },
\end{align}
which allows to integrate it over and gives
\begin{align}
	\frac{d\sigma}{ d x_B \, dQ^2 \, d \varphi_{\ell} \, d |t| \, d \varphi }
	= \frac{x_B \, y^2 \, \overline{ |\M|^2}}{(4\pi)^5 Q^4 \sqrt{1 + \gamma^2}}	
	= \frac{\alpha_e^3 \, x_B \, y^2}{(4\pi)^2 Q^4 \sqrt{1 + \gamma^2}} \overline{ \left| \frac{\M}{e^3} \right|^2}.
\end{align}
The last expression agrees with the Eq.~(22) of \refcite{Belitsky:2001ns} if we simply trade $\varphi_{\ell}$ for $\varphi_S$.
However, as we have seen in Appendix~\ref{ssec:jacobian}, it is associated with a nontrivial Jacobian, so that the correct result is
\begin{align}
	\frac{d\sigma}{ d x_B \, dQ^2 \, d \varphi_S \, d |t| \, d \varphi }
	= \frac{\alpha_e^3 \, x_B \, y^2}{(4\pi)^2 Q^4 \sqrt{1 + \gamma^2}} 
		\pp{ \frac{\cos\theta_q - \alpha_L}{1 - \sin^2\theta_q \sin^2\varphi_S} }
		\overline{ \left| \frac{\M}{e^3} \right|^2}.
\end{align}
where $\alpha_L$ is given in \eq{eq:alpha-L}, and $\theta_q$ is given in \eq{eq:theta-l} as a function of $(x_B, y, Q^2)$.
This result applies to a generic Breit frame, unaffected by any longitudinal boost along the $z$ axis.

The $\varphi_S$ dependence in this extra Jacobian factor $J$ is nontrivial.
To examine its impacts, we consider the hard scattering regime with $Q \gg m$.
In this case, $\theta_q$ is a small angle, and we can Taylor expand $J$ in $\gamma$,
\begin{align}
	J = 1 - \gamma \sqrt{1-y} \, \frac{P_N}{S_T} \cos\varphi_S 
		- \frac{\gamma^2}{2} (1-y) \cos 2\varphi_S
		+ \order{\gamma^3}.
\end{align}
(As mentioned in Appendix~\ref{ssec:jacobian}, we use $\varphi_S$ as a kinematic variable and thus consider the Jacobian factor $J$ 
only when $S_T \neq 0$, so the occurrence of $S_T$ in the denominator should not be a worry.)
The same $J$ factor exists in DIS and SIDIS, but there 
we work at the LP of $\gamma$ in $\left| \M \right|^2$, 
so taking $J$ to 1 does not affect the accuracy.

However, for the exclusive photon electroproduction, this $J$ affects both BH and DVCS.
As we have argued from the power counting in Secs.~\ref{sec:dvcs-sdhep} and \ref{sec:calc-sdhep},
the BH squared is at LP, its interference with the DVCS is at NLP, and the DVCS squared is at NNLP.
Since one tends to work up to NNLP to include the DVCS squared,
this Jacobian affects all of these three parts:
\begin{itemize}
\item When $P_N \neq 0$, $J$ contributes a $\cos\varphi_S$ component at NLP, which enters the BH squared and interference part.
\item Even when $P_N = 0$, $J$ gives a $\cos 2\varphi_S$ component at NNLP,
which can contribute through the BH square and contaminate the $\varphi_S$ modulations in the interference part and DVCS squared.
\end{itemize}

The SDHEP formulation of the kinematics takes a two-stage description in \sec{ssec:sdhep} 
and avoids any nontrivial Jacobian factor associated with both $\phi_S$ and $\phi$.
So it is natural to ask whether a similar two-stage method works for the Breit frame to remove the $J$ factor.
The answer is very likely to be no.
The reason is that a two-stage description works for the SDHEP framework 
because $\phi_S$ and $\phi$ are associated with two separated stages, 
with $\phi_S$ solely responsible for the nucleon diffraction;
Its only connection to the hard $2\to2$ scattering is via the $A^*$ state.
In contrast, in the Breit frame formulation, $\bm{s}_{\perp}$ and $\varphi$ are involved 
in the same collision subprocess between $N(p)$ and $\gamma^*_{ee}(q)$.
Even if one may consider using $\varphi_\l$ instead of $\varphi_S$,
the $\varphi$ distribution depends on the azimuthal direction $\varphi_S$ of $\bm{s}_{\perp}$, 
which in turn is dependent on $\varphi_\l$ in the complicated form as \eq{eq:tan-phis-phil}
to render a harmonic analysis of $\varphi_\l$ useless in practice.
The physical reason, again, is the lack of scale separation in the Breit frame formulation.
The separation at the virtual state $\gamma^*_{ee}(q)$ 
is not associated with a power counting in the same sense as the $A^*(\Delta)$ in the SDHEP framework.

Aside from the $J$ factor, we notice that in the Breit frame, 
the $s_{\perp}$ and $s_{\parallel}$ depend on $\varphi_S$ in a complicated way, cf.~\eqs{eq:ST-sol}{eq:SL-sol}.
This can make the use of $\varphi_S$ modulations difficult.
The target spin asymmetries in the Breit frame do not directly correspond to those measured in experiments.
We refer to \refcite{Diehl:2005pc} for a detailed discussion of their relations.

\section{Comparison of the calculation results in the Breit frame and SDHEP frame}
\label{app:comp-ret}

In this Appendix, we calculate the square of the photon electroproduction amplitude in a covariant approach. 
The results will be Lorentz-invariant expressions.
Then specifying a frame (and a choice of the lightlike vector $n$) will adapt them to specific forms.
In this way, we can verify the equivalence between the SDHEP formalism and the Breit frame formulation.

We follow the setup in \sec{sec:review} and separate the whole amplitude into BH and DVCS by \eq{eq:dvcs-bh-amp}.
The BH amplitude $\M^{\rm BH}$ is given in Eqs.~\eqref{eq:BH-amplitude}--\eqref{eq:EM-form-factor}.
The DVCS amplitude $\M^{\rm DVCS}$ is given in \eqs{eq:DVCS-amp}{eq:compton-amp}.
At twist 2 and LO, the virtual Compton tensor $T^{\mu\nu}$ in \eq{eq:compton-amp} is factorized into GPD moments,
\begin{align}
	T^{\mu\nu}
	&= \frac{1}{2 P \cdot n} \bar{u}(p', \lambda_N') 
		\cc{
			\bb{ \H(\xi, t) \gamma \cdot n - \E(\xi, t) \frac{i \sigma^{n \Delta} }{2m} } \tau_1^{\mu\nu}
			- \bb{ \wt{\H}(\xi, t) \gamma \cdot n \gamma_5 - \wt{\E}(\xi, t) \frac{\gamma_5 \Delta \cdot n}{2m} } \wt{\tau}_1^{\mu\nu}
		}
		u(p, \lambda_N),
\label{eq:dvcs-T-LP-moments}
\end{align}
which is the same as \eq{eq:dvcs-T-LP} but expressed in terms of the GPD moments defined in \eq{eq:dvcs-GPD-form-factors}.
\footnote{Our GPD moments $(\H, \E, \Ht, \Et)$ are defined using the kernel $1 / (x - \xi + i \epsilon)$ for the C-even GPD combinations.
These differ from the BMK convention by a $-1$ factor (see Eqs.~(9) and (11) in \refcite{Belitsky:2001ns}). 
That is, $(\H, \E, \Ht, \Et)_{\rm here} = -(\H, \E, \Ht, \Et)_{\rm BMK}$.}
The tensors $\tau_1^{\mu\nu}$ and $\wt{\tau}_1^{\mu\nu}$ are defined in \eqs{eq:p-even-tau1}{eq:p-odd-tau1}.
At this stage, we keep the vector $n$ to be general and will specify it later in specific frames.

\begin{figure}[htbp]
	\centering
	\begin{tabular}{ccc}
		\includegraphics[scale=0.6]{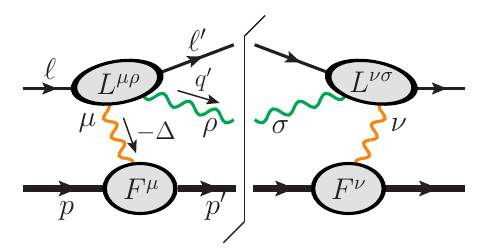} &
		\includegraphics[scale=0.6]{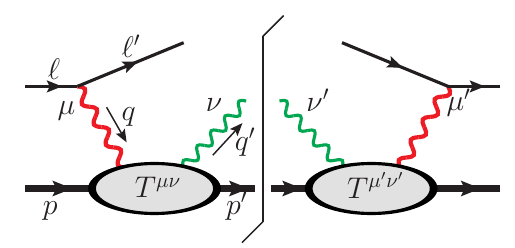} &
		\includegraphics[scale=0.6]{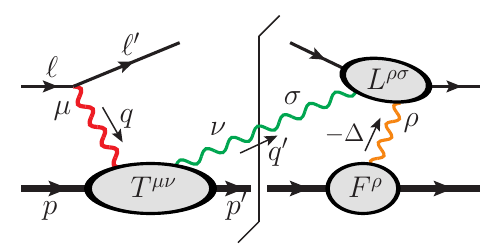} \\
		(a) & (b) & (c)
	\end{tabular}
	\caption{Cut diagrams representing the three parts in \eq{eq:amp2-3-parts}:
		(a) BH amplitude squared,
		(b) DVCS amplitude squared,
		and
		(c) the DVCS-BH interference.}
\label{fig:cov-amps-cut}
\end{figure}

The amplitude squared, with full spin dependence included, is given by
\begin{align}
	\overline{|\M|^2} 
	& = \rho^N_{\lambda_N \bar{\lambda}_N} \rho^e_{\lambda_e \bar{\lambda}_e}
		\pp{ \M^{\rm BH} + M^{\rm DVCS} }_{\lambda_N \lambda_e, \{\alpha'\} }
		\pp{ \M^{\rm BH} + M^{\rm DVCS} }^*_{\bar{\lambda}_N \bar{\lambda}_e, \{\alpha'\} }
	\nn\\
	& = \overline{|\M^{\rm BH}|^2} + \overline{|\M^{\rm DVCS}|^2} + 2 \Re{\M^{\rm DVCS} \M^{{\rm BH}*}},
\label{eq:M2-3-parts}
\end{align}
where repeated indices are summed over, and $\{ \alpha' \}$ stands for helicities of the final-state particles.
In the second step of \eq{eq:M2-3-parts} we defined
the three parts constituting the whole amplitude squared,
\begin{align}
	\overline{|\M^{\rm BH}|^2}
	&= \rho^N_{\lambda_N \bar{\lambda}_N} \rho^e_{\lambda_e \bar{\lambda}_e}
		\M^{\rm BH}_{\lambda_N \lambda_e, \{\alpha'\} } \bigpp{ \M^{\rm BH}_{\bar{\lambda}_N \bar{\lambda}_e, \{\alpha'\} } }^*,
	\nn\\
	\overline{|\M^{\rm DVCS}|^2}
	&= \rho^N_{\lambda_N \bar{\lambda}_N} \rho^e_{\lambda_e \bar{\lambda}_e}
		\M^{\rm DVCS}_{\lambda_N \lambda_e, \{\alpha'\} } \bigpp{ \M^{\rm DVCS}_{\bar{\lambda}_N \bar{\lambda}_e, \{\alpha'\} } }^*,
	\nn\\
	2 \Re{\M^{\rm DVCS} \M^{{\rm BH}*}}
	&= 2 \Re{ \rho^N_{\lambda_N \bar{\lambda}_N} \rho^e_{\lambda_e \bar{\lambda}_e}
		\M^{\rm DVCS}_{\lambda_N \lambda_e, \{\alpha'\} } \bigpp{ \M^{\rm BH}_{\bar{\lambda}_N \bar{\lambda}_e, \{\alpha'\} } }^*
		},
\label{eq:amp2-3-parts}
\end{align}
which are shown as cut diagrams in \fig{fig:cov-amps-cut}.

The spin sum or average in terms of the helicity amplitudes and density matrices can be performed in a covariant way.
First, unpolarized spin sum or average is frame independent, so 
\begin{itemize}
	\item for the final-state electron, we replace $\sum_{\lambda_e'} u(\l', \lambda_e') \, \bar{u}(\l', \lambda_e')$ by $\gamma \cdot \l'$;
	\item for the final-state nucleon, we replace $\sum_{\lambda_N'} u(p', \lambda_N') \, \bar{u}(p', \lambda_N')$ by $\gamma \cdot p' + m$; and
	\item for the final-state photon, we replace $\sum_{\lambda'} \epsilon^{\mu *}_{\lambda'}(q') \, \epsilon^{\nu}_{\lambda'}(q')$ by $-g^{\mu\nu}$.
\end{itemize}
Then, the massless electron helicity $P_e$ is Lorentz invariant, so we can make the replacement,
\beq
	\sum_{\lambda_e, \bar{\lambda}_e} u(\l, \lambda_e) \, \rho^e_{\lambda_e \bar{\lambda}_e} \, \bar{u} (\l, \bar{\lambda}_e)
	=\frac{\slash{\l}}{2} \pp{1 - P_e \gamma_5}.
\eeq
Finally, the initial-state proton spin average can be done with a covariant spin vector $S_{\mu}$,
\beq
	\sum_{\lambda_N, \bar{\lambda}_N} u(p, \lambda_N) \, \rho^N_{\lambda_N \bar{\lambda}_N} \, \bar{u} (p, \bar{\lambda}_N)
	= \frac{1}{2}(\slash{p} + m) \pp{1 + \gamma_5 \Slash{S}}.
\eeq
After these replacements, the amplitude square is a Lorentz-invariant expression. 

In the following, we calculate the three parts in \eq{eq:amp2-3-parts} one by one.

\subsection{BH squared}
\label{ssec:breit-amp2-bh}
Using \eq{eq:BH-amplitude}, we can write the BH amplitude squared as
\beq[eq:covariant-BH2]
	\overline{|\M^{\rm BH}|^2}
		= \frac{e^6}{t^2} \, L_{\rm BH}^{\mu\nu}(\l, \l', \Delta)  W_{{\rm BH}, \mu\nu}(P, \Delta),
\eeq
where we defined the leptonic tensor,
\begin{align}
	L_{\rm BH}^{\mu\nu}
	&= \sum_{\lambda_e, \bar{\lambda}_e, \lambda_e', \lambda'} 
		\rho^e_{\lambda_e \bar{\lambda}_e} 
		L_{\lambda_e\lambda_e'}^{\mu\rho} \, L_{\bar{\lambda}_e\lambda_e'}^{\nu\sigma *} \, \epsilon^*_{\lambda', \rho}(q') \, \epsilon_{\lambda', \sigma}(q') 	\nn\\
	&= \Tr\cc{ \Slash{\l'}
		\bb{ \gamma^{\mu} \frac{\gamma\cdot (\l - q')}{(\l - q')^2} \gamma^{\rho} 
			+ \gamma^{\rho} \frac{\gamma\cdot (\l + \Delta)}{(\l + \Delta)^2} \gamma^{\mu} 
		}
		\frac{\slash{\l} (1 - P_e \gamma_5)}{2} 
		\bb{ \gamma^{\sigma} \frac{\gamma\cdot (\l - q')}{(\l - q')^2} \gamma^{\nu} 
			+ \gamma^{\nu} \frac{\gamma\cdot (\l + \Delta)}{(\l + \Delta)^2} \gamma^{\sigma} 
		}
	} (-g_{\rho\sigma}),
\end{align}
and the hadronic tensor,
\begin{align}
	W_{\rm BH}^{\mu\nu}
	&= \sum_{\lambda_N, \bar{\lambda}_N, \lambda_N'} \rho^N_{\lambda_N \bar{\lambda}_N} 
		F_{\lambda_N\lambda_N'}^{\mu} F_{\bar{\lambda}_N \lambda_N'}^{\nu *} \nn\\
	&= \Tr\cc{ (\Slash{p'} + m) \bb{ F_1(t) \gamma^{\mu} - F_2(t) \frac{i \sigma^{\mu\Delta}}{2m} }
			\frac{\slash{p} + m}{2} (1 + \gamma_5 \Slash{S})
			\bb{ F_1(t) \gamma^{\nu} + F_2(t) \frac{i \sigma^{\nu\Delta}}{2m} }
		}.
\end{align}
To simplify expressions in the following, we define the two propagators from the BH amplitude,
\beq[eq:bh-props]
	\P_1 = \frac{(\l - q')^2}{Q^2} = 1 - \frac{2\l \cdot \Delta}{Q^2}, \quad
	\P_2 = \frac{(\l + \Delta)^2}{Q^2} = \frac{t + 2\l \cdot \Delta}{Q^2},
	\quad
	\P_1 + \P_2 = 1 + \frac{t}{Q^2}.
\eeq

The $L_{\rm BH}$ and $W_{\rm BH}$ are connected by the virtual photon of momentum $\Delta$. They both satisfy the Ward identity,
\beq
	L_{\rm BH}^{\mu\nu} \Delta_{\mu} = L_{\rm BH}^{\mu\nu} \Delta_{\nu} 
	= W_{\rm BH}^{\mu\nu} \Delta_{\mu} = W_{\rm BH}^{\mu\nu} \Delta_{\nu} = 0.
\eeq
Therefore, we can write them as gauge-invariant structures.
The leptonic tensor is decomposed as
\begin{align}
	\bb{ \frac{8}{\P_1(\phi) \P_2(\phi)} }^{-1} L_{\rm BH}^{\mu\nu}
	&= \frac{ (\l \cdot \Delta)^2 + (\l' \cdot \Delta)^2 }{Q^4} \, \wt{g}^{\mu\nu}
		+ \, \frac{t}{Q^4} \pp{ \, \wt{\l}^{\mu} \wt{\l}^{\nu} + \wt{\l'}{}^{\mu} \wt{\l'}{}^{\nu} } \nn\\
	& -i P_e \bb{ \frac{1}{2 Q^2} \pp{ 1 + \frac{t^2}{Q^4} } 
			\frac{ \epsilon^{\mu \l \l' \Delta} \wt{\Delta}_{t}^{\nu} - \epsilon^{\nu \l \l' \Delta} \wt{\Delta}_{t}^{\mu} }{ \Delta_t^2 }
			- \frac{1}{Q^4} \pp{ 1 - \frac{t}{Q^2} } \pp{ \epsilon^{\mu \l \l' \Delta} \, \wt{q}^{\nu} - \epsilon^{\nu \l \l' \Delta} \, \wt{q}^{\mu} }
		},
\end{align}
where we defined
\beq[eq:cov-gauge-inv-proj]
	\wt{g}^{\mu\nu} = g^{\mu\nu} - \frac{\Delta^{\mu} \Delta^{\nu}}{\Delta^2}, \quad
	\wt{A}^{\mu} = \wt{g}^{\mu\nu} A_{\nu} = A^{\mu} - \frac{\Delta\cdot A}{\Delta^2} \Delta^{\mu}.
\eeq
The $\Delta_t$ is defined as the transverse component of $\Delta$ on the lightcone basis spanned by $\l$ and $\l'$,
\beq
	\Delta_t = \Delta - \frac{(\Delta \cdot \l) \l' + (\Delta \cdot \l') \l}{\l \cdot \l'}, 
\eeq
which has the properties,
\beq
	\Delta_t \cdot \l = \Delta_t \cdot \l' = 0, \qquad
	\Delta_t^2 = t - \frac{4 (\Delta \cdot \l) (\Delta \cdot \l')}{ Q^2 }.
\eeq
The hadronic tensor is given by
\begin{align}
	W_{\rm BH}^{\mu\nu}
	&= t (F_1 + F_2)^2 \, \wt{g}^{\mu\nu} + 4 \pp{ F_1^2 - \frac{t}{4m^2} F_2^2 } P^{\mu} P^{\nu} 
		+ \frac{\Delta \cdot S}{2m} F_1 F_2 \, ( i \epsilon^{\mu\nu P \Delta} )
			- 2 m (F_1 + F_2) \pp{ F_1  + \frac{t}{4m^2} F_2 } \, ( i \epsilon^{\mu\nu S \Delta} )
			\nn\\
	&\hspace{1em}
			+ \frac{t}{2m} F_2 (F_1 + 2 F_2) \, 
			i \pp{ \epsilon^{\mu\nu P S} + \frac{\epsilon^{\mu P \Delta S} \Delta^{\nu} - \epsilon^{\nu P \Delta S} \Delta^{\mu} }{t} }.
\end{align}

In both $L_{\rm BH}^{\mu\nu}$ and $W_{\rm BH}^{\mu\nu}$, the spin-independent parts are symmetric with respect to $(\mu, \nu)$,
while the spin-dependent parts are antisymmetric.
Thus, when they contract, those two parts do not mix. By writing 
\beq[eq:LW-BH]
	L_{\rm BH} \cdot W_{\rm BH} = L_{\rm BH} \cdot W_{\rm BH} |_{\rm unp.} - P_e \, L_{\rm BH} \cdot W_{\rm BH} |_{\rm pol.},
\eeq 
the unpolarized part is
\begin{align}
	\bb{ \frac{32 m^2}{\P_1 \P_2} }^{-1} L_{\rm BH} \cdot W_{\rm BH} |_{\rm unp.}
	&= \frac{t}{m^2} \pp{ F_1^2 - \frac{t}{4m^2} F_2^2 } \frac{(\l \cdot P)^2 + (\l' \cdot P)^2 }{Q^4} 	\nn\\
	&\quad
		+ \bb{ \pp{1 -  \frac{t}{4m^2} } \pp{ F_1^2 - \frac{t}{4m^2} F_2^2 } + \frac{t}{2 m^2} (F_1 + F_2)^2 } \frac{(\l \cdot \Delta)^2 + (\l' \cdot \Delta)^2 }{Q^4},
\label{eq:cov-amp2-bh-0}
\end{align}
and the polarized part is
\begin{align}
	\bb{ \frac{32 m^2 (F_1 + F_2) }{\P_1 \P_2} }^{-1} L_{\rm BH} \cdot W_{\rm BH} |_{\rm pol.}
	&= \frac{\Delta \cdot S}{m} \bb{\pp{ F_1 + \frac{t}{4m^2} F_2 } \frac{ (\l \cdot \Delta)^2 - (\l' \cdot \Delta)^2 }{Q^4}
		+ \frac{t}{2 m^2} F_2  \frac{ (\l \cdot P)(\l \cdot \Delta) - (\l' \cdot P)(\l' \cdot \Delta) }{Q^4}
		}	\nn\\
	&\quad
		- \frac{t}{m \, Q} \pp{ F_1 + \frac{t}{4m^2} F_2 }
		 \frac{ (\l \cdot S)(\l \cdot \Delta) - (\l' \cdot S)(\l' \cdot \Delta) }{Q^3}.
\label{eq:cov-amp2-bh-1}
\end{align}

\subsubsection{Specified to the Breit frame}
\label{sssec:check-amp2-bh-bmk}
The Breit frame kinematics has been discussed in Appendix~\ref{ssec:breit-kins}.
From that, we see that both propagators in \eq{eq:bh-props} depend on $\varphi$, though the same invariant $\l \cdot \Delta$.
Following the convention of \refcite{Belitsky:2001ns}, we write $\P_{1,2}$ as
\beq[eq:P12-breit-explicit]
	\P_1 = - \frac{1}{y(1 + \gamma^2)} \pp{ J + 2 K \cos\varphi }, \quad
	\P_2 = 1 + \frac{t}{Q^2} + \frac{1}{y(1 + \gamma^2)} \pp{ J + 2 K \cos\varphi },
\eeq
where $J$ and $K$ are
\begin{align}
	J &= \pp{ 1 - y - \frac{1}{2} y \gamma^2 } \pp{1 + \frac{t}{Q^2} } - (1 - x_B) (2 - y) \frac{t}{Q^2}, \nn\\
	K &= \frac{p_T}{Q} \sqrt{(1 + \gamma^2) \pp{ 1 - y - \frac{1}{4} y^2 \gamma^2 } },
\end{align}
with $p_T$ given in \eq{eq:breit-pT}.
At large $Q$, they scale as $J \sim (1 - y)$ and $K \sim \sqrt{1 - y} \, (p_T / Q)$.

Then using Eqs.~\eqref{eq:covariant-BH2} and \eqref{eq:LW-BH}--\eqref{eq:cov-amp2-bh-1}, 
we obtain the BH amplitude squared in the Breit frame,
\begin{align}
	\overline{|\M^{\rm BH}|^2} \big|_{\rm BMK}
	&= \frac{e^6}{x_B^2 y^2 (1 + \gamma^2)^2 \, t \, \P_1 \P_2 } 
	\Big\{ \Sigma_{UU, 0}^{\rm BH} + \Sigma_{UU, 1}^{\rm BH} \cos\varphi + \Sigma_{UU, 2}^{\rm BH} \cos2\varphi 
		+ s_{\parallel} P_e \pp{ \Sigma_{LL, 0}^{\rm BH} + \Sigma_{LL, 1}^{\rm BH} \cos\varphi }
		\nn\\
	&\hspace{4em}
		+ s_{\perp} P_e \bb{ 
			\Sigma_{TL, 0}^{\rm BH} \cos(\varphi - \varphi_S)
			+ \Sigma_{TL, 1}^{\rm BH} \cos\varphi \cos(\varphi - \varphi_S)
			+ \Sigma_{TL, 2}^{\rm Int.} \sin\varphi \sin(\varphi - \varphi_S)
		}
	\Big\},
\label{eq:bmk-bh}
\end{align}
which agrees with the BMK results in the Eqs.~(25) and (35)--(42) of \refcite{Belitsky:2001ns} exactly for all polarization cases.
Specifically, the polarization asymmetries here are related to quantities in \refcite{Belitsky:2001ns} by
\begin{align}
	\cc{ \Sigma_{UU, 0}^{\rm BH}, \Sigma_{UU, 1}^{\rm BH}, \Sigma_{UU, 2}^{\rm BH} }
	&= \cc{ c_{0, {\rm unp}}^{\rm BH} , c_{1, {\rm unp}}^{\rm BH}, c_{2, {\rm unp}}^{\rm BH} }_{\rm BMK}, \nn\\
	\cc{ \Sigma_{LL, 0}^{\rm BH}, \Sigma_{LL, 1}^{\rm BH} }
	&= \frac{1}{\lambda \Lambda}\cc{ c_{0, {\rm LP}}^{\rm BH} , c_{1, {\rm LP}}^{\rm BH} }_{\rm BMK}, \nn\\
	\cc{ \Sigma_{TL, 0}^{\rm BH}, \Sigma_{TL, 1}^{\rm BH}, \Sigma_{TL, 2}^{\rm BH} }
	&= \frac{1}{\lambda}\cc{ c_{0, {\rm TP}}^{\rm BH} , c_{1, {\rm TP}}^{\rm BH}, -s_{1, {\rm TP}}^{\rm BH} }_{\rm BMK},
\end{align}
where the notations on the right-hand side are all defined in \refcite{Belitsky:2001ns}.
The minus sign for $s_{1, {\rm TP}}^{\rm BH}$ is because the $\sin(\varphi)$ in the Eq.~(42) of \refcite{Belitsky:2001ns} 
is in fact our $\sin(\varphi_S - \varphi)$ here.

\subsubsection{Specified to the SDHEP frame}
\label{sssec:check-amp2-bh-sdhep}
The SDHEP-frame kinematics has been specified in \eqs{eq:sdhep-p-p'-s}{eq:sdhep-p+} for the nucleon diffraction 
and \eq{eq:bh-kins} for the $2\to2$ hard scattering.
This kinematics is very simple for the BH. In particular, we have no $\phi$ dependence in either propagator of \eq{eq:bh-props},
\begin{align}
	\P_1 \P_2 = - \frac{1}{1 - t / \hat{s}} \frac{\tan^2(\theta/2)}{\cos^2(\theta/2)}.
\end{align}

The $\phi$ dependence comes from $\l' \cdot P$ and $\l' \cdot S$ in \eqs{eq:cov-amp2-bh-0}{eq:cov-amp2-bh-1}.
By explicit calculation, we obtain the BH amplitude squared in the SDHEP frame,
\begin{align}
	\overline{|\M^{\rm BH}|^2} \big|_{\rm SDHEP} & = \pp{ \frac{2e^3 m }{t} }^2 
		\bb{ \bar{\Sigma}^{\rm BH}_{UU} + \bar{\Sigma}^{\rm BH}_{UU, 1} \cos\phi + \bar{\Sigma}^{\rm BH}_{UU, 2} \cos2\phi
		+ P_e P_N (\bar{\Sigma}^{\rm BH}_{LL, 0} + \bar{\Sigma}^{\rm BH}_{LL, 1} \cos\phi) 	\right.\nn\\
		&\hspace{9em}\left.
		+ P_e S_T ( \bar{\Sigma}^{\rm BH}_{TL, 0} \cos\phi_S + \bar{\Sigma}^{\rm BH}_{TL, 1} \cos\phi_S \cos\phi + \bar{\Sigma}^{\rm BH}_{TL, 2} \sin\phi_S \sin\phi)
		},
\label{eq:bh-amp2-sdhep}
\end{align}
where we used $\bar{\Sigma}$ to distinguish from the same notations in \eq{eq:bmk-bh}.
These dimensionless polarization parameters are
\bse\label{eq:bh-amp2-sdhep-pol}\begin{align}
	\bar{\Sigma}^{\rm BH}_{UU} &
		= \bb{ \frac{1 - t / \hat{s}}{\sin^2(\theta/2)} + \frac{\sin^2(\theta/2)}{1 - t / \hat{s}} \pp{ 1 + \frac{t^2}{\hat{s}^2} \cot^4(\theta/2) } }
			\bb{ \frac{\Delta_T^2}{2 m^2} \pp{\frac{1+\xi}{\xi}}^2 \pp{ F_1^2 + \frac{-t}{4m^2} F_2^2 } 
				+ \frac{-t}{m^2} (F_1 + F_2)^2 }	\nn\\
		& \hspace{1em}
			+ \frac{t^2}{\hat{s} \, m^2} \frac{1 + \cos\theta}{1 - t / \hat{s}} \bb{ \frac{1}{\xi^2} \pp{ F_1^2 + \frac{-t}{4m^2} F_2^2 } - (F_1 + F_2)^2 }, \\
	\bar{\Sigma}^{\rm BH}_{UU, 1} & 
		= \frac{-t}{m \sqrt{\hat{s}} } \frac{\Delta_T}{m} \frac{\sin\theta}{1 - t / \hat{s}} \pp{1 - \frac{t}{\hat{s}} \cot^2(\theta/2) }
				\pp{\frac{1 + \xi}{\xi^2}}
				\pp{ F_1^2 + \frac{-t}{4 m^2} F_2^2 },	\\
	\bar{\Sigma}^{\rm BH}_{UU, 2} & 
		= \frac{2t}{\hat{s}}  \frac{1 + \cos\theta}{1 - t / \hat{s}} 
				\pp{ 1 - \frac{1 - \xi^2}{\xi^2} \frac{-t}{4m^2} }
				\pp{ F_1^2 + \frac{-t}{4 m^2} F_2^2 },	\\
	\bar{\Sigma}^{\rm BH}_{LL, 0} & 
		= \bb{ \frac{1 - t / \hat{s}}{\sin^2(\theta/2)} - \frac{\sin^2(\theta/2)}{1 - t / \hat{s}} \pp{ 1 - \frac{t^2}{\hat{s}^2} \cot^4(\theta/2) } }
			(F_1 + F_2) \bb{ F_1 \pp{ \frac{-t}{\xi m^2} - \frac{4 \xi}{1 + \xi} } + \frac{-t}{m^2} \, F_2},	\\
	\bar{\Sigma}^{\rm BH}_{LL, 1} & 
		= - \pp{ \frac{-t}{m \sqrt{\hat{s}} } } \frac{\Delta_T}{m} \frac{\sin\theta}{1 - t / \hat{s}} \pp{1 + \frac{t}{\hat{s}} \cot^2(\theta/2) }
				\pp{F_1 + F_2}
				\pp{ \frac{1 + \xi}{\xi} F_1 + F_2 },	\\
	\bar{\Sigma}^{\rm BH}_{TL, 0} &
		= -\frac{2 \Delta_T}{m} \bb{ \frac{1 - t / \hat{s}}{\sin^2(\theta/2)} - \frac{\sin^2(\theta/2)}{1 - t / \hat{s}} \pp{ 1 - \frac{t^2}{\hat{s}^2} \cot^4(\theta/2) } }
			(F_1 + F_2) \bb{ F_1 - \frac{1 + \xi}{\xi} \, \frac{-t}{4 m^2} \, F_2 }, \\
	\bar{\Sigma}^{\rm BH}_{TL, 1} &
		= \frac{-t}{m \sqrt{\hat{s}} } \frac{2\sin\theta}{1 - t / \hat{s}} \pp{1 + \frac{t}{\hat{s}} \cot^2(\theta/2) }
			(F_1 + F_2) \bb{ F_1 + \pp{ \frac{\xi}{1 + \xi} - \frac{-t}{4 \xi m^2} } F_2 }, \\
	\bar{\Sigma}^{\rm BH}_{TL, 2} &
		= \frac{-t}{m \sqrt{\hat{s}} } \frac{2\sin\theta}{1 - t / \hat{s}} \pp{1 + \frac{t}{\hat{s}} \cot^2(\theta/2) }
			(F_1 + F_2) \bb{ F_1 - \frac{-t}{4 m^2} F_2 }.
\end{align}\ese
Note that we have not made any approximations to the BH amplitude, in contrast to \eq{eq:BH-helicity-amplitudes}.
So compared with the BH part of \eq{eq:dvcs-LP-NLP-xsec}, 
we now have a new azimuthal modulation $\cos2\phi$, 
as expected from the interference of the two $\gamma^*_T$ helicity states.

The $(\bar{\Sigma}^{\rm BH}_{UU}$, $\bar{\Sigma}^{\rm BH}_{LL, 0}$, $\bar{\Sigma}^{\rm BH}_{TL, 0})$ start at the LP of $\sqrt{-t / \hat{s}}$,
and agree with the 
$(\Sigma^{\rm LP}_{UU}$, $\Sigma^{\rm LP}_{UU} A^{\rm LP}_{LL}$, $\Sigma^{\rm LP}_{UU} A^{\rm LP}_{TL})$ 
in \eq{eq:dvcs-pol-LP} at the LP.

The $(\bar{\Sigma}^{\rm BH}_{UU, 1}$, $\bar{\Sigma}^{\rm BH}_{LL, 1}$, $\bar{\Sigma}^{\rm BH}_{TL, 1}$, $\bar{\Sigma}^{\rm BH}_{TL, 2})$ 
start at the NLP of $\sqrt{-t / \hat{s}}$. 
They were organized in \sec{ssec:bh-dvcs} into the NLP part of the total amplitude squared together with the BH-DVCS interference.
It is easy to see that the NLP parts of 
$(\bar{\Sigma}^{\rm BH}_{UU, 1}$, $\bar{\Sigma}^{\rm BH}_{LL, 1}$, $\bar{\Sigma}^{\rm BH}_{TL, 1}$, $\bar{\Sigma}^{\rm BH}_{TL, 2})$ 
agree with BH terms of 
$\Sigma^{\rm LP}_{UU} \cdot (A^{\rm NLP}_{UU}$, $A^{\rm NLP}_{LL}$, $A^{\rm NLP}_{TL, 1}$, $A^{\rm NLP}_{TL, 2})$ 
from \eqs{eq:dvcs-pol-NLP}{eq:pol-asys}.

\subsection{DVCS squared}
\label{ssec:breit-amp2-dvcs}

Although by the power counting argument laid out in Secs.~\ref{sec:dvcs-sdhep} and \ref{sec:calc-sdhep}, 
the DVCS amplitude square formally counts as NNLP and should be considered together with twist-3 GPDs,
we may still calculate it here simply as a cross check between the two frames. 
It is obtained from \eq{eq:DVCS-amp},
\begin{align}
	\overline{|\M^{\rm DVCS}|^2}
	&= \pp{ \frac{e^3}{Q^2} }^2 \, \Tr\bb{ (\gamma \cdot \l') \gamma_{\mu} \frac{\gamma \cdot \l}{2} (1 - P_e \gamma_5) \gamma_{\mu'} } 
		\pp{ -g_{\nu \nu'} } \bb{ \rho^N_{\lambda_N \bar{\lambda}_N} \, T^{\mu\nu}_{\lambda_N \lambda_N'}  \bigpp{ T^{\mu'\nu'}_{\bar{\lambda}_N \lambda_N'} }^* } \nn\\
	&\equiv \frac{e^6}{Q^4} L_{{\rm DVCS}, \mu\mu'} W_{\rm DVCS}^{\mu\mu'},
\label{eq:cov-dvcs2}
\end{align}
where we defined the leptonic tensor,
\begin{align}
	L_{\rm DVCS}^{\mu\mu'}
	&= \Tr\bb{ (\gamma \cdot \l') \gamma^{\mu} \frac{\gamma \cdot \l}{2} (1 - P_e \gamma_5) \gamma^{\mu'} }	\nn\\
	&= - Q^2 \pp{ g^{\mu\mu'} + \frac{q^{\mu} q^{\mu'}}{Q^2} } + 4 \pp{ \l^{\mu} - \frac{1}{2} q^{\mu} } \pp{ \l^{\mu'} - \frac{1}{2} q^{\mu'} }
		-2 i P_e \, \epsilon^{\mu\mu' \l q},
\end{align}
and the hadronic tensor,
\begin{align}
	W_{\rm DVCS}^{\mu\mu'}
	&= \pp{ -g_{\nu \nu'} } 
		\bb{ \rho^N_{\lambda_N \bar{\lambda}_N} \, 
			T^{\mu\nu}_{\lambda_N \lambda_N'}  
			\bigpp{ T^{\mu'\nu'}_{\bar{\lambda}_N \lambda_N'} }^* 
		}	\nn\\
	&= \pp{ -g_{\nu \nu'} } \frac{1}{(2 P \cdot n)^2} 
		\Tr\cc{ 
			\bb{ \pp{ \H \gamma \cdot n - \E \frac{i \sigma^{n \Delta}}{2m} } \tau_1^{\mu\nu}
				- \pp{ \Ht \gamma \cdot n \gamma_5 - \Et \frac{\gamma_5 \Delta \cdot n}{2m} } \wt{\tau}_1^{\mu\nu}
			}
			\frac{\slash{p} + m}{2} \pp{1 + \gamma_5 \Slash{S}}
			\right.\nn\\
			&\left.\hspace{9.7em} \times
			\bb{ \pp{ \H^* \gamma \cdot n + \E^* \frac{i \sigma^{n \Delta}}{2m} } \tau_1^{\mu' \nu'}
				- \pp{ \Ht^* \gamma \cdot n \gamma_5 + \Et^* \frac{\gamma_5 \Delta \cdot n}{2m} } ( - \wt{\tau}_1^{\mu' \nu'} )
			}
			(\gamma \cdot p' + m) 
		} \nn\\
	&\equiv \pp{ -g_{\nu \nu'} } \bb{
			A_{++} \tau_1^{\mu\nu} \tau_1^{\mu' \nu'} 
			+ A_{--} \wt{\tau}_1^{\mu\nu} ( - \wt{\tau}_1^{\mu' \nu'} )
			+ A_{+-} \tau_1^{\mu\nu} ( - \wt{\tau}_1^{\mu' \nu'} )
			+ A_{-+} \wt{\tau}_1^{\mu\nu} \tau_1^{\mu' \nu'} 
		}.
\end{align}
We have organized the $W_{\rm DVCS}^{\mu\mu'}$ as 4 GPD structures associated with 4 tensor contractions. 
The latter give
\begin{align}
	\tau_+^{\mu \mu'}
	&= \pp{ -g_{\nu \nu'} } \tau_1^{\mu\nu} \tau_1^{\mu' \nu'} 
	= \pp{ -g_{\nu \nu'} } \wt{\tau}_1^{\mu\nu} ( - \wt{\tau}_1^{\mu' \nu'} )
	= - \pp{ g_{\mu\mu'} + \frac{q_{\mu} q_{\mu'}}{Q^2} } 
		+ \frac{4 Q^2}{(Q^2 + t)^2} \pp{ \Delta^{\mu} + \frac{\Delta \cdot q}{Q^2} q^{\mu} } \pp{ \Delta^{\mu'} + \frac{\Delta \cdot q}{Q^2} q^{\mu'} },
	\nn\\
	\tau_-^{\mu \mu'}
	&= \pp{ -g_{\nu \nu'} } \tau_1^{\mu\nu} ( - \wt{\tau}_1^{\mu' \nu'} )
	= \pp{ -g_{\nu \nu'} } \wt{\tau}_1^{\mu\nu} \tau_1^{\mu' \nu'} 
	= \frac{2 i}{Q^2 + t} \epsilon^{\mu \mu' q \Delta},
\end{align}
so that we reorganize $W_{\rm DVCS}^{\mu\mu'}$ as
\begin{align}
	W_{\rm DVCS}^{\mu\mu'}
	&= \pp{ A_{++} + A_{--} } \tau_+^{\mu \mu'} + \pp{ A_{+-} + A_{-+} } \tau_-^{\mu \mu'}.
\end{align}
The coefficients $\pp{ A_{+\pm} + A_{-\mp} }$ define the GPD combinations that enter the DVCS amplitude squared, giving
\begin{align}
	A_{++} + A_{--}
	&= (1 - \xi^2) \pp{ | \H |^2 + | \Ht |^2 } - 2 \xi^2 \Re\bigpp{ \H \E^* + \Ht \Et^* } 
		- \pp{ \xi^2 + \frac{t}{4m^2} } | \E |^2
		- \frac{\xi^2 \, t}{4m^2} | \Et |^2	\nn\\
	& \hspace{1em}
		+ \frac{\epsilon^{n P \Delta S}}{m (n \cdot P)} \Im\bigpp{ \H \E^* - \xi \Ht \Et^* }
		- \frac{m^2 n^2}{(n \cdot P)^2} \bb{ \frac{-t}{4m^2} | \H + \E |^2 + \pp{1 - \frac{t}{4m^2} } | \Ht |^2 }, \nn\\
	A_{+-} + A_{-+}
	&= \frac{2m (n \cdot S)}{n \cdot P} 
		\cc{ \xi \Re\bb{ (\H + \E) \pp{ \Ht^* + \frac{t}{4m^2} \Et^* } } 
			- \Re\bb{ \pp{ \H + \frac{t}{4m^2} \E } \Ht^* } 
		}	\nn\\
	& \hspace{1em}
		- \frac{\Delta \cdot S}{m} 
		\cc{ (1 + \xi) \Re\bigpp{ \xi \H \Et^* - \E \Ht^*} + \xi^2 \Re\bigpp{ \E \Et^* }
			+ \frac{m^2 n^2}{(n \cdot P)^2} \Re\bigbb{ (\H + \E) \Ht^* }
		}.
\label{eq:cov-dvcs2-A4}
\end{align}
Both of them contain a term that is proportional to $n^2$ and is power suppressed. 
This vanishes for lightlike $n$, but not in the BMK choice of $n \propto R$. 
The contraction of the leptonic and hadronic tensors is then determined by
\begin{align}
	L_{{\rm DVCS}, \mu\mu'} W_{\rm DVCS}^{\mu\mu'}
	&= \pp{ A_{++} + A_{--} } \bigpp{ L_{{\rm DVCS}, \mu\mu'} \, \tau_+^{\mu \mu'} } 
		+ \pp{ A_{+-} + A_{-+} } \bigpp{ L_{{\rm DVCS}, \mu\mu'} \, \tau_-^{\mu \mu'} }		\nn\\
	&= 2 Q^2
		\bb{ \pp{ A_{++} + A_{--} } \frac{ \P_1^2 + \P_2^2 }{(\P_1 + \P_2)^2} 
			+ P_e \pp{ A_{+-} + A_{-+} } \frac{ \P_1 - \P_2 }{\P_1 + \P_2} 
		}.
\label{eq:cov-LW-dvcs2}
\end{align}

\subsubsection{Specified to the Breit frame}
\label{sssec:check-amp2-dvcs-bmk}
The DVCS squared is the simplest part to be calculated in the Breit frame because there is no $\varphi$ dependence from $\P_1$ and $\P_2$.
To the LP of $(\sqrt{-t}, m) / Q$, it is straightforward to adapt \eq{eq:cov-LW-dvcs2} to the the Breit-frame kinematics,
\begin{align}
	\overline{|\M^{\rm DVCS}|^2} \big|_{\rm BMK}
	&= \frac{e^6}{y^2 Q^2} \bb{ \Sigma_{UU}^{\rm DVCS} + s_{\parallel} P_e \Sigma_{LL}^{\rm DVCS} 
		+ s_{\perp} P_e \Sigma_{TL}^{\rm DVCS} \cos(\varphi - \varphi_S) 
		+ s_{\perp} \Sigma_{TU}^{\rm DVCS} \sin(\varphi - \varphi_S)
	},
\label{eq:cov-dvcs2-bmk}
\end{align}
As \eq{eq:bmk-bh}, it is natural to organize the $\varphi_S$ dependence in the Breit frame 
as trigonometric functions $ \cos(\varphi - \varphi_S) $ and $\sin(\varphi - \varphi_S)$.
The four polarization coefficients are given at LP by
\begin{align}
	\Sigma_{UU}^{\rm DVCS}
	& = \frac{2 \bigpp{ 1 + (1 - y)^2 }}{(2 - x_B)^2} 
		\cc{ 4(1 - x_B) \bigpp{ |\H|^2 + | \Ht |^2 } 
			- 2 x_B^2 \Re\bigpp{ \H \E^* + \Ht \Et^* }
			- \bb{ x_B^2 + (2 - x_B)^2 \frac{t}{4m^2} } | \E |^2
			- \frac{x_B^2 t}{4m^2} | \Et |^2
		}, \nn\\
	\Sigma_{LL}^{\rm DVCS}
	& = \frac{2 y (2 - y)}{(2 - x_B)^2} 
		\cc{ 8(1 - x_B) \Re\bigpp{ \H \Ht^* } 
			- 2 x_B^2 \Re\bigpp{ \H \Et^* + \Ht \E^* }
			- x_B \bb{ x_B^2 + (2 - x_B) \frac{t}{2m^2} } \Re\bigpp{ \E \Et^* }
		}, \nn\\
	\Sigma_{TL}^{\rm DVCS}
	& = \frac{Q K}{m \sqrt{1 - y}} \frac{2 y (2 - y)}{(2 - x_B)^2} 
		\cc{ 2x_B \Re\bigpp{ \H \Et^* } 
			- 2 (2 - x_B) \Re\bigpp{ \E \Ht^* }
			+ x_B^2 \Re\bigpp{ \E \Et^* }
		}, \nn\\
	\Sigma_{TU}^{\rm DVCS}
	& = \frac{Q K}{m \sqrt{1 - y}} \frac{4 \bigpp{ 1 + (1 - y)^2 }}{(2 - x_B)^2} 
		\bb{ (2 - x_B) \Im\bigpp{ \H \E^* }
			- x_B \Im\bigpp{ \Ht \Et^* }
		},
\label{eq:cov-dvcs2-pol}
\end{align}
where we have used $n \propto R$ and taken $\xi \simeq x_B / (2 - x_B)$ to the LP. 
The $K$ in \eq{eq:cov-dvcs2-pol} should also be taken to the LP,
\beq[eq:bmk-LP-K]
	\frac{Q K}{m \sqrt{1 - y}}
	\simeq x_B \sqrt{\frac{1-x_B}{x_B^2} \frac{-t}{m^2} - 1}.
\eeq

These agree with the results in \refcite{Belitsky:2001ns}, namely, in their Eqs.~(26) and (43)--(51),
to the LP for the DVCS amplitude squared,
with errors suppressed by $\order{ (\sqrt{-t}, m) / Q }$ or $\alpha_S$ for the gluon transversity.

We have also verified that using other choices of $n$, such as $q'$ and $\l$, does not affect the above results.

\subsubsection{Specified to the SDHEP frame}
\label{sssec:check-amp2-dvcs-sdhep}
Using the SDHEP kinematics, with $n \propto \l$, we have 
\begin{align}
	&Q^2 = (\hat{s} - t)\cos^2(\theta / 2), \quad
	\l \cdot \Delta = \frac{\hat{s} - t}{2}, \quad
	n^2 = 0, \quad
	n \cdot S = P_N \frac{1 + \xi}{2 \xi}  \frac{\sqrt{ \hat{s} / 2} }{m}, \nn\\
	& \epsilon^{n P \Delta S} = \Delta_T S_T \sin\phi_S \frac{1 + \xi}{2 \xi} \sqrt{ \frac{\hat{s} }{ 2} } , \quad
	\frac{\Delta \cdot S}{2m} = - \frac{\Delta_T}{2m} S_T \cos\phi_S - P_N \pp{ \frac{\xi}{1 + \xi} - \frac{t}{4m^2} } .
\end{align}
Substituting these into \eq{eq:cov-LW-dvcs2}, we obtain $\overline{|\M^{\rm DVCS}|^2}$ in the SDHEP frame,
\begin{align}
	\overline{|\M^{\rm DVCS}|^2} \big|_{\rm SDHEP}
	&= \frac{2 e^6}{\hat{s}} 
		\bb{ \bar{\Sigma}_{UU}^{\rm DVCS} + S_T \, \bar{\Sigma}_{TU}^{\rm DVCS} \sin\phi_S
			+ P_N P_e \, \bar{\Sigma}_{LL}^{\rm DVCS}
			+ S_T P_e \, \bar{\Sigma}_{TL}^{\rm DVCS} \cos\phi_S
		}.
\label{eq:cov-dvcs2-sdhep}
\end{align}
The azimuthal modulations take rather different forms from \eq{eq:cov-dvcs2-bmk},
but the polarization asymmetries are given by the same GPD combinations as \eq{eq:cov-dvcs2-pol},
as we should have expected from \eq{eq:cov-dvcs2-A4},
\begin{align}
	\bar{\Sigma}_{UU}^{\rm DVCS}
	&= \bb{ \frac{1 + \sin^4(\theta/2)}{\cos^6(\theta/2)} }
		\bb{ (1 - \xi^2) \bigpp{ | \H |^2 + | \Ht |^2 } - 2 \xi^2 \Re\bigpp{\H \E^* + \Ht \Et^*} - \pp{ \xi^2 + \frac{t}{4m^2} } | \E |^2 - \frac{\xi^2 t}{4m^2} | \Et |^2 },	\nn\\
	\bar{\Sigma}_{TU}^{\rm DVCS}
	&= \bb{ \frac{1 + \sin^4(\theta/2)}{\cos^6(\theta/2)} }
		\bb{ \frac{ \Delta_T }{m} (1 + \xi) \Im\bigpp{ \H \E^* - \xi \Ht \Et^* } }, \nn\\
	\bar{\Sigma}_{LL}^{\rm DVCS}
	&= \bb{ \frac{1 - \sin^4(\theta/2)}{\cos^6(\theta/2)} }
		\bb{ 2(1 - \xi^2) \Re\bigpp{ \H \Ht^* } - 2 \xi^2 \Re\bigpp{\H \Et^* + \Ht \E^*} - 2\xi \pp{ \frac{\xi^2}{1 + \xi} + \frac{t}{4m^2} } \Re\bigpp{\E \Et^*} }, \nn\\
	\bar{\Sigma}_{TL}^{\rm DVCS}
	&= \bb{ \frac{1 - \sin^4(\theta/2)}{\cos^6(\theta/2)} }
		\pp{ - \frac{ \Delta_T }{m} } \bb{ (1 + \xi) \Re\bigpp{ \xi \H \Et^* - \Ht \E^* } + \xi^2 \Re\bigpp{ \E \Et^* } }.
\end{align}
where we have neglected terms suppressed by $\order{t / \hat{s}}$.

It is easy to verify that starting with the helicity amplitudes in \eq{eq:dvcs-helicity-amplitudes}, we can arrive at the same result as above.
We leave this exercise to the reader.

\subsection{DVCS-BH interference}
\label{ssec:breit-amp2-int}

For the interference term in \eq{eq:amp2-3-parts}, we first organize it as
\begin{align}
	I = 2 \Re{\M^{\rm DVCS} \M^{{\rm BH}*}}
	= \frac{2e^6}{Q^2 t} \Re\pp{ L^{\rho} W_{\rho} + \wt{L}^{\rho} \wt{W}_{\rho} }.
\label{eq:interf-I}
\end{align}
We have defined two leptonic vectors, $L^{\rho}$ and $\wt{L}^{\rho}$, 
which are obtained by combining \eqs{eq:BH-lepton-tensor}{eq:DVCS-amp} 
and contracting with $\tau_1^{\mu\nu}$ or $\wt{\tau}_1^{\mu\nu}$ which are separated from the hadronic sector,
\begin{align}
	\bigcc{ L^{\rho}, \wt{L}^{\rho} }
	&= \cc{ \tau_1^{\mu\nu}, \wt{\tau}_1^{\mu\nu} }
		\Tr{ (\gamma \cdot \l') \gamma_{\mu} \frac{\gamma \cdot \l}{2} (1 - P_e \gamma_5) 
		\bb{ \gamma^{\sigma} \frac{\gamma\cdot (\l - q')}{(\l - q')^2} \gamma^{\rho} 
			+ \gamma^{\rho} \frac{\gamma\cdot (\l + \Delta)}{(\l + \Delta)^2} \gamma^{\sigma} 
		} 
		} (-g_{\nu\sigma}).
\end{align}
The two hadronic vectors are given, respectively, by the unpolarized GPD and polarized GPD,
\begin{align}
	W_{\rho}
	&= \frac{1}{2 P \cdot n} 
		\Tr\cc{ 
			\bb{ \H \gamma \cdot n - \E \frac{i \sigma^{n \Delta}}{2m} }
			\frac{\slash{p} + m}{2} \pp{1 + \gamma_5 \Slash{S}}
			\bb{ F_1(t) \gamma^{\rho} + F_2(t) \frac{i \sigma^{\rho\Delta}}{2m} }
			(\gamma \cdot p' + m) 
		},
	\nn\\
	\wt{W}_{\rho}
	&= \frac{-1}{2 P \cdot n} 
		\Tr\cc{ 
			\bb{ \Ht \gamma \cdot n \gamma_5 - \Et \frac{\gamma_5 \Delta \cdot n}{2m} }
			\frac{\slash{p} + m}{2} \pp{1 + \gamma_5 \Slash{S}}
			\bb{ F_1(t) \gamma^{\rho} + F_2(t) \frac{i \sigma^{\rho\Delta}}{2m} }
			(\gamma \cdot p' + m) 
		}.
\end{align}
They are connected by the virtual photon $\gamma^*(\Delta)$ and satisfy the Ward identity,
\beq
	\Delta_{\rho} \cc{ L^{\rho}, \wt{L}^{\rho}, W^{\rho}, \wt{W}^{\rho} } = 0.
\eeq
This allows us to write them as gauge-invariant forms.

Using the same definition from \eq{eq:cov-gauge-inv-proj}, we organize the leptonic vectors as
\begin{align}
	L^{\rho}
	&= -2 \bb{ \frac{Q^2 - t}{Q^2 + t} \pp{ \P_1^2 + \P_2^2 } + \frac{t}{Q^2} \pp{ 1 + \frac{t}{Q^2} } } \wt{\l}^{\rho}
		+ \frac{2 Q^2}{Q^2 + t} \bb{ (1 - \P_1) (1 - 2 \P_1 \P_2) + \frac{t}{Q^2} (1 + \P_1) } \wt{q}^{\rho}	\nn\\
	&\hspace{1em}
		+ 4 i P_e \, \frac{\P_1 - \P_2}{\P_1 + \P_2} \frac{\epsilon^{\rho \l q \Delta}}{Q^2},
	\nn\\
	\wt{L}^{\rho}
	&= - 2 P_e \bb{ (\P_1 - \P_2) \wt{\l}^{\rho} + \pp{ 1 - \P_1 - \frac{2 \P_1 \P_2}{\P_1 + \P_2} } \wt{q}^{\rho} }
		+ 4 i \pp{ 1 - \frac{2 \P_1 \P_2}{\P_1 + \P_2} } \frac{\epsilon^{\rho \l q \Delta}}{Q^2}.
\label{eq:cov-int-l}
\end{align}
The hadronic vectors are organized as
\begin{align}
	W^{\rho}
	&= 2 P^{\rho} \bb{ \pp{F_1 \H - \frac{t}{4m^2} F_2 \E } + \frac{i}{2m} \frac{\epsilon^{n P \Delta S}}{2 P \cdot n} \pp{ F_1 \E - F_2 \H } }
		+ \frac{t}{2 P \cdot n} \, \wt{n}^{\rho} (F_1 + F_2) (\H + \E)	\nn\\
	&\hspace{1em}
		- \frac{i \Delta \cdot S}{4m (P \cdot n)} \epsilon^{\rho n P \Delta} \pp{ F_1 \E + F_2 \H }
		+ \frac{i}{2m} \epsilon^{\rho P \Delta S} \pp{ F_1 \E - F_2 \H - 2 \xi F_2 \E }
		- \frac{i \, t}{2 m (P \cdot n)} \pp{ \epsilon^{\rho n P S} - \frac{\Delta^{\rho}}{t} \epsilon^{n P \Delta S} }
			(F_2 \E)	\nn\\
	&\hspace{1em}
		- \frac{i \, m}{2 (P \cdot n)} \epsilon^{\rho n \Delta S} \bb{ \pp{ F_1 + \frac{t}{4m^2} F_2 } (\H + \E) + (F_1 + F_2) \pp{ \H + \frac{t}{4m^2} \E } },
	\nn\\
	\wt{W}^{\rho}
	&= 2 \xi m \, \wt{S}^{\rho} (F_1 + F_2) \pp{ \Ht + \frac{t}{4m^2} \Et }
		- 2 P^{\rho} \bb{ \frac{\Delta \cdot S}{2 m}  \pp{ \xi F_1 \Et - (1 + \xi) F_2 \Ht }
			+ \frac{m (n \cdot S)}{P \cdot n}\pp{ F_1 + \frac{t}{4m^2} F_2 } \Ht 
		}	\nn\\
	&\hspace{1em}
		- \pp{ \frac{m (\Delta \cdot S)}{P \cdot n} \wt{n}^{\rho} + \frac{i \epsilon^{\rho n P \Delta}}{P \cdot n} }
			(F_1 + F_2) \Ht.
\label{eq:cov-int-h}
\end{align}
Combining them is then straightforward.

\subsubsection{Specified to the Breit frame}
\label{sssec:check-amp2-int-bmk}
While the BH amplitude in the covariant calculation is exact, the DVCS amplitude is only accurate at the LP, with corrections of order $\order{\sqrt{-t} / Q}$.
So similarly to \sec{sssec:check-amp2-dvcs-bmk}, 
we adapt the kinematics to the Breit frame, take $n \propto R$, and keep only the LP part of the BH-DVCS interference,
\begin{align}
	I \big|_{\rm BMK} &= \frac{8 K e^6}{x y^3 t \, \P_1 \P_2 } 
	\Big\{ \pp{ \Sigma_{UU}^{\rm Int.} + s_{\parallel} P_e \, \Sigma_{LL}^{\rm Int.} } \cos\varphi 
		+ \pp{ P_e \, \Sigma_{UL}^{\rm Int.} + s_{\parallel} \, \Sigma_{LU}^{\rm Int.} } \sin\varphi 
		\nn\\
	&\hspace{6em}
		+ s_{\perp} \bb{ 
			\Sigma_{TU, 1}^{\rm Int.} \sin\varphi \cos(\varphi - \varphi_S)
			+ \Sigma_{TU, 2}^{\rm Int.} \cos\varphi \sin(\varphi - \varphi_S)
		}
		\nn\\
	&\hspace{6em}
		+ s_{\perp} P_e \bb{ 
			\Sigma_{TL, 1}^{\rm Int.} \cos\varphi \cos(\varphi - \varphi_S)
			+ \Sigma_{TL, 2}^{\rm Int.} \sin\varphi \sin(\varphi - \varphi_S)
		}
	\Big\}.
\label{eq:bmk-int}
\end{align}
Similarly to \eq{eq:dvcs-M2-NLP}, this also has eight polarization asymmetries. They are given at LP by
\begin{align}
	\begin{pmatrix}
		\Sigma_{UU}^{\rm Int.} \\
		\Sigma_{UL}^{\rm Int.}
	\end{pmatrix}
	= \begin{pmatrix}
		Y \, \hat{M}_1 \cdot \Re V_\F  \\
		- \bar{Y}  \, \hat{M}_1 \cdot \Im V_\F 
	\end{pmatrix},
	\quad
	&\begin{pmatrix}
		\Sigma_{LL}^{\rm Int.} \\
		\Sigma_{LU}^{\rm Int.}
	\end{pmatrix}
	= \frac{x_B}{2}
	\begin{pmatrix}
		\bar{Y} \, \hat{M}_2 \cdot \Re V_\F  \\
		- Y  \, \hat{M}_2 \cdot \Im V_\F 
	\end{pmatrix},
	\nn\\
	\begin{pmatrix}
		\Sigma_{TL, 1}^{\rm Int.} \\
		\Sigma_{TU, 1}^{\rm Int.}
	\end{pmatrix}
	= x_B \frac{m \sqrt{1-y}}{K Q}
	\begin{pmatrix}
		\bar{Y} \, \hat{M}_3 \cdot \Re V_\F  \\
		- Y  \, \hat{M}_3 \cdot \Im V_\F 
	\end{pmatrix},
	\quad
	&\begin{pmatrix}
		\Sigma_{TL, 2}^{\rm Int.} \\
		\Sigma_{TU, 2}^{\rm Int.}
	\end{pmatrix}
	= x_B \frac{m \sqrt{1-y}}{K Q}
	\begin{pmatrix}
		\bar{Y} \, \hat{M}_4 \cdot \Re V_\F  \\
		Y  \, \hat{M}_4 \cdot \Im V_\F 
	\end{pmatrix}
\end{align}
where we defined $(Y, \, \bar{Y}) = \pp{ 1 + (1-y)^2, \, y (2 - y) }$.
The $V_\F$ is defined in the same way as in \eq{eq:dvcs-pol-M},
and $\hat{M}_i$ is the $i$-th row vector of the matrix $M$ in \eq{eq:coef-matrix},
but with $\xi$ replaced by $\hat{\xi} = x_B / (2 - x_B)$ to the LP.
Again, we note that $K$ should be taken as \eq{eq:bmk-LP-K}.

These results agree with \refcite{Belitsky:2001ns} in their Eqs.~(26) and (43)--(51) to the LP of $(\sqrt{-t}, m) / Q$.
Since the $n$-dependent terms in \eq{eq:cov-int-h} are automatically power suppressed, 
using other choices of $n$, such as $q'$ and $\l$, does not affect the LP result,
as we have explicitly verified.

\subsubsection{Specified to the SDHEP frame}
\label{sssec:check-amp2-int-sdhep}
Since there is no $\phi$ dependence in $\P_1$ and $\P_2$, 
the calculation of \eqs{eq:cov-int-l}{eq:cov-int-h} in the SDHEP frame is straightforward.
Taking $n \propto \l$, we explicitly verify that \eq{eq:interf-I} agrees with the 
BH-DVCS interference part in \eqs{eq:dvcs-M2-NLP}{eq:dvcs-pol-NLP},
up to power-suppressed corrections.

\section{Covariant formulation of the SDHEP observables}
\label{app:covariant}

Apart from $t$, all the SDHEP observables $(t, \xi, \phi_S, \theta, \phi)$ are specifically defined in the diffractive or the SDHEP frame,
but one can still formulate them as Lorentz invariant expressions. This procedure relies on expressing the frame axes 
$(T^{\mu}, X^{\mu}, Y^{\mu}, Z^{\mu})$ in terms of the Lorentz covariant momentum vectors.

First, as remarked in \sec{ssec:trans-boost}, the diffractive frame can differ by an arbitrary boost along the $\hat{z}_{D}$,
which does not affect the values of $\xi$ and $\phi_S$, so we simply choose it as the target rest frame.
The time axis vector $T_{D}^{\mu}$ is simply proportional to $p^{\mu}$, normalized such that $T_{D}^2 = 1$.
The $Z_{D}^{\mu}$ is along the negative direction of $\l^{\mu}$, 
constructed using both $\l^{\mu}$ and $T_{D}^{\mu}$ with the orthonormal condition $T_{D} \cdot Z_{D} = 0$ and $Z_{D}^2 = -1$.
\begin{align}
	T_{D}^{\mu} = \frac{p^{\mu}}{m}, \quad
	Z_{D}^{\mu} = \frac{1}{m} \pp{ p^{\mu} - \frac{2 m^2}{s - m^2} \l^{\mu} },
\label{eq:lab-tz}
\end{align}
This frame naturally has $n = (T_{D} - Z_{D}) / \sqrt{2} \propto \l$, so we define
\beq[eq:xi-cov-def]
	\xi = \frac{\Delta \cdot \l}{2 P \cdot \l}.
\eeq
Similarly, the $X_{D}^{\mu}$ is fixed as the unit vector along the direction of $\Delta$, satisfying $X_{D} \cdot T_{D} = X_{D} \cdot Z_{D} = 0$,
\begin{align}
	X_{D}^{\mu} = \frac{1}{\sqrt{t_0 - t} } \sqrt{ \frac{1 + \xi}{1 - \xi} }
		\bb{ \Delta^{\mu} - \frac{2 \xi}{1 + \xi} \, p^{\mu} - \pp{ t - \frac{4 \xi m^2}{1 + \xi} } \frac{\l^{\mu}}{s - m^2} },
\label{eq:lab-x}
\end{align}
where $t_0 = - 4\xi^2 m^2 / (1 - \xi^2)$.
Finally, the $Y_{D}^{\mu}$ is determined as the remaining vector in the four-dimension space,
\beq
	Y_{D}^{\mu} = \epsilon^{T_D X_D Z_D \mu} = - \epsilon^{\mu\nu\rho\sigma} T_{D, \nu} X_{D, \rho} Z_{D, \sigma}
	= \frac{2}{\sqrt{t_0 - t} } \sqrt{ \frac{1 + \xi}{1 - \xi} } \frac{\epsilon^{\mu p p' \l}}{s - m^2}.
\eeq
These vectors satisfy 
\beq[eq:lab-complete]
	g^{\mu\nu} = T_{D}^{\mu} T_{D}^{\nu} - X_{D}^{\mu} X_{D}^{\nu} - Y_{D}^{\mu} Y_{D}^{\nu} - Z_{D}^{\mu} Z_{D}^{\nu}
\eeq
and take the simple forms in the target rest frame,
\beq[eq:lab-specific]
	 T_{D}^{\mu} = (1, 0, 0, 0)_{\rm c}, \quad
	 X_{D}^{\mu} = (0, 1, 0, 0)_{\rm c}, \quad
	 Y_{D}^{\mu} = (0, 0, 1, 0)_{\rm c}, \quad
	 Z_{D}^{\mu} = (0, 0, 0, 1)_{\rm c}.
\eeq

In this frame, the nucleon spin vector is simply 
$S^{\mu} = (0, \bm{S}_T, P_N)_{\rm c} = (0, S_T \cos\phi_S, S_T \sin\phi_S, P_N)_{\rm c}$
with the values of $S_T$ and $P_N$ to be given as an experimental input.
We can then express $\phi_S$ as
\beq
	S \cdot X_{D} = - S_T \cos\phi_S, \quad
	S \cdot Y_{D} = - S_T \sin\phi_S,
\eeq
so that
\beq
	\cos\phi_S = - \frac{1}{S_T}
		\frac{1}{\sqrt{t_0 - t} } \sqrt{ \frac{1 + \xi}{1 - \xi} }
		\bb{ \Delta\cdot S - \pp{ t - \frac{4 \xi m^2}{1 + \xi} } \frac{\l\cdot S}{s - m^2} }, \quad
	\sin\phi_S = - \frac{1}{S_T} 
		\frac{2}{\sqrt{t_0 - t} } \sqrt{ \frac{1 + \xi}{1 - \xi} } \frac{\epsilon^{S p p' \l}}{s - m^2},
\eeq
where we have used $p \cdot S = 0$.

Similarly, the SDHEP frame, chosen as the c.m.\ frame of the $2\to2$ hard collision, can also be expressed in a covariant form.
The time axis vector $T_{S}^{\mu}$ can be simply chosen as
\beq
	T_{S}^{\mu} = \frac{\Delta^{\mu} + \l^{\mu}}{\sqrt{\hat{s}}},
\eeq
where using the $\xi$ defined in \eq{eq:xi-cov-def},
\beq[eq:shat-sdhep]
	\hat{s} = (\Delta + \l)^2 = t + 2 \Delta \cdot \l = t + \frac{2\xi (s - m^2)}{1 + \xi},
\eeq 
Then we construct $Z_{S}^{\mu}$ along the direction of $\Delta$ and $X_{S}^{\mu}$ along the negative direction of $P$,
\beq
	Z_{S}^{\mu} = \frac{1}{\sqrt{\hat{s}}} \pp{ \Delta^{\mu} - \frac{\hat{s} + t}{\hat{s} - t} \, \l^{\mu} }, \quad
	X_{S}^{\mu} 
		= \frac{-2 \xi}{\sqrt{(1 - \xi^2) (t_0 - t)}} \pp{ P^{\mu} - \frac{\Delta^{\mu}}{2\xi} + \frac{t}{s - m^2} \frac{1 + \xi}{2 \xi^2} \l^{\mu} }.
\eeq
Then the $Y_{S}^{\mu}$ is simply
\begin{align}
	Y_{S}^{\mu} &= \epsilon^{T_{S} X_{S} Z_{S} \mu} 
		= - \epsilon^{\mu\nu\rho\sigma} T_{{S}, \nu} X_{{S}, \rho} Z_{{S}, \sigma} 
		= \sqrt{ \frac{1 + \xi}{1 - \xi} } \frac{2 \epsilon^{\mu \l P \Delta}}{ (s-m^2) \sqrt{t_0 - t} }.
\end{align}
The $(T_{S}, X_{S}, Y_{S}, Z_{S})$ satisfy the same property as \eq{eq:lab-complete} 
and take the same forms as \eq{eq:lab-specific} in the SDHEP frame.

One shall first notice that in the SDHEP frame, we also have $n = (T_{S} - Z_{S}) / \sqrt{2} \propto \l$, so 
the $\xi$ is the same as \eq{eq:xi-cov-def}, confirming our conclusion in \sec{ssec:trans-boost}.

In the SDHEP frame, the $\l'$ can be written as
\beq
	\l^{\prime \mu} 
	= \l^{\prime 0} \pp{ T_{S}^{\mu} + \sin\theta \cos\phi \, X_{S}^{\mu} + \sin\theta \sin\phi \, Y_{S}^{\mu} + \cos\theta \, Z_{S}^{\mu} },
\eeq
where $\l^{\prime 0} = \l' \cdot T_{S} = \sqrt{\hat{s}} / 2$.
The angles $(\theta, \phi)$ of $\l'$ can then be expressed in a Lorentz covariant form,
\begin{align}
	\cos\theta = - \frac{\l' \cdot Z_{S}}{\l' \cdot T_{S}}, \quad
	\sin\theta \cos\phi = - \frac{\l' \cdot X_{S}}{\l' \cdot T_{S}}, \quad
	\sin\theta \sin\phi = - \frac{\l' \cdot Y_{S}}{\l' \cdot T_{S}},
\end{align}
which gives
\begin{align}
	\cos\theta & = - 2 \frac{(\hat{s} - t) \Delta \cdot \l' - (\hat{s} + t) \l \cdot \l'}{\hat{s} (\hat{s} - t) }, \nn\\
	\cos\phi & =	\frac{\xi (s - m^2) \pp{ 2 \xi P \cdot \l' - \Delta \cdot \l'} + (1 + \xi) (\l \cdot \l') t
			}{ (1 + \xi)\sqrt{(1 - \xi^2) \, \hat{s}\, (t_0 - t) (\Delta \cdot \l' - t / 2)}}, \nn\\
	\sin\phi & = - \frac{2 \xi \epsilon^{\l' \l P \Delta}}{ \sqrt{(1 - \xi^2) \, \hat{s}\, (t_0 - t) (\Delta \cdot \l' - t / 2)} }.
\end{align}

\newpage
\bibliographystyle{apsrev}
\bibliography{reference}

\end{document}